\documentclass[superscriptaddress,prb,amsmath,twocolumn,amssymb]{revtex4}

\usepackage{dcolumn}
\usepackage{amsfonts}
\usepackage{amssymb}
\usepackage{amsmath}
\usepackage{amsthm}
\usepackage{latexsym}
\usepackage{epsfig}
\usepackage{color}
\usepackage{lscape}
\usepackage{multirow}
\usepackage{braket}
\usepackage{bbm}
\usepackage{mathrsfs}
\usepackage{fancyhdr}
\usepackage{lastpage}
\usepackage{soul}
\usepackage{subfigure}
\usepackage{longtable}
\usepackage[noend]{algpseudocode}
\usepackage{algorithmicx, algorithm}
\usepackage{diagbox}
\usepackage{bm}
\usepackage{placeins}

\newtheorem{theorem}{Theorem}       
\newtheorem{lemma}{Lemma}

\newcommand{\be}{\begin{equation}}
\newcommand{\ee}{\end{equation}}
\newcommand{\ba}{\begin{array}}
\newcommand{\ea}{\end{array}}
\newcommand{\bea}{\begin{eqnarray}}
\newcommand{\eea}{\end{eqnarray}}

\makeatletter
\def\@author@present@script#1#2{%
  \begingroup
  \gdef\comma@space{\textsuperscript{,\,}}%
  \doauthor{#1}{#2}{}%
  \endgroup
  \advance\@tempcnta\m@ne
}
\makeatother

\renewcommand{\bibpreamble}{
  ~~~$^\dagger$ These authors contributed equally to this work.
  \par
  ~~~$^\ast$ Electronic address: qiangxiaogang@gmail.com
    \par
  ~~~$^\sharp$ Electronic address: wqliu@stdu.edu.cn
  \par
  ~~~$^\ddagger$ Electronic address: weizhaohui@gmail.com
  \par
  ~~~$^\S$ Electronic address: zqyin@bit.edu.cn
}

\begin{document}

\let\oldaddcontentsline\addcontentsline
\renewcommand{\addcontentsline}[3]{}

\title{Universal Quantum Particle Sampling with a Programmable Photonic Processor}

\author{Xiaogang Qiang$^{1,  \dagger, \ast}$}
\affiliation{National Innovation Institute of Defense Technology, AMS, 100071 Beijing, China}

\author{Lijinzhi Lin$^{2,  \dagger}$}
\affiliation{Department of Computer Science and Technology, Tsinghua University, Beijing 100084, China.}

\author{Wenqiang Liu$^{3, 6, \sharp}$}
\affiliation{Department of Mathematics and Physics, Shijiazhuang Tiedao University, Shijiazhuang 050043, China}

\author{Chuanyi Lu$^{1}$}
\affiliation{National Innovation Institute of Defense Technology, AMS, 100071 Beijing, China}

\author{Shixin Ma$^{1}$}
\affiliation{National Innovation Institute of Defense Technology, AMS, 100071 Beijing, China}

\author{Peixiang Li$^{1}$}
\affiliation{National Innovation Institute of Defense Technology, AMS, 100071 Beijing, China}

\author{Yang Wang$^{1}$}
\affiliation{National Innovation Institute of Defense Technology, AMS, 100071 Beijing, China}

\author{Jie Huang$^{1}$}
\affiliation{National Innovation Institute of Defense Technology, AMS, 100071 Beijing, China}

\author{Zhaohui Wei$^{4,5,\ddagger}$}
\affiliation{Yau Mathematical Sciences Center, Tsinghua University, Beijing 100084, China.}
\affiliation{Yanqi Lake Beijing Institute of Mathematical Sciences and Applications, Beijing 101407, China.}

\author{Zhangqi Yin$^{6,7,\S}$}
\affiliation{Center for Quantum Technology Research and Key Laboratory of Advanced Optoelectronic Quantum Architecture and Measurements (MOE), School of Physics, Beijing Institute of Technology, Beijing 100081, China}
\affiliation{Beijing Key Laboratory of Quantum Matter State Control and Ultra-Precision Measurement Technology, School of Physics, Beijing Institute of Technology, Beijing 100081, China}

\date{\today}

\begin{abstract}

Particle-exchange statistics govern multiparticle quantum interference among indistinguishable particles. Sampling-based quantum advantage has so far been explored primarily through bosonic interference, while the potential of more general exchange statistics as a resource for quantum sampling remains largely unexplored. Here we present a universal quantum particle sampling framework that unifies bosonic, fermionic and anyonic sampling with the continuously tunable exchange phase and particle indistinguishability. 
We show that classically reproducing the sampling distribution is intractable for generic anyonic sampling in the fully indistinguishable regime, extending the complexity-theoretic foundations of boson sampling to a broader family of particle statistics. 
Experimentally, we develop an entanglement-based architecture that renders effective particle-exchange statistics programmable, with independent tunability of the exchange phase and particle indistinguishability. Using a large-scale silicon photonic processor comprising more than 2,000 integrated components, we demonstrate universal three-particle sampling across bosonic, anyonic and fermionic statistics, and statistically distinguish the measured distributions from uniform, fully distinguishable and alternative particle-statistics hypotheses. Building on this programmable sampling capability, we use quantum particle sampling to estimate nonequilibrium quantum work distributions for particles with various exchange statistics, revealing how exchange statistics reshape quantum thermodynamic fluctuations. These results establish particle-exchange statistics as a programmable resource for quantum sampling and provide a platform for exploring generalized particle statistics, computational complexity and statistics-dependent many-body physics.

\end{abstract}

\maketitle

\setcounter{equation}{0}
\setcounter{figure}{0}
\setcounter{table}{0}
\setcounter{page}{1}
\setcounter{section}{0}
\makeatletter
\renewcommand{\theequation}{\arabic{equation}}
\renewcommand{\thefigure}{\arabic{figure}}
\renewcommand{\bibnumfmt}[1]{[#1]}
\renewcommand{\citenumfont}[1]{#1}
\normalsize

\section*{I. Introduction}

Quantum sampling has emerged as a leading route to demonstrating quantum computational advantage~\cite{aaronson2011computational,bouland2019ComplexityVerification,arute2019quantum,gao2025establishing,madsen2022quantum,liu2026gaussian}. 
In these protocols, a quantum processor draws samples from an output probability distribution generated by a high-dimensional many-body evolution, exploiting interference patterns that are hard to reproduce classically without requiring fault-tolerant universal quantum computation. 
Random circuit sampling and photonic quantum sampling have been mainly realized on 
superconducting and optical platforms, respectively~\cite{arute2019quantum,gao2025establishing,madsen2022quantum,liu2026gaussian}. 
Boson sampling, in particular, relates the output probabilities of indistinguishable bosons propagating through a linear interferometer to matrix permanents, establishing a complexity-theoretic foundation for classically hard sampling with intermediate-scale photonic systems~\cite{aaronson2011computational,bouland2019ComplexityVerification}.

Quantum sampling based on multiparticle interference has focused primarily on bosonic statistics, while the potential of more general particle-exchange
statisitcs as a sampling resource remains largely unexplored. Exchange statistics govern the coherent addition of many-particle amplitudes:
bosonic paths add symmetrically, fermionic paths antisymmetrically, and anyonic exchange in low-dimensional systems contributes nontrivial fractional 
phases~\cite{leinaas1977theory,wilczek1982quantum,haldane1991fractional,nakamura2020direct}. 
Transforming this intrinsic particle property into a programmable degree of freedom would extend quantum sampling beyond the bosonic case. Bosonic, fermionic and anyonic sampling can thus be viewed as members of a broader family governed by particle statistics. 
Fermionic sampling amplitudes are associated with determinants and efficiently computable, whereas anyonic sampling with fractional exchange phases produce more general phase-weighted sums over multiparticle interference paths. 
Partial distinguishability determines the coherent visibility of multiparticle interference and controls the crossover from fully coherent quantum sampling to distinguishable classical-particle propagation~\cite{moylett2018quantum, oh2025recent}. 
Particle exchange statistics and indistinguishability therefore substantially affect the sampling complexity, and yet existing theories and experiments have 
largely treated these regimes separately~\cite{kuo2022boson,oszmaniec2022fermion,moylett2018quantum}.

Experimental access to anyonic statistics has followed two complementary routes. Fractional quantum Hall devices probe specific anyon models and quasiparticle interference~\cite{leinaas1977theory,wilczek1982quantum,haldane1991fractional,nakamura2020direct}, whereas quantum simulators emulate anyonic statistics and few-particle braiding processes in engineered systems~\cite{nakamura2020direct,google2023non}. 
Entanglement-based photonic schemes have also shown that correlated measurements of entangled photons across identical copies of a process can reproduce bosonic, fermionic and intermediate exchange statistics~\cite{sansoni2012twoparticle,matthews2013observing,qiang2021implementing}.
These advances establish controllable signatures of exchange statistics, but generally target specific braiding observables or few-particle correlations rather than a unified sampling framework that combines continuously tunable exchange phase, particle indistinguishability, and arbitrary linear transformations. Implementing an experimental system capable of such capability is also highly challenging. Programmable integrated photonic processors provide the high-dimensional transformations and multiphoton control needed to address this gap~\cite{carolan2015universal, qiang2021implementing, qiang2018large}.

\begin{figure*}[!t]
    \centering
    \includegraphics[width=1.0\textwidth]{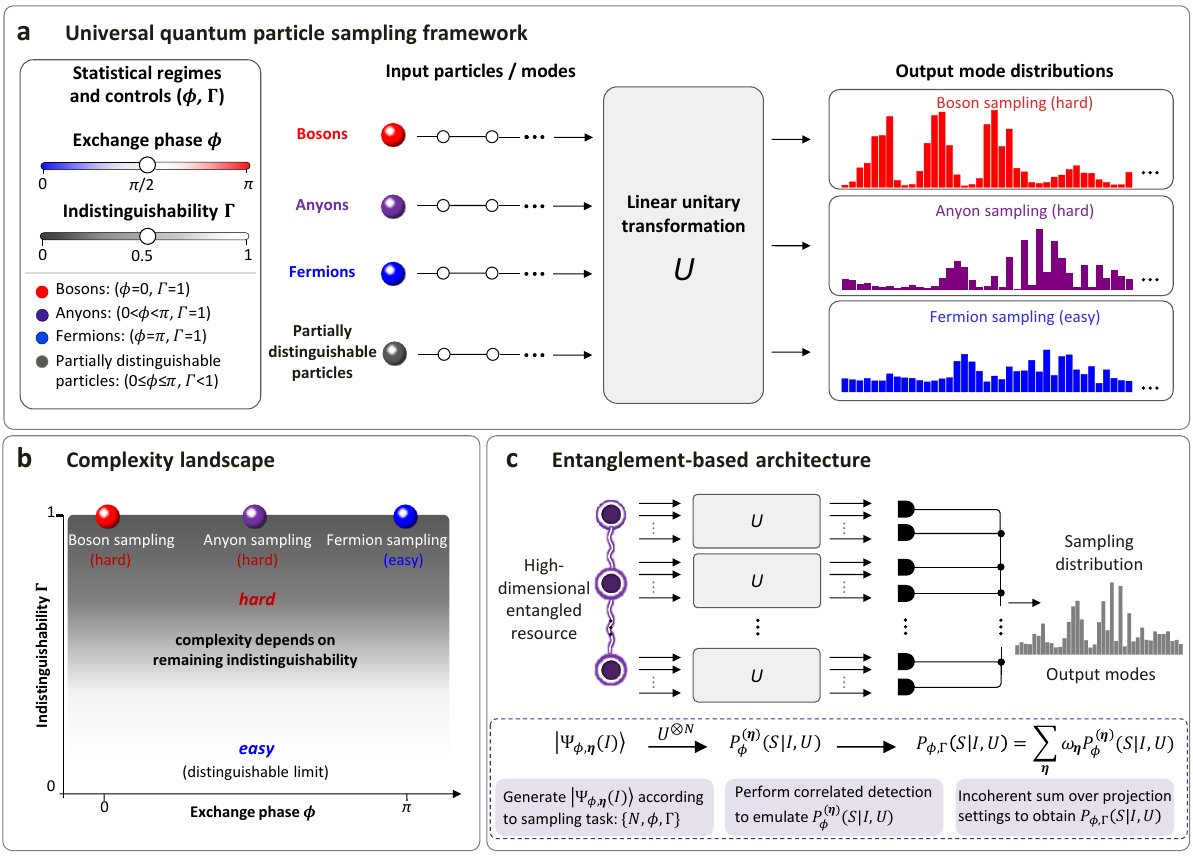}
    \caption{
    \textbf{Universal quantum particle sampling framework.}
    \textbf{a,} Schematic of the universal quantum particle sampling framework. 
    Input particles are described by two programmable parameters: the exchange phase $\phi$ and the global indistinguishability $\Gamma$. After a linear unitary transformation $U$, different particle statistics give rise to distinct output distributions. 
    \textbf{b,} The complexity landscape of quantum particle sampling. Fully indistinguishable bosons, $(\phi,\Gamma)=(0,1)$, and generic fully indistinguishable anyons, $0<\phi<\pi$ and $\Gamma=1$, correspond to classically hard sampling regimes, whereas the fully indistinguishable fermions, $(\phi,\Gamma)=(\pi,1)$, is efficiently simulable through determinants. 
    The distinguishable-particle limit $\Gamma=0$ is classically tractable, while the complexity of partially distinguishable regimes depends on the remaining indistinguishability.
    \textbf{c,} Entanglement-based architecture for implementing universal quantum particle sampling. 
    A high-dimensional entangled resource coherently links multiple copies of the same linear transformation $U$. 
    Correlated, projection-resolved detection across corresponding output modes generates the desired many-particle sampling distribution (see Section S3). 
    }
    \label{fig:uqps_framework}
\end{figure*}

Here we present a universal quantum particle sampling framework that unifies bosonic, fermionic and anyonic sampling within a single theoretical structure. 
The framework is parameterized by an exchange phase $\phi$ and a global indistinguishability parameter $\Gamma$, defined as the mean pairwise 
indistinguishability of the particles. The choices $\phi=0$ and $\phi=\pi$ recover bosonic and fermionic interference, respectively, whereas $0 < \phi < \pi$ describes fractional exchange statistics. We further prove that, under standard complexity-theoretic assumptions~\cite{aaronson2011computational}, generic anyon sampling in the fully indistinguishable regime is classically intractable, extending the complexity foundations of boson sampling to a broader space of particle-exchange statistics.

To realize this framework, we implement a programmable processor for universal quantum particle sampling on a large-scale silicon photonic chip. 
The photonic processor uses an entanglement-based architecture, which enables continuous tunability of the exchange phase $\phi$ and global indistinguishability $\Gamma$. Together with the reconfigurable linear-optical networks, we experimentally demonstrate universal three-particle sampling across bosonic, anyonic and fermionic statistics. The sampled distributions are also statistically validated against uniform, fully distinguishable, bosonic, fermionic and alternative-anyonic hypotheses. With this programmable sampling capability, we further employ quantum particle sampling to estimate nonequilibrium work distributions for bosonic, fermionic and anyonic particles, revealing how exchange statistics reshape quantum thermodynamic fluctuations~\cite{gong2014interference,myers2021thermodynamics}. 
Together, these results establish particle-exchange statistics as a programmable resource for quantum sampling and provide a photonic platform for exploring generalized particle statistics, computational complexity and statistics-dependent many-body physics.

\section*{II. Universal quantum particle sampling framework and complexity}

We formulate universal quantum particle sampling with two physical parameters that are tied to particle species: the phase acquired under exchange particles, described by an exchange phase $\phi \in [0, \pi]$, and the coherence between different particle assignments, described by the global indistinguishability $\Gamma \in [0,1]$ (Fig. 1\textbf{a}). Here $\Gamma$ is defined as the mean of the pairwise indistinguishability $\gamma_{\alpha \beta}=|\braket{\chi_{\alpha} | \chi_{\beta}}|^2$ of particles with the pure internal states $\ket{\chi_\alpha}$ and $\ket{\chi_\beta}$. $\Gamma$ provides a compact coordinate for visualizing the overall quantum-to-classical crossover, where $\Gamma=1$ corresponds to fully indistinguishable particles, and $\Gamma = 0$ corresponds to mutually distinguishable particles.

We first consider the fully indistinguishable regime. For $N$ particles injected into a linear network $U$, 
let $I=\{i_1,\ldots,i_N\}$ and $S=\{s_1,\ldots,s_N\}$ denote ordered input and output configurations, and let $A=U_{S,I}$ be the corresponding $N \times N$ submatrix of $U$. The exchange-statistics-weighted many-particle amplitude is then
\begin{align}
\mathcal{A}_{\phi}(S|I,U) = \sum_{\sigma\in S_N}
e^{i\phi \ell(\sigma)}
\prod_{j=1}^{N}
A_{j,\sigma(j)} ,
\end{align}
where $S_N$ is the permutation group and $\ell(\sigma)$ is the inversion number of $\sigma$, equivalently the minimum number of adjacent transpositions required to transform $\sigma$ into the natural ordering. 
The corresponding ordered-output probability is
\begin{align}
P_{\phi}(S|I,U)=
\left|
\mathcal{A}_{\phi}(S|I,U)
\right|^2 .
\end{align}
Probabilities for occupation-number outcomes are obtained by combining equivalent ordered configuration with the appropriate normalization, as described in Section S1. 

Equation (1) makes explicit how exchange statistics reshape multiparticle interference. For $\phi =0$, all permutation amplitudes enter the coherent sum with 
the same phase and $\mathcal{A}_{0}={\rm Per}(A)$, recovering boson sampling. For $\phi =\pi$, the exchange factor becomes $e^{i\pi \ell(\sigma)} = {\rm sgn(\sigma)}$, yielding $\mathcal{A}_{\pi}={\rm det}(A)$ and the fermionic sampling limit. 
Intermediate exchange phases, $0<\phi<\pi$, define the Abelian anyonic sampling model considered here through an exchange-phase-weighted matrix function interpolating between the permanent and determinant. 
Considering that Abelian exchange statistics is represented by a scaler phase~\cite{greiter2024fractional} and in non-Abelian braiding exchanges implement generally noncommuting unitary transformations within a multi-anyon fusion space~\cite{nayak2008non}, we focus here on the sampling of Abelian anyons.

Partial distinguishability suppresses the coherence between different permutation contributions. We first resolve the particles' internal states into mutually orthogonal projection settings $\boldsymbol{\eta}$, each defining a projection-resolved distribution $P_{\phi}^{(\boldsymbol{\eta})}(S|I,U)$. The partially distinguishable distribution is then obtained by incoherently summing over these mutually exclusive projection settings, weighted by their respective probabilities. Within this description, the projection setting weights encode the pairwise indistinguishablities $ \gamma_{\alpha \beta} $, while $\Gamma$ denotes the average of $\gamma_{\alpha \beta}$ and thus provides a convenient one-dimensional measure of the overall degree of multiparticle indistinguishability. A more general description of partial distinguishability in terms of the full pairwise internal-state overlap matrix is given in Section S1.

The resulting parameter space contains several distinct complexity regimes (Fig. 1\textbf{b}). At $(\phi, \Gamma) = (0,1)$, the model reduces to standard boson sampling, whose classical intractability is associated with matrix permanents under standard complexity-theoretic assumptions. At $(\phi, \Gamma) = (\pi,1)$, the amplitudes reduce to determinants and are efficiently computable, providing a natural tractable reference. At $\Gamma=0$, coherence between different particle assignments is absent, rendering the exchange phase unobservable and recovering the efficiently simulable distinguishable-particle limit.

For fully indistinguishable anyonic particles, we show that classically reproducing the sampling distribution is intractable for generic exchange phases~\cite{aaronson2011computational}, since the output probabilities can also be related to the computation of matrix permanents (see Section S2). 
This extends the complexity-theoretic foundations of boson sampling from the bosonic case to a broader family of Abelian exchange statistics. For $0<\Gamma<1$, however, the complexity depends on the complete distinguishability structure, its scaling with particle number and the allowed approximation error. 
The hardness may be robust against sufficiently small distinguishability noise, although this remains an open theoretical question.

\begin{figure*}[t]
    \centering
    \includegraphics[width=1.0\textwidth]{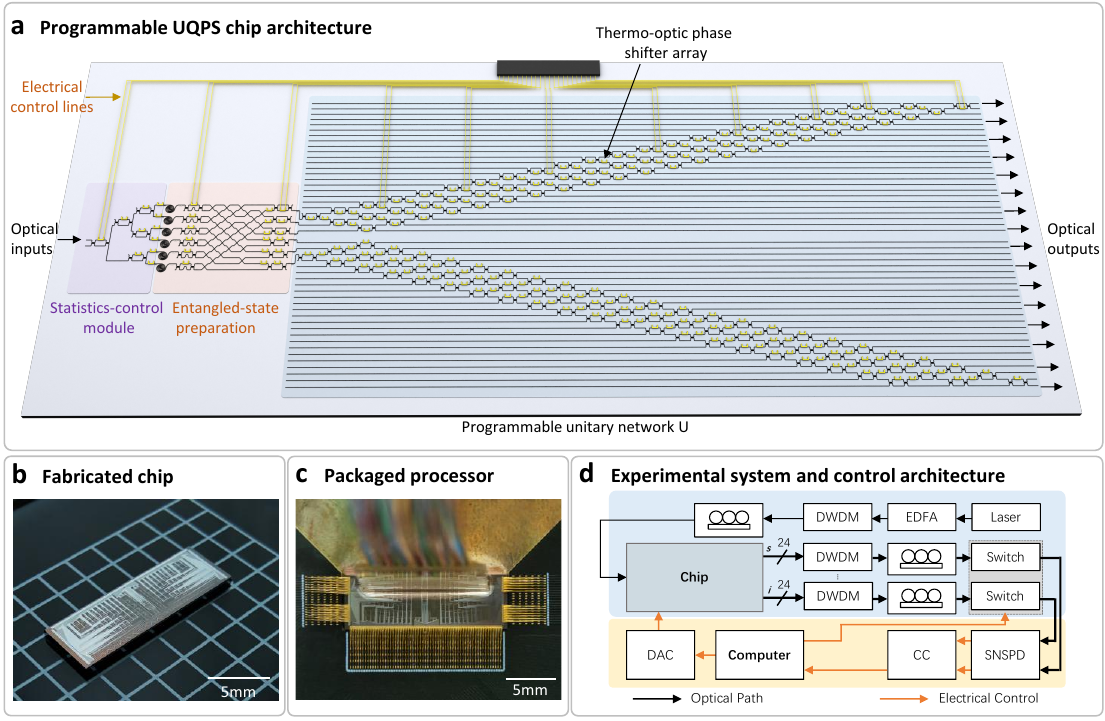}
    \caption{
    \textbf{Programmable silicon photonic processor for universal quantum particle sampling.}
    \textbf{a,} Schematic of the programmable silicon photonic chip. 
    The processor integrates entangled-state generation, statistics-control module and two large-scale reconfigurable unitary networks $U$. 
    Thermo-optic phase shifters configure the optical transformation and the particle-statistics parameters used in the sampling experiments.
    \textbf{b,} Photograph of the fabricated silicon photonic chip. 
    \textbf{c,} Photograph of the packaged silicon photonic chip.  
    The chip is mounted on a printed circuit board, with a fiber array providing optical input and output coupling and wire-bonded electrical connections enabling external voltage-driven programming.
    \textbf{d,} Schematic of the experimental system and control architecture.     
    Optical pumping and conditioning modules drive the photonic chip, while software-defined settings $(U,\phi,\Gamma)$ are converted into multi-channel electrical control signals. Output photons are detected by a superconducting nanowire single-photon detector array, time tagged and processed through coincidence counting to obtain the sampling distribution. For further details of the device and system see Section S4.
    }
    \label{fig:chip_system}
\end{figure*}

\section*{III. Programmable photonic processor for universal quantum particle sampling}

We implement the universal sampling framework by mapping the permutation sum in Eq. (1) onto correlated measurements of a high-dimensional entangled state evolving through identical copies of the programmed transformation $U$ (Fig. 1\textbf{c}). This architecture builds on entanglement-based emulation of exchange symmetry~\cite{qiang2021implementing,matthews2013observing,sansoni2012twoparticle}, and extends it from selected bosonic or fermionic correlation signatures to complete, programmable sampling distributions. It provides an implementation of projection-resolved sampling, where the projection settings are measured separately, and their probabilities can subsequently be combined to obtain the corresponding partially distinguishable-particle distribution. The linear transformation $U$, exchange phase $\phi$ and global indistinguishability $\Gamma$ can therefore be configured independently within a common photonic architecture.

For $N$ particles with input configuration $I=\{i_1,…,i_N\}$, we prepare the statistics-encoded permutation state
\begin{align}
|\Psi_{\phi}^{(N)} (I)\rangle=
\frac{1}{\sqrt{N!}}
\sum_{\sigma\in S_N}
e^{i\phi\ell(\sigma)}
\prod_{r=1}^{N}
\hat{a}_{i_{\sigma(r)}}^{(r)\dagger}
|0\rangle ,
\end{align}
where $r$ labels an identical copy of $U$, and $\hat{a}_{i}^{(r)\dagger}$ creates a photon in input mode $i$ of copy $r$. In the fully indistinguishable regime, 
$\Gamma=1$, all permutation components remain coherent. After propagation through $U^{\otimes N}$, an $N$-fold correlated detection event at 
$S=\{s_1,\ldots,s_N\}$, with one output selected from each copy, has an amplitude proportional to $\mathcal{A}_{\phi}(S|I,U)$. The relative phases of the permutation components therefore program $\phi$, while the measured correlated probabilities reproduce the target sampling distribution after normalization and the appropriate aggregation of equivalent ordered outputs.

Partial distinguishability is implemented through state preparation and projection-resolved detection rather than by changing the intrinsic exchange statistics of the photons. Orthogonal internal-state projection settings $\boldsymbol{\eta}$ are measured independently, yielding projection-setting-conditioned distributions $P_{\phi}^{(\boldsymbol{\eta})}(S|I,U)$. Their incoherent sum produces the desired partially distinguishable distribution. By programming the relative weights and coherences of these projection settings, we control the pairwise indistinguishabilities $\gamma_{\alpha \beta}$ and hence their global mean $\Gamma$. The fully coherent setting recovers $\Gamma =1$, whereas suppressing cross-projection-setting coherence drives the distribution towards the distinguishable-particle limit. The explicit mapping between the entangled-state parameters, projection settings and $ \gamma_{\alpha \beta}$ is derived in Section S3.

The photonic processor comprises two functional modules: programmable high-dimensional entangled-state generation and reconfigurable linear-optical transformation (Fig. 2\textbf{a}). Here we use a compressed implementation of the entanglement-based scheme for the three-particle experiments. 
For each output of an ordered output triple $(r, p, q)$, say $q$, 
the contribution associated with one copy of the ideal three-particle transformation is compiled into the amplitudes of a two-photon six-dimensional entangled state. For an input configuration $I=(j,k,l)$, this state is 
\begin{align}
\ket{\psi_q} = 
(\alpha_0^{(q)} \hat{a}_k^\dag \hat{b}_j^\dag &+\alpha_1^{(q)} \hat{a}_j^\dag \hat{b}_k^\dag +\alpha_2^{(q)} \hat{a}_l^\dag \hat{b}_j^\dag \\ \nonumber
 &+\alpha_3^{(q)} \hat{a}_j^\dag \hat{b}_l^\dag+\alpha_4^{(q)} \hat{a}_l^\dag \hat{b}_k^\dag+\alpha_5^{(q)} \hat{a}_k^\dag \hat{b}_l^\dag )\ket{0},
\end{align}
where $\hat{a}^\dag$ and $\hat{b}^\dag$ act on the two physical photons propagating through the remaining two copies of $U$. The coefficients $\alpha_n^{(q)}$ incorporate the transfer amplitudes associated with the fixed index $q$, the programmed exchange phase and the selected projection setting; their explicit expressions are given in Section S3. 
Two-photon coincidences measured after the two copies of $U$ yield the $q$-resolved slice of the target three-particle distribution, up to a known normalization. Repeating the experiment over all $q$ reconstructs the complete distribution. The compression therefore changes the hardware encoding, rather than the target sampling distribution.

The $15\times5 {\rm mm}^2$ silicon chip integrates more than 2,000 photonic and electrical components, including six spontaneous four wave mixing sources, six on-chip filters, 297 thermo-optic phase shifters, 298 multimode-interferometer beam splitters, grating couplers, waveguide crossings, length-matching optical delay lines, together with the electrical routing structures required for programmable operation (Fig. 2\textbf{b},\textbf{c}). The source module prepares the required high-dimensional entangled states, while the transformation module implements the reconfigurable networks used for correlated sampling. Further details of the fabrication and component-level characterization of the chip are provided in Section S4.

The chip is fibre-array coupled for optical input and output and wire-bonded to a printed circuit board for multichannel electrical control. External voltage-driving electronics configure both the entangled-state generation and the programmable networks, while superconducting nanowire single-photon detectors and time-tagging electronics record the correlated output events (Fig. 2\textbf{d}). A software control layer compiles the target transformation, exchange phase, indistinguishability setting and projection setting into heater voltages and acquisition sequences. Independent calibration confirms high-fidelity implementation of the programmed linear transformations and supports rapid reconfiguration (up to $\sim$kHz) between sampling tasks, providing the experimental basis for the universal particle-sampling demonstrations described below.

\begin{figure*}[htbp!] 
\centering
\includegraphics[width=1.0\textwidth]{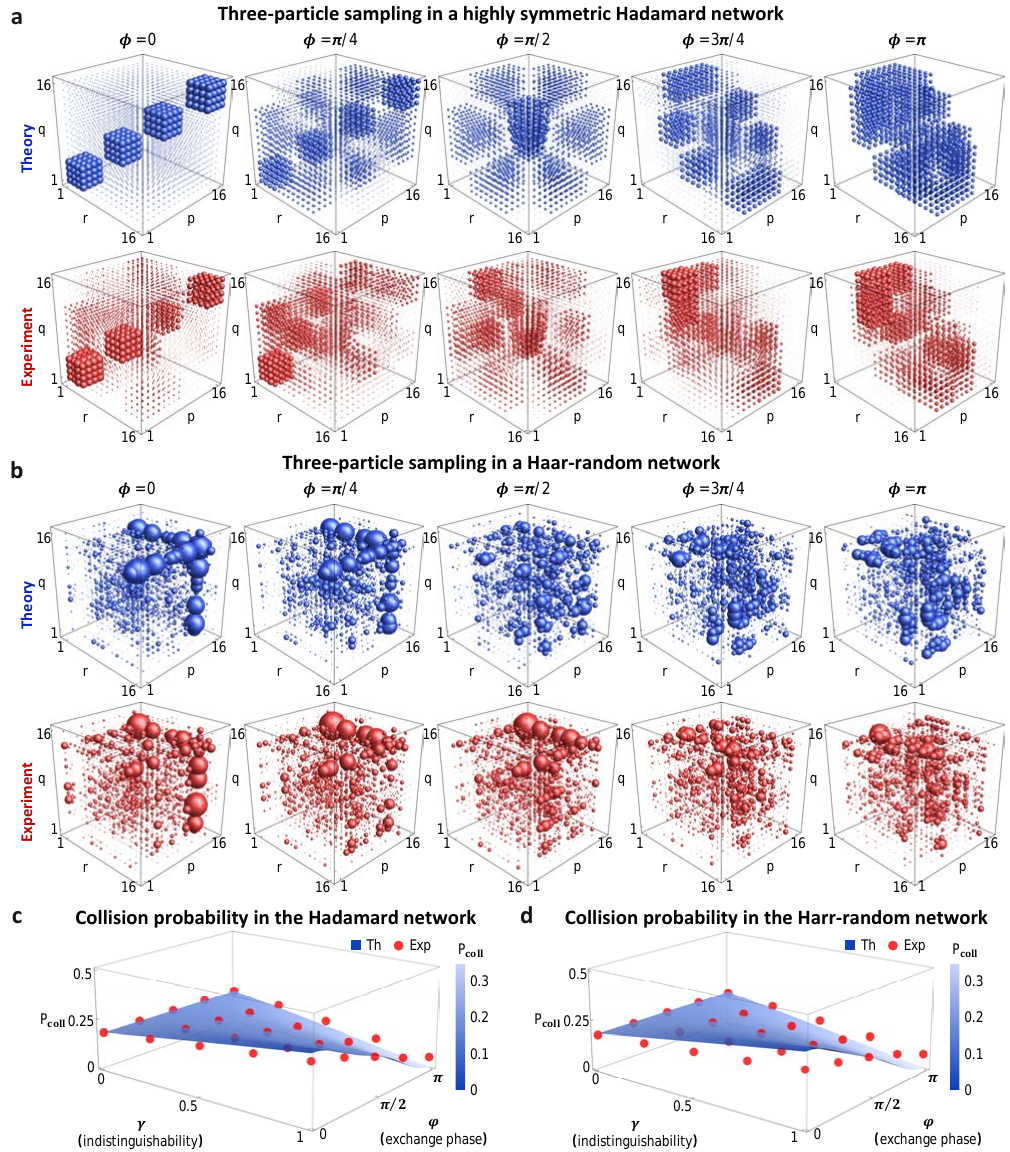}
\caption{\textbf{Experimental demonstration of universal quantum particle sampling and statistical phase transitions.} 
\textbf{a,} Three-particle sampling distributions in a 16-dimensional Hadamard network $H_{16}$ at full indistinguishability for exchange phases $\phi\in\{0,\pi/4,\pi/2,3\pi/4,\pi\}$. Theoretical predictions (blue) and experimental results (red) show the continuous transition from bosonic bunching to fermionic anti-bunching, with experimental fidelities $F=\{0.9668,0.9441,0.9530,0.9504,0.8891\}$, respectively. Sphere sizes indicate output probabilities. 
\textbf{b,} Corresponding sampling distributions in a 16-dimensional Haar-random network $U_{\mathrm{Haar}, {16}}$, demonstrating the same exchange-statistics transition with experimental fidelities $F=\{0.9445,0.9278,0.9280,0.9272,0.9091\}$, respectively. 
\textbf{c,} Collision probability $P_{\mathrm{coll}}$ versus exchange phase $\phi$ and particle indistinguishability $\Gamma$ in the $H_{16}$ network. The blue surface represents the theoretical prediction, while red points denote experimental measurements at the 25 combinations of $\Gamma \in \{0,0.25,0.5,0.75,1\}$ and $\phi \in \{0,\pi/4,\pi/2,3\pi/4,\pi\}$.
\textbf{d,} Corresponding collision probabilities in the $U_{\rm Haar, 16}$ network, demonstrating joint control of $\phi$ and $\Gamma$ in a generic random network.}
\label{fig3}
\end{figure*}

\section*{IV. Demonstrating universal quantum particle sampling with programmable statistics}

Using this processor, we next systematically demonstrate universal quantum particle sampling through four sets of experiments: continuous programming of the exchange phase $\phi$ in the fully indistinguishable regime, independent tuning of the global indistinguishability $\Gamma$, scaling to high-dimensional random networks, and statistical validation against physically motivated alternative models.

We first isolate the effect of exchange statistics in the fully indistinguishable limit. To resolve the phase-dependent evolution of the complete output distribution, we implement three-particle sampling in two representative 16-mode interferometers: a highly symmetric Hadamard network $H_{16}$ and a Haar-random unitary $U_{\rm Haar, 16}$. By programming the statistics-encoded input state, we set $\phi\in\{0,\pi/4,\pi/2,3\pi/4,\pi\}$, continuously connecting bosonic, fractional and fermionic exchange statistics. The measured output distributions closely follow the corresponding theoretical predictions for both interferometers (Fig. 3\textbf{a},\textbf{b}). As $\phi$ increases, the probability landscape evolves continuously from bosonic bunching through structured anyonic interference to fermionic antibunching. The same evolution is observed in both the symmetric and random networks, showing that the programmable statistics are not tied to a particular interferometer symmetry. Additional measurements across further network dimensions and random instances are reported in Section S5.

We next tune the global indistinguishability $\Gamma$ to probe the crossover from coherent quantum interference to distinguishable-particle propagation. We summarize this transition using the collision probability $P_{\rm coll}$, defined as the total probability of outcomes containing at least one multiply occupied output mode. Figure 3\textbf{c}, \textbf{d} shows $P_{\rm coll}$ as a joint function of $\phi$ and $\Gamma$ for $H_8$ and $U_{\rm Haar, 8 }$. At $\Gamma = 1$, the collision probability varies markedly with $\phi$, reflecting the transition from bosonic bunching to fermionic antibunching. As $\Gamma$ decreases, the phase-dependent contrast gradually diminishes as coherence between different permutation contributions is suppressed. In the distinguishable-particle limit, $\Gamma\rightarrow0$, the collision probabilities for all exchange phases converge to the same classical baseline. These measurements directly separate the roles of the two programmable parameters: $\phi$ determines the exchange-statistics-dependent interference structure, whereas $\Gamma$ determines the visibility with which that structure appears in the sampling distribution.

To showcase scalability, we implement three-particle sampling in a 24-mode Haar-random interferometer, spanning a Fock space of dimension 2600. The measured distribution agrees closely with the theoretical target, reaching a statistical fidelity of 99.34\% (Fig. 4\textbf{a}). We further implement more than 350 sampling instances spanning different network dimensions, input configurations and programmed exchange statistics. For each instance, we compare the measured and target distributions using the statistical fidelity $F=\sum_i\sqrt{p_i q_i}$ and the total variation distance $D=\frac{1}{2}\sum_i |p_i-q_i|$, where $p_i$ and $q_i$ denote the experimental and theoretical probabilities, respectively. 
Across the full benchmark set, the processor maintains consistently high fidelities and low total variation distances (Fig. 4\textbf{b}), demonstrating stable programmable operation across both network size and particle statistics. Extended measurements over additional networks, inputs and parameter settings are provided in Section S5.

At scales where the target distribution becomes classically intractable, exhaustive certification is itself unavailable. We therefore apply a suite of complementary statistical tests~\cite{aaronson2014bosonsampling,spagnolo2014experimental,bentivegna2015bayesian,dai2020bayesian} designed to reject the most relevant alternative models. 
To provide a consistent benchmark, all five statistical tests are applied to the same representive 24-mode Haar-random sampling instance with input and exchange phase. We first rejects a uniform sampler using the row-norm-estimator (RNE) test~\cite{aaronson2014bosonsampling} (Fig. 4\textbf{c}), and then distinguish the measured samples from the fully distinguishable-particle model using a likelihood-ratio (LR) test~\cite{spagnolo2014experimental} (Fig. 4\textbf{d}). Bayesian model comparisons~\cite{bentivegna2015bayesian,dai2020bayesian} further discriminate the fractional-statistics data from the bosonic (Fig. 4\textbf{e}), fermionic (Fig. 4\textbf{f}), and an alternative fractional-statistics distribution (Fig. 4\textbf{g}). This final comparison shows that the processor resolves specific fractional exchange phases, rather than merely producing distributions intermediate between the bosonic and fermionic limits. Additional networks and exchange phases are examined in Section S6.

Together, these experiments demonstrate independent programmability of exchange phase and global indistinguishability, scalable operation in up to 2,600-dimensional many-particle configuration space, and statistical validation of fractional-statistics sampling. The processor therefore implements universal quantum particle sampling across particle statistics, coherence regimes and network dimensions, rather than a sampling task specialized to a single particle species or interferometer.

\begin{figure*}[htbp!]
\centering
\includegraphics[width=1.0\textwidth]{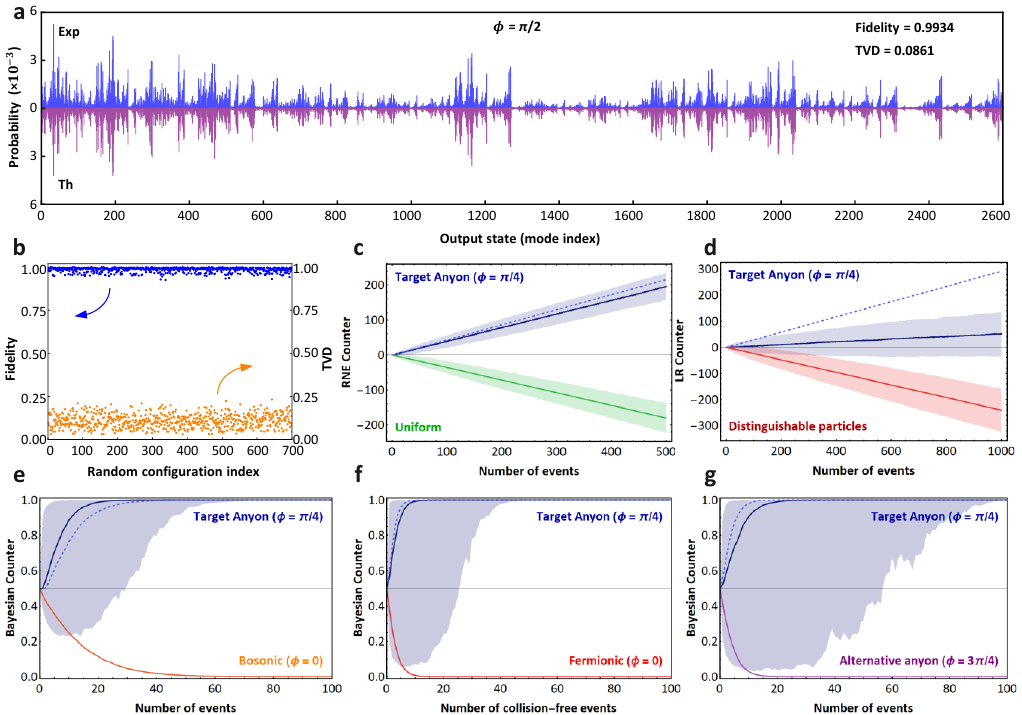}
\caption{
\textbf{Large-scale universal quantum particle sampling and statistical validation of fractional exchange statistics.} 
\textbf{a,} Measured and ideal theoretical output probability distributions for three-particle sampling in a 24-dimensional Haar-random network at the ideal quantum limit ($\Gamma = 1$), shown for a representative exchange phase $\phi =\pi/2$. 
\textbf{b,} Statistical fidelity (blue) and total variation distance (TVD; orange) between the measured and theoretical probability distributions over 350 randomly generated sampling configurations spanning different network settings and exchange phases. The consistently high fidelities and low TVDs demonstrate robust sampling performance across a broad range of programmed configurations.
\textbf{c,} Row-norm-estimator (RNE) validation of the target anyonic distribution at $\phi=\pi/4$ against the uniform-distribution hypothesis. 
\textbf{d,} Likelihood-ratio (LR) validation of the target anyonic distribution at $\phi=\pi/4$ against the fully distinguishable-particle hypothesis. 
\textbf{e--g,} Bayesian model comparisons of the target anyonic distribution at ($\phi=\pi/4$) against, \textbf{e}, the bosonic distribution ($\phi=0$), \textbf{f}, the fermionic distribution ($\phi=\pi$), and \textbf{g}, an alternative anyonic distribution at ($\phi=3\pi/4$).  
For the tests in \textbf{c}--\textbf{g}, the experimental solid curve shows the pointwise median over Poisson-resampled and randomly permuted experimental realizations. The target-anyon dashed curve and the alternative-model solid curve show the corresponding pointwise median trajectories. The shaded experimental and alternative-model regions denote the pointwise $95\%$ quantile intervals. }
\label{fig4}
\end{figure*}

\begin{figure*}[t]
   \centering
    \includegraphics[width=0.98\textwidth]{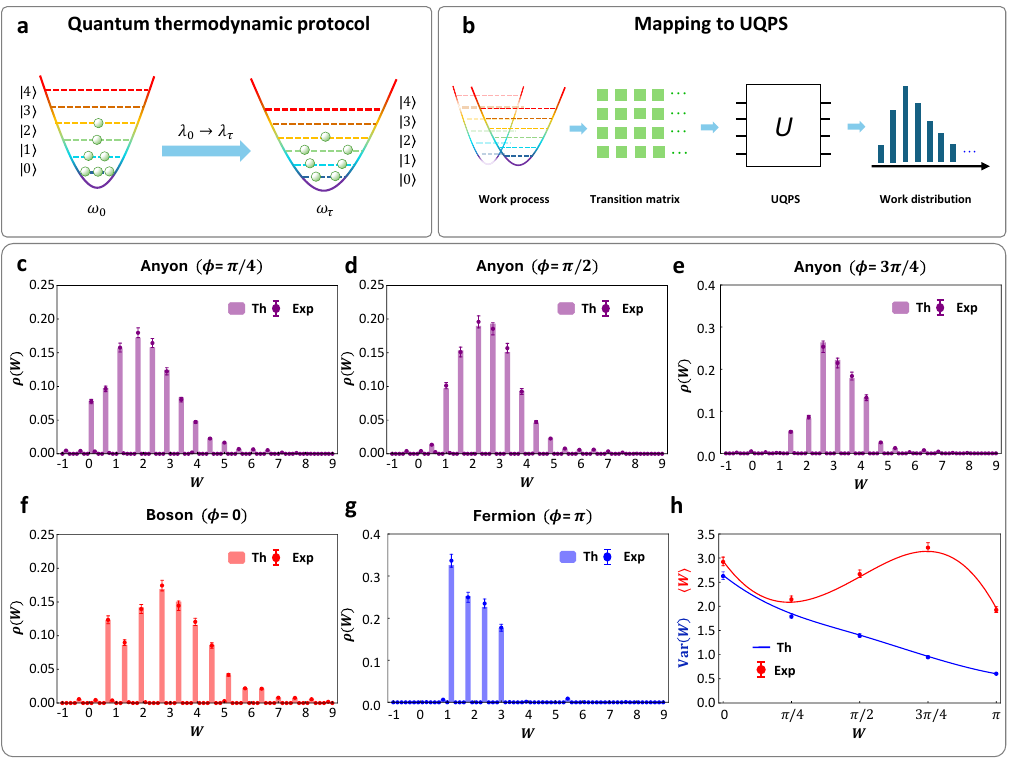}
    \caption{
    \textbf{Quantum work distribution estimation using universal quantum particle sampling.}
    \textbf{a,} Quantum thermodynamic protocol for estimating the work distribution.
    Particles initially prepared in a thermal state at inverse temperature $\beta$
    undergo a driven evolution from parameter $\lambda_0$ to $\lambda_\tau$ (for a one-dimensional harmonic oscillator, these correspond to the initial and final angular frequencies $\omega_0$ and $\omega_\tau$, respectively).
    Projective energy measurements before and after the evolution define the stochastic work, whose realizations yield the work distribution.
    \textbf{b,} Mapping of the quantum thermodynamic process to universal quantum particle sampling (UQPS).
    The single-particle transition matrix is encoded into the programmable unitary transformation $U$, while the exchange phase $\phi$ and indistinguishability $\gamma$ determine the many-particle transition probabilities.
    Sampling from the UQPS processor thus yields energy-transition samples that are converted into the work distribution $\rho(W)$.
    \textbf{c--g,} Particle-statistics-dependent work distributions at fixed inverse temperature $\beta=0.2$ and driving time $\tau=1$.
    Work distributions are shown for boson $(\phi=0)$, anyon $(\phi=\pi/4,\pi/2,3\pi/4)$ and fermion $(\phi=\pi)$.
    Bars represent theoretical values and points with error bars denote experimental results. \textbf{h,} Extracted thermodynamic mean work $\langle W\rangle=\sum_W\rho(W)W$ and work fluctuation Var($W$) as a function of exchange phase $\phi$. Solid lines show theoretical fits and error bars represent experimental data.
    }
    \label{fig:work_distribution}
\end{figure*}

\section*{V. Work-distribution estimation with universal particle sampling}

Having established programmable quantum particle sampling, we further use the device to estimate an important physical quantity that is intrinsically sensitive to particle statistics: the quantum work distribution~\cite{talkner2007fluctuation, jarzynski2015quantum,liu2024efficiently}. Work distributions characterize energy exchange in nonequilibrium quantum processes and are commonly defined through two projective energy measurements. For many identical particles, the work distribution depends not only on the single-particle driving protocol, but also on multiparticle interference between exchange-related transition paths. Universal quantum particle sampling therefore provides a natural route to access work statistics beyond the bosonic regime.

We consider an $N$-particle system initially prepared in a thermal state of a Hamiltonian $H(\lambda_0)$, where $\lambda_0$ is an initial externally controlled work parameter at time $t=0$. The system is detached from the heat bath and driven unitarily to a final Hamiltonian $H(\lambda_\tau)$ at $t=\tau$ (see Fig. \ref{fig:work_distribution}\textbf{a}). During this process, the work distribution can be written as
\begin{align}
\rho(W)=
\sum_{i,f}
P_{\phi}(i)
P_{\phi}(i\rightarrow f)
\delta
\left[
W-\left(E_f^{\tau}-E_i^{0}\right)
\right],
\end{align}
where $P_{\phi}(i)$ is the initial thermal occupation probability, and $E_i^{0}$ and $E_f^{\tau}$ are the corresponding many-particle energies before and after the drive. $P_{\phi}(i\rightarrow f)$ is the many-particle transition probability from the initial $i$th eigenstate to the final $j$th eigenstate.
If $\Lambda^{(I,F)}$ denotes the single-particle transition matrix between the initial and final eigenstates, then the corresponding many-particle transition probability contains a coherent sum over particle permutations,
\begin{align}
P_{\phi}(i\rightarrow f)=
\bigg|\sum_{\sigma\in S_N}
e^{i\phi\ell(\sigma)}
\prod_{j=1}^{N}
\Lambda^{(I,F)}_{j,{\sigma(j)}}\bigg|^2,
\end{align}
with occupation-dependent normalization factors included for repeated energy levels. Thus, the transition probability has the same statistics-weighted structure as the sampling probability $P_{\phi}(S|I,U)$ introduced above.  The work distribution can therefore be estimated by programming the particle sampler with the transition matrix $\Lambda^{(I,F)}$, sampling the output events, and grouping the observed transitions according to their work values (see Fig. \ref{fig:work_distribution}\textbf{b}). 

We implement this protocol using a driven one-dimensional harmonic oscillator as a model nonequilibrium quantum system. The chip is programmed with the transition matrices generated by initial angular frequency $\omega_0$, final frequency $\omega_\tau$, and driving durations $\tau$, while the initial thermal occupations are set by the inverse temperature $\beta$. For each pair of parameters $(\beta,\tau)$, we estimate the work distribution for bosonic, fermionic and anyonic statistics using the same integrated particle-sampling processor (see Figs. \ref{fig:work_distribution}\textbf{c-g}). We further extract the mean work \(\langle W\rangle=\sum_W\rho(W)W\) and work fluctuation $\mathrm{Var}(W)$ (see Fig. \ref{fig:work_distribution}\textbf{h}). The observed reduction in work fluctuation from bosons through anyons to fermions demonstrates that quantum exchange statistics can suppress thermodynamic work fluctuations: the Pauli-exclusion-like repulsive effect arising from nontrivial exchange phases narrows the probability distribution of the extracted work.
The measured distributions agree with theoretical predictions, demonstrating that sampling-based work estimation can be extended from bosonic interference to a unified set of exchange statistics (see additional results in Section S7).

The experiment reveals two distinct controls on the work distribution. The inverse temperature $\beta$ determines which initial energy configurations contribute to the sampling problem. Large $\beta$ prepares a colder ensemble dominated by low-energy configurations, whereas smaller $\beta$ increases the occupation of excited states and gives access to a broader set of many-particle transitions. The driving time $\tau$ controls the degree of nonadiabaticity. In the adiabatic limit of large $\tau$, the distribution is concentrated near transitions that follow the instantaneous eigenstates. As $\tau$ is reduced, nonadiabatic transitions become more pronounced and the work distribution broadens (see additional results in Section S7).

Particle statistics reshape these distributions in a way that cannot be captured by single-particle dynamics alone. Bosonic exchange enhances transition pathways associated with symmetric multiparticle interference, fermionic exchange suppresses configurations forbidden by antisymmetry, and fractional exchange phases generate intermediate anyonic work statistics. The ability to switch among these cases on the same chip demonstrates that the processor can be used not only as a sampling device, but also as a programmable simulator of statistics-dependent nonequilibrium many-body physics.

This work-distribution estimation application shows that the same probability structure that makes universal particle sampling sensitive to exchange statistics also determines the many-particle transition probabilities entering $\rho(W)$. Work-distribution estimation therefore provides a physically motivated use case for programmable exchange statistics, opening a route to studying quantum thermodynamics, transport and many-body interference beyond the bosonic regime.

\section*{VI. Discussion}

We have established universal quantum particle sampling framework and promoted particle exchange statistics to a programmable degree of freedom. The exchange phase $\phi$ determines the phase structure of the permutation amplitudes, whereas the global indistinguishability $\Gamma$ controls the coherence retained between them. For fully indistinguishable particles, we prove classical intractability for generic Abelian exchange phases under standard complexity-theoretic assumptions, extending the complexity landscape of quantum sampling beyond the bosonic case. Experimentally, entanglement acts as a statistics transducer that maps programmable photonic amplitudes onto the permutation weights of the target sampling model. Combined with reconfigurable linear-optical transformations, this enables the network $U$, exchange phase $\phi$ and global indistinguishability $\Gamma$ to be configured within a common silicon photonic processor.

Using this processor, we demonstrated a continuous evolution from bosonic bunching through fractional-statistics interference to fermionic antibunching, controlled the crossover from coherent multiparticle interference to distinguishable-particle propagation, and implemented three-particle sampling in random interferometers of up to 24 modes, corresponding to a 2,600-dimensional Fock space. A suite of complementary statistical tests rejects uniform, fully distinguishable, bosonic and fermionic alternative models and distinguishes samples generated at different fractional exchange phases. These results show that the measured distributions resolve the phase structure of anyonic interference, rather than simply occupying an intermediate region between bosonic and fermionic distributions. As a representative application of this programmable sampling capability, we further use quantum particle sampling to estimate statistics-dependent nonequilibrium work distributions, linking programmable exchange statistics to quantum thermodynamic fluctuations.

The present experiments shown a proof of principle demonstration for programmable particle-statistics sampling; reaching quantum computational advantage will require brighter photon sources, scalable preparation of high-dimensional entangled states, lower optical loss, photon-number-resolving detection and validation protocols that remain informative when exhaustive classical comparison is unavailable. 
Scaling the entanglement-based architecture will require efficient preparation of $N$-photon high-dimensional permutation-entangled states. Recent graph-based boson-subtraction~\cite{chin2024creating,chin2024shortcut} and identical-particle-interferometry schemes~\cite{piccolini2025robust, mahdavipour2024generation} provide explicit heralded or measurement-conditioned constructions for $N$-partite $N$-level symmetric and antisymmetric states, as well as broader classes of multipartite qudit entanglement, offering possible routes towards larger statistics-encoded photonic resources. 

The broader opportunity is to introduce particle statistics as an additional design variable for quantum computation and simulation. The enlarged control space enables particle statistics to be optimized together with the interferometer as part of the computational design, potentially extending existing sampling-based approaches to molecular spectroscopy~\cite{huh2015boson,huh2017vibronic}, graph-theoretic problems~\cite{arrazola2018using, bradler2018gaussian, schuld2020measuring}, point processes~\cite{jahangiri2020point}, molecular modelling~\cite{banchi2020molecular} and quantum machine learning~\cite{hoch2025quantum, zhang2026enhancing}. 
Further extensions could incorporate more general internal-state Gram matrices, larger multiparticle entangled resources and adaptive sampling protocols. Extending the present scalar-phase model to non-Abelian braiding would additionally require matrix-valued transformations acting on fusion-space degrees of freedom~\cite{nayak2008non}. More broadly, programming how multiparticle amplitudes combine makes particle-exchange statistics a configurable element of quantum computational design.

\noindent\textbf{Data access statement:} The data that support the findings of this study are available from the corresponding author upon reasonable request.

\noindent\textbf{Acknowledgements:} This work was supported by the National Natural Science Foundation of China (Grant No. 12341103, 12505028, 12441502, 62272259 and 62332009). 
X.Q. acknowledges support from National Natural Science Foundation of China (12341103) and the National Young Talents Program. 
W.L. acknowledges support from National Natural Science Foundation of China (12505028), and Science Research Project of Hebei Education Department (QN2025054). 
L.L. and Z.W. acknowledges support from National Natural Science Foundation of China (62272259, 62332009), Beijing Natural Science Foundation (Z220002), and Beijing Science and Technology Planning Project (Z25110100810000).
Z.Y. acknowledges support from National Natural Science Foundation of China (12441502) and the Beijing Institute of Technology Research Fund Program (2024CX01015).

\noindent\textbf{Author contributions:} X.Q., W.L., Z.W. and Z.Y. conceived and designed the project. X.Q., W.L. and Z.Y. developed the univeral quantum particle sampling framework. L.L. and Z.W. provided the complexity theory proofs. W.L. and Z.Y. developed the work-distribution estimation protocol. X.Q. designed and implemented the silicon photonic processor. X.Q., C.L., P.L., Y.W., and J.H. built the experimental setup. X.Q., C.L., P.L., S.M. and W.L. carried out the experiments and analyzed the data. X.Q., W.L., L.L., C.L., Z.W., and Z.Y. wrote the manuscript, with contributions from all other authors. X.Q. led the project. All authors discussed the results and approved the final manuscript.

\clearpage


\pagebreak
\newpage
\widetext
\begin{center}
\section*{\large{Supplementary Information}}
\end{center}
\setcounter{equation}{0}
\setcounter{figure}{0}
\setcounter{table}{0}
\setcounter{page}{1}
\setcounter{section}{0}
\makeatletter
\renewcommand{\theequation}{S\arabic{equation}}
\renewcommand{\thefigure}{S\arabic{figure}}
\renewcommand{\bibnumfmt}[1]{[#1]}
\renewcommand{\citenumfont}[1]{#1}
\normalsize

\let\addcontentsline\oldaddcontentsline

\tableofcontents

\newpage

\section{Universal quantum particle sampling framework}
\label{sec:supp_UQPS_framework}

We formulate universal quantum particle sampling framework with two physical parameters that are 
tied to particle species: the phase acquired under exchange particles, described by an exchange phase $\phi\in[0,\pi]$, 
and the coherence between different particle assignments, described by the global indistinguishability $\Gamma \in [0,1]$. 

\emph{\textbf{Particle exchange statistics.}} 
A wave-function acquires a phase $\phi$ when interchanging two indistinguishable particles, which is characterized by the creation operator $a^{\dag} _j$ and annihilation operator $a_j$ (of mode $j$) obeying the commutation relations
\begin{align}
a^{\dag}_j a^{\dag}_k -e^{i \phi} a^{\dag}_k a^{\dag}_j &= 0, \\
a_j a_k -e^{i \phi} a_k a_j &= 0, \\
a_j a^{\dag}_k -e^{i \phi} a^{\dag}_k a_j &= \delta_{j,k} 
\label{eq:supp_commutation_relation}
\end{align}
for $j<k$. For bosonic particles, $\phi=0$, for fermionic particles, $\phi = \pi$, and for Abelian anyonic particles, they undergo a fractional phase $0< \phi <\pi$ with $j<k$.  

In this work, anyonic sampling refers to an effective Abelian exchange-phase model. Consider the case of $N$ particles injected into a linear network, the exchange weight 
\begin{equation}
w_{\phi}(\sigma) = e^{i\phi\ell(\sigma)},
\label{eq:supp_exchange_weight}
\end{equation}
is assigned for a permutation $\sigma\in S_N$, where $\ell(\sigma)$ is the inversion number of $\sigma$, equivalently the
minimum number of adjacent transpositions required to reach $\sigma$ from
the reference ordering. The bosonic and fermionic limits are recovered at
$\phi=0$ and $\phi=\pi$, respectively, whereas $0<\phi<\pi$ describes
fractional Abelian exchange statistics~\cite{greiter2024fractional}. In non-Abelian braiding, exchanges implement generally noncommuting unitary transformation within a multi-anyon fusion space~\cite{nayak2008non}, which lies outside the scope of the present framework.

\emph{\textbf{Particle indistinguishability.}} For $N$ particles with normalized internal states
$\{|\eta_{\alpha}\rangle\}_{\alpha=1}^{N}$, we can define the internal-state
Gram matrix
\begin{equation}
G_{\alpha\beta}
=
\langle\eta_{\beta}|\eta_{\alpha}\rangle .
\label{eq:supp_Gram_matrix}
\end{equation}
The pairwise indistinguishability between particles $\alpha$ and $\beta$
is quantified by
\begin{equation}
\gamma_{\alpha\beta}
=
|G_{\alpha\beta}|^{2},
\label{eq:supp_pairwise_indistinguishability}
\end{equation}
and $0\leq\gamma_{\alpha\beta}\leq1$. The global indistinguishability parameter $\Gamma$ is defined as the mean pairwise indistinguishability,
\begin{equation}
\Gamma
=
\frac{2}{N(N-1)}
\sum_{\alpha<\beta}
\gamma_{\alpha\beta}.
\label{eq:supp_global_indistinguishability}
\end{equation}
Therefore, $\Gamma=1$ corresponds to fully indistinguishable particles,
whereas $\Gamma=0$ corresponds to mutually distinguishable particles.
The scalar parameter $\Gamma$ provides a compact coordinate for displaying the
overall quantum-to-classical crossover. It is noteworthy that a general partially
distinguishable distribution depends on the complete Gram
matrix $G$.

\subsection{Universal quantum sampling of fully indistinguishable particles}
\label{sec:supp_fully_indistinguishable}

Consider that $N$ particles are injected into an $M$-mode linear network described
by a unitary matrix $U$, with the transformation as
\begin{equation}
\hat{a}_{i}^{\dagger}
\longrightarrow
\sum_{s=1}^{M}
U_{si}\hat{b}_{s}^{\dagger}.
\label{eq:supp_mode_transformation}
\end{equation}
Define the $N$-particle input occupation state $|\bm{\nu}\rangle$ and the $N$-particle output occupation state $|\bm{\mu}\rangle$ as below: 
\begin{equation}
|\bm{\nu}\rangle
=
\prod_{i=1}^{M}
\frac{
1
}{
\sqrt{\nu_i!}
}(\hat{a}_{i}^{\dagger})^{\nu_i}
|0\rangle,
\qquad
\sum_{i=1}^{M}\nu_i=N.
\label{eq:supp_input_fock_state}
\end{equation}
\begin{equation}
|\bm{\mu}\rangle
=
\prod_{s=1}^{M}
\frac{
1
}{
\sqrt{\mu_s!}
}(\hat{b}_{s}^{\dagger})^{\mu_s}
|0\rangle,
\qquad
\sum_{s=1}^{M}\mu_s=N.
\label{eq:supp_output_fock_state}
\end{equation}
The input and output occupations can be equivalently expanded into ordered mode lists as
\begin{equation}
\bm{i}=(i_1,\ldots,i_N),
\qquad
\bm{s}=(s_1,\ldots,s_N),
\end{equation}
in which each input mode $i$ appears $\nu_i$ times and each output mode
$s$ appears $\mu_s$ times. The exchange-phase-weighted amplitude for the
ordered configuration is
\begin{equation}
\mathcal{A}_{\phi}(\bm{s}|\bm{i},U)
=
\sum_{\sigma\in S_N}
e^{i\phi\ell(\sigma)}
\prod_{a=1}^{N}
U_{s_a,i_{\sigma(a)}},
\label{eq:supp_ordered_amplitude}
\end{equation}
and the corresponding Fock-basis probability is
\begin{equation}
P_{\phi}(\bm{\mu}|\bm{\nu},U)
=
\frac{
1
}{
\prod_{i=1}^{M}\nu_i!
\prod_{s=1}^{M}\mu_s!
}
\left|
\mathcal{A}_{\phi}(\bm{s}|\bm{i},U)
\right|^{2}.
\label{eq:supp_fock_probability}
\end{equation}
Equation~\eqref{eq:supp_fock_probability} follows directly by expanding
the transformed input state and projecting it onto the target output
occupation. Each permutation $\sigma$ specifies one assignment of the
ordered input particles to the ordered output modes, contributes the
product of single-particle amplitudes in
Eq.~\eqref{eq:supp_ordered_amplitude}, and acquires the exchange factor
$e^{i\phi\ell(\sigma)}$. Repeated input and output occupations generate
equivalent labelings, whose multiplicities are removed by the factorial
normalization in Eq.~\eqref{eq:supp_fock_probability}.

Defining $U^{(\bm{\mu},\bm{\nu})}$ as the $N\times N$ matrix obtained by
repeating output row $s$ of $U$ $\mu_s$ times and input column $i$ of $U$
$\nu_i$ times, we express the exchange-phase-weighted permutation sum using the $q$-permanent~\cite{LAL19981} with $q= e^{i \phi}$, 
\begin{equation}
\operatorname{Per}_{\phi}(A)
\equiv
\sum_{\sigma\in S_N}
e^{i\phi\ell(\sigma)}
\prod_{a=1}^{N}
A_{a,\sigma(a)}.
\label{eq:supp_phi_permanent}
\end{equation}
The sampling probability can therefore be written as
\begin{equation}
P_{\phi}(\bm{\mu}|\bm{\nu},U)
=
\frac{
1
}{
\prod_i\nu_i!
\prod_s\mu_s!
}\left|
\operatorname{Per}_{\phi}
\left(
U^{(\bm{\mu},\bm{\nu})}
\right)
\right|^{2}.
\label{eq:supp_phi_permanent_probability}
\end{equation}
It is obtained that at $\phi=0$,
\begin{equation}
\operatorname{Per}_{0}(A) = \operatorname{Per}(A),
\end{equation}
and Eq.~\eqref{eq:supp_fock_probability} reduces to the boson-sampling
distribution; at $\phi=\pi$,
\begin{equation}
\operatorname{Per}_{\pi}(A)
=
\det(A),
\end{equation}
Eq.~\eqref{eq:supp_fock_probability} reduces to the fermion-sampling distribution, according to $e^{i\pi\ell(\sigma)} = (-1)^{\ell(\sigma)} = \operatorname{sgn}(\sigma)$. For $0<\phi<\pi$,
Eq.~\eqref{eq:supp_phi_permanent} defines the Abelian anyonic sampling
model used throughout this work.

\subsection{Projection-resolved quantum sampling and partial distinguishability}
\label{sec:projection_resolved_sampling}

The projection-resolved formulation is independent of the dimensionality of
the particles' internal Hilbert space. We therefore consider normalized input
internal states $\{|\eta_\alpha\rangle\}_{\alpha=1}^{N}$ and output projection
states $\{|\xi_a\rangle\}_{a=1}^{N}$, written in an orthonormal basis
$\{|\mu\rangle\}_{\mu=1}^{d_{\mathrm{int}}}$ as
\begin{equation}
|\eta_\alpha\rangle
=
\sum_{\mu=1}^{d_{\mathrm{int}}}
c_{\mu\alpha}|\mu\rangle,
\qquad
|\xi_a\rangle
=
\sum_{\mu=1}^{d_{\mathrm{int}}}
d_{\mu a}|\mu\rangle .
\label{eq:general_internal_states}
\end{equation}
Their projection overlap is
\begin{equation}
C_{a\alpha}
=
\langle\xi_a|\eta_\alpha\rangle .
\label{eq:projection_coefficient}
\end{equation}

For a collision-free ordered input configuration
$\bm{i}=(i_1,\ldots,i_N)$ and an ordered output configuration
$\bm{s}=(s_1,\ldots,s_N)$, the projection-resolved amplitude is
\begin{equation}
\mathcal{A}_{\phi,C}^{\mathrm{proj}}
(\bm{s}|\bm{i},U)
=
\sum_{\sigma\in S_N}
e^{i\phi\ell(\sigma)}
\prod_{a=1}^{N}
C_{a,\sigma(a)}
U_{s_a,i_{\sigma(a)}} .
\label{eq:projection_resolved_amplitude}
\end{equation}
The corresponding ordered probability is
\begin{equation}
P_{\phi,C}^{\mathrm{proj}}
(\bm{s}|\bm{i},U)
=
\frac{1}{N!}
\left|
\mathcal{A}_{\phi,C}^{\mathrm{proj}}
(\bm{s}|\bm{i},U)
\right|^2 .
\label{eq:projection_resolved_probability}
\end{equation}

The factor $C_{a,\sigma(a)}$ is the amplitude for projecting the internal state
carried by input particle $\sigma(a)$ onto the internal-state channel selected
at ordered output position $a$. By contrast,
$e^{i\phi\ell(\sigma)}$ controls the exchange-statistics phase associated with
the corresponding particle assignment. If all particles occupy the same
internal state and all outputs are projected onto that same state, then
$C_{a\alpha}=1$, and the fully indistinguishable distribution is recovered.

A single choice of
$\boldsymbol{\xi}=(\xi_1,\ldots,\xi_N)$ defines one projection-resolved
setting. This should be distinguished from conventional partially
distinguishable sampling in which the output internal degrees of freedom are
not resolved. In the unresolved case, the ordered distribution is
\begin{equation}
P_{\phi,G}^{\mathrm{partial}}
(\bm{s}|\bm{i},U)
=
\frac{1}{N!}
\sum_{\sigma,\tau\in S_N}
e^{i\phi[\ell(\sigma)-\ell(\tau)]}
\prod_{a=1}^{N}
G_{\sigma(a),\tau(a)}
U_{s_a,i_{\sigma(a)}}
U_{s_a,i_{\tau(a)}}^* ,
\label{eq:partial_G_distribution}
\end{equation}
where
\begin{equation}
G_{\alpha\beta}
=
\langle\eta_\beta|\eta_\alpha\rangle
\label{eq:internal_Gram_matrix}
\end{equation}
is the internal-state Gram matrix.

Let $\mathcal{B}$ be a complete orthonormal basis of the internal space and let
\begin{equation}
\boldsymbol{\xi}
=
(\xi_1,\ldots,\xi_N)
\in\mathcal{B}^{\otimes N}
\label{eq:complete_projection_sector}
\end{equation}
label a complete output projection setting. The completeness relation gives
\begin{equation}
\sum_{\boldsymbol{\xi}\in\mathcal{B}^{\otimes N}}
\prod_{a=1}^{N}
C_{a,\sigma(a)}
C_{a,\tau(a)}^*
=
\prod_{a=1}^{N}
G_{\sigma(a),\tau(a)} .
\label{eq:projection_completeness}
\end{equation}
Consequently,
\begin{equation}
P_{\phi,G}^{\mathrm{partial}}
(\bm{s}|\bm{i},U)
=
\sum_{\boldsymbol{\xi}\in\mathcal{B}^{\otimes N}}
P_{\phi,\boldsymbol{\xi}}^{\mathrm{proj}}
(\bm{s}|\bm{i},U) .
\label{eq:partial_as_sector_sum}
\end{equation}

The sum is performed at the probability level because different internal-state
projection settings are mutually orthogonal and experimentally exclusive. This establishes the direct connection between projection-resolved
sampling and the conventional Gram-matrix description of partial
distinguishability. 
The internal-state dimension is not fixed by the projection-resolved
formalism. To obtain a symmetric and continuous interpolation between fully distinguishable and
fully indistinguishable three-particle sampling, we therefore use the
three-dimensional target internal space
\begin{equation}
\mathcal{H}_{\mathrm{int}}
=
\operatorname{span}
\{|1\rangle,|2\rangle,|3\rangle\}.
\label{eq:three_dimensional_internal_space}
\end{equation}

For the three-particle experiments, we choose the symmetric Gram-matrix family
\begin{equation}
G^{(\Gamma)}
=
\begin{pmatrix}
1&t&t\\
t&1&t\\
t&t&1
\end{pmatrix},
\qquad
t=\sqrt{\Gamma},
\label{eq:symmetric_Gamma_Gram}
\end{equation}
so that
\begin{equation}
|G_{12}^{(\Gamma)}|^2
=
|G_{13}^{(\Gamma)}|^2
=
|G_{23}^{(\Gamma)}|^2
=
\Gamma .
\label{eq:equal_pairwise_indistinguishability}
\end{equation}
Thus, the global indistinguishability defined as the mean pairwise
indistinguishability is exactly $\Gamma$. An explicit realization of this Gram matrix is
\begin{subequations}
\label{eq:three_internal_states_Gamma}
\begin{align}
|\eta_1(\Gamma)\rangle
&=
|1\rangle,
\\
|\eta_2(\Gamma)\rangle
&=
t|1\rangle+
\sqrt{1-t^2}|2\rangle,
\\
|\eta_3(\Gamma)\rangle
&=
t|1\rangle
+
t\sqrt{\frac{1-t}{1+t}}|2\rangle
+
\sqrt{\frac{(1-t)(1+2t)}{1+t}}|3\rangle .
\end{align}
\end{subequations}
These states satisfy
\begin{equation}
\langle\eta_\beta(\Gamma)|\eta_\alpha(\Gamma)\rangle
=
G_{\alpha\beta}^{(\Gamma)} ,
\qquad
\alpha, \beta \in \{1,2,3 \}.
\label{eq:states_reproduce_Gram}
\end{equation}

At $\Gamma=0$, the three internal states are mutually orthogonal and the
particles are fully distinguishable. At $\Gamma=1$, the three states become
identical and the fully indistinguishable limit is recovered. Intermediate
values of $\Gamma$ provide a symmetric path in which all three particle pairs
have the same pairwise indistinguishability. The scalar $\Gamma$ therefore
uniquely specifies the family used in the main-text experiments, although a
general partially distinguishable distribution still depends on the complete
Gram matrix.

\section{ Complexity-theoretic proof for anyonic sampling}

Let $B\in C^{m \times m}$, $k\in C$, and $k \neq 0$. Define the fermionant of $B$ to be

	\begin{align*}
		\mathrm{Ferm}_{k}(B)=\sum_{\sigma \in S_m}k^{\ell (\sigma)}\prod_{j=1}^{m}B_{j,\sigma(j)},
	\end{align*}
	where $\ell (\sigma)$ is the number of inverse pairs of the permutation $\sigma$, and $S_m$ is the group of all permutations of $m$ objects.
	Also define the \textit{value} of a permutation $\sigma$ as
	\begin{align*}
		v_B(\sigma)=k^{\ell (\sigma)}\prod_{j=1}^{m}B_{j,\sigma(j)},
	\end{align*}
    so that $\mathrm{Ferm}_{k}(B)=\sum_{\sigma \in S_m}v_B(\sigma)$. 
	
	Let $m=4(n+n^2)$. We now show that, for any $n\times n$ matrix $A$, computing its permanent can be reduced to computing the fermionant of an $m\times m$ matrix $B$. This establishes that computing the fermionant is at least as hard as computing the permanent.

    \begin{theorem}
		\label{thm:complexityofanyon}
    For an arbitrary $A\in\mathbb{C}^{n\times n}$, there exists a polynomial-time reduction that constructs a matrix $B\in\mathbb{C}^{m\times m}$, with $m=4(n+n^2)$, such that 
    \begin{align*}
    \mathrm{Ferm}_{k}(B)=c(k,n)\mathrm{Perm}(A),
    \end{align*}
    where {$\mathrm{Perm}(A)$ is the permanent of $A$, and} $c(k,n)$ is a constant depending only on $k$ and $n$.
    \end{theorem}

    \begin{proof}
	For convenience, the rows and columns of $B$ will be indexed with the labels ordered as following:
	\begin{itemize}
		\item
			$R_1,\cdots,R_n$;
		\item
			$C_1,\cdots,C_n$;
		\item
			$X_{i,j,t}$ in lexicographic order of $(i,j,t)$, where $i,j\in[n],t\in[4]$;
		\item
			$C'_n,\cdots,C'_1$;
		\item
			$R'_n,\cdots,R_1$.
	\end{itemize}
	Note that there are $4(n+n^2)$ rows/columns in total,
    hence $m=4(n+n^2)$.

	We now construct $B$ as following:
	\begin{enumerate}
		\item
			Set
			\begin{align*}
				B_{R_i,X_{i,j,1}}= & A_{i,j}, \\
				B_{C_j,X_{i,j,2}}= &
				B_{X_{i,j,3},R'_i}=
				B_{X_{i,j,4},C'_j}=
				B_{R'_i,R_{i+1}}=
				B_{C'_i,C_{i+1}}=1,
			\end{align*}
			where {$i,j\in[n]\equiv\{1,2,...,n\}$} and we identify $R_{n+1}$ and $C_{n+1}$ with $C_1$ and $R_1$ respectively.
		\item
			For $i,j\in[n]$, {the nonzero entries of the $4\times 4$ submatrix $B_{X_{i,j,*},X_{i,j,*}}$ are} set as following:
			\begin{align*}
				\begin{pmatrix}
					1 & & 1 & \\
					& 1 & & 1 \\
					-k^{-1} & & 1 & \\
					& -k^{-1} & & 1 \\
				\end{pmatrix}.
			\end{align*}
		\item
			Set all unspecified entries of $B$ to be $0$.
	\end{enumerate}

    In the following, we evaluate $\mathrm{Ferm}_{k}(B)$ via the values of permutations.
	Consider a permutation {$\sigma \in S_m$} that only contains the specified entries of $B$.
	Suppose there is $i,j$ such that $\sigma(R_i)=X_{i,j,1}$ but $\sigma(C_j)\neq X_{i,j,2}$.
	Then we have $\sigma(X_{i,j,1})=X_{i,j,3}$ and $\sigma(X_{i,j,3})=R'_i$.
	If $\sigma(X_{i,j,2})=X_{i,j,2}$ and $\sigma(X_{i,j,4})=X_{i,j,4}$,
	then consider another permutation $\sigma'=(X_{i,j,2}\ X_{i,j,4})\sigma$.
	Since the swap introduces precisely one extra inverse pair into $\sigma$ and uses a $-k^{-1}$ entry,
	we have $v_B(\sigma')=(-k^{-1}k)v_B(\sigma)=-v_B(\sigma)$, hence $v_B(\sigma')+v_B(\sigma)=0$.
	The other case is $\sigma(X_{i,j,2})=X_{i,j,4}$ and $\sigma(X_{i,j,4})=X_{i,j,2}$,
	which is when the roles of $\sigma$ and $\sigma'$ are reversed in the first case.

	Similarly, suppose there is $i,j$ such that $\sigma(R_i)\neq X_{i,j,1}$ but $\sigma(C_j)=X_{i,j,2}$.
	We have $\sigma(X_{i,j,2})=X_{i,j,4}$ and $\sigma(X_{i,j,4})=C'_j$.
	If $\sigma(X_{i,j,1})=X_{i,j,1}$ and $\sigma(X_{i,j,3})=X_{i,j,3}$,
	then an argument similar as above gives $v_B(\sigma')+v_B(\sigma)=0$
	where $\sigma'=(X_{i,j,1}\ X_{i,j,3})\sigma$.

	{Based on the above discussions, it suffices to consider $\sigma$ such that $\sigma(R_i)=X_{i,j,1}$ if and only if $\sigma(C_j)=X_{i,j,2}$,
	namely those $\sigma\in S_m$ for which there exists a corresponding permutation $\tau_{\sigma}\in S_n$ such that $\sigma(R_i)=X_{i,\tau_{\sigma}(i),1}$ and $\sigma(C_{\tau_{\sigma}(i)})=X_{i,\tau_{\sigma}(i),2}$.}

	Hence we now consider the converse: we fix $\tau\in S_n$ and let $\sigma_{\tau}\in S_m=S_{4(n+n^2)}$ be given by
	\begin{align*}
		\sigma_{\tau}(R_i)= & X_{i,\tau(i),1}, \\
		\sigma_{\tau}(C_{\tau(i)})= & X_{i,\tau(i),2}, \\
		\sigma_{\tau}(X_{i,j,1})= &
		\begin{cases}
            X_{i,j,3}, & \tau(i)=j, \\
			X_{i,j,1}, & \tau(i)\neq j, \\
		\end{cases} \\
		\sigma_{\tau}(X_{i,j,2})= &
		\begin{cases}
			X_{i,j,4}, & \tau(i)=j, \\
			X_{i,j,2}, & \tau(i)\neq j, \\
		\end{cases} \\
		\sigma_{\tau}(X_{i,j,3})= &
		\begin{cases}
			R'_i, & \tau(i)=j, \\
            X_{i,j,3}, & \tau(i)\neq j, \\
		\end{cases} \\
		\sigma_{\tau}(X_{i,j,4})= &
		\begin{cases}
			C'_j, & \tau(i)=j, \\
            X_{i,j,4}, & \tau(i)\neq j, \\
		\end{cases} \\
		\sigma_{\tau}(C'_{i})= & C_{i+1}, \\
		\sigma_{\tau}(R'_{i})= & R_{i+1}.
	\end{align*}
	It could be shown from straightforward counting in Lemma \ref{lem:inversion-counting}
	that {the number of inverse pairs for the permutation $\sigma_{\tau}$ is}
	\begin{align*}
		\ell (\sigma_{\tau})\equiv & 16n^3+4n^2-8n+1,
	\end{align*}
	hence
	\begin{align*}
		v_B(\sigma_{\tau})= & k^{16n^3+4n^2-8n+1}\prod_{i=1}^{n}A_{i,\tau(i)}.
	\end{align*}
	By the placement on the non-zero entries in $B$, all other permutations that contribute to $\mathrm{Ferm}_k(B)$
	can be obtained via arbitrarily applying swaps $(X_{i,j,1}\ X_{i,j,3})$ or $(X_{i,j,2}\ X_{i,j,4})$ where $\tau(i)\neq j$.
	That is, the Fermionant reads
	\begin{align*}
		\mathrm{Ferm}_k(B)= & \sum_{\tau\in S_n}\left[\sum_{b,c}v_B\left(\prod_{\substack{i,j\\\tau(i)\neq j}}\left((X_{i,j,1}\ X_{i,j,3})^{{b(i,j)}}(X_{i,j,2}\ X_{i,j,4})^{{c(i,j)}}\right)\sigma_{\tau}\right)\right],
	\end{align*}
    {where $b,c:\{(i,j):i,j\in[n],\tau(i)\neq j\}\to\{0,1\}$ range over all such functions.}
	Notice that if $\tau(i)\neq j$ then $\sigma_{\tau}$ permutes the consecutive four elements $X_{i,j,*}$,
	hence any swaps among them are independent with all other elements.
	Therefore, the Fermionant can be factorized as
	\begin{align*}
		\mathrm{Ferm}_k(B)= & k^{16n^3+4n^2-8n+1}\sum_{\tau\in S_n}\left[\prod_{\substack{i,j\\\tau(i)\neq j}}\sum_{b,c\in\{0,1\}}\Delta(b,c)\right]\left[{\prod_{t=1}^{n}A_{t,\tau(t)}}\right],
	\end{align*}
	where $\Delta(b,c)$ is the extra factors toward the value $v_B$ caused by applying the swaps $(X_{i,j,1}\ X_{i,j,3})^b$ and $(X_{i,j,2}\ X_{i,j,4})^c$.
	A case-by-case analysis gives
	\begin{align*}
		\Delta(0,0)=1, \Delta(0,1)=\Delta(1,0)=k^3\cdot (-k^{-1})=-k^2, \Delta(1,1)=k^4\cdot (-k^{-1})^2=k^2,
	\end{align*}
	hence the Fermionant evaluates to
	\begin{align*}
		\mathrm{Ferm}_k(B)= & k^{16n^3+4n^2-8n+1}\sum_{\tau\in S_n}\left[\prod_{\tau(i)\neq j}(1-k^2)\right]\left[{\prod_{t=1}^{n}A_{t,\tau(t)}}\right] \\
		= & k^{16n^3+4n^2-8n+1}(1-k^2)^{n^2-n}\sum_{\tau\in S_n}\left[{\prod_{t=1}^{n}A_{t,\tau(t)}}\right] \\
		= & k^{16n^3+4n^2-8n+1}(1-k^2)^{n^2-n}\mathrm{Perm}(A).
	\end{align*}
    \end{proof}
    
	\begin{lemma}
		\label{lem:inversion-counting}
		$\ell(\sigma_{\tau})\equiv 16n^3+4n^2-8n+1$.
	\end{lemma}

	\begin{proof}
        {Recall that $\sigma_{\tau}\in S_m$. We use the following thought experiment to facilitate the calculation of $\ell(\sigma_{\tau})$. We arrange the $m$ indices, in the prescribed order, in two identical and aligned rows. We then connect each element in the upper row to the corresponding position in the lower row to which it is mapped by the permutation $\sigma_{\tau}$ with a solid line. For convenience, we refer to the points in the upper row as starting points and those in the lower row as ending points. In this construction, each intersection produced by these lines corresponds to an inversion pair. Thus, the total number of intersections gives the value of $\ell(\sigma_{\tau})$.}

        {Define the function $g$ as $g(R_i)=R'_i$ and $g(C_j)=C'_j$. Let $S\equiv \{R_1,\cdots,R_n,C_1\cdots,C_n\}$ and $S'\equiv g(S)$. First consider the inversion pairs (i.e., intersections) produced by the lines whose starting points lie in $S$ and those whose ending points lie in $S'$.}
        Suppose $s,t\in S$ with $s<t$ following the order defined as indices of the matrix $B$.
        Note that $g(s)>g(t)$.
        If $\sigma_{\tau}(s)<\sigma_{\tau}(t)$, then $\sigma_{\tau}^{-1}(g(s))<\sigma_{\tau}^{-1}(g(t))$, thus we have exactly one inversion pair $(\sigma_{\tau}^{-1}(g(s)),\sigma_{\tau}^{-1}(g(t)))$.
        If $\sigma_{\tau}(s)>\sigma_{\tau}(t)$, then we have one inversion pair $(s,t)$, while $\sigma_{\tau}^{-1}(g(s))>\sigma_{\tau}^{-1}(g(t))$. {In other words, for any $s,t\in S$, the two lines starting from $s$ and $t$ cross if and only if the two lines ending at $g(s)$ and $g(t)$ do not cross, and vice versa. Thus the inversion pair count for this part} is $2n(2n-1)/2$.

        Now consider the {lines starting from the bundle of indices $R_i,C_{\tau(i)},X_{i,\tau(i),3},X_{i,\tau(i),4}$. We have $\sigma_{\tau}(R_i)=X_{i,\tau(i),1}$, $\sigma_{\tau}(C_{\tau(i)})=X_{i,\tau(i),2}$}, $\sigma_{\tau}(X_{i,\tau(i),3})=R'_i$, and $\sigma_{\tau}(X_{i,\tau(i),4})=C'_{\tau(i)}$. If a block $X_{i',j',*}$ has $j'\neq\tau(i')$, then {the lines starting and ending within this block produce 8 inversion pairs with this bundle. On the other hand, if $j'=\tau(i')$, then the lines starting and ending within $X_{i',j',*}$ produce 4 inversion pairs with this bundle (for $i\neq i'$)}. Hence the inversion pair contribution from this part is $8n(n^2-n)+4n(n-1)$.

        {Then consider the lines whose starting points lie in $S'$. Note that each such line crosses all the lines starting from $S$, thus contributing $4n^2$ inversion pairs. In addition, each such line crosses all the lines starting from the blocks $X_{i',j',*}$, contributing $8n^3$ inversion pairs. Finally, there are $(2n-1)(2n-2)/2$ intersections among the lines starting from $S'$.}

        We finally sum all the contributions for
        \begin{align*}
            \ell(\sigma_{\tau})= & 2n(2n-1)/2+8n(n^2-n)+4n(n-1)+4n^2+8n^3+(2n-1)(2n-2)/2 \\
            = & 16n^3+4n^2-8n+1.
        \end{align*}
	\end{proof}

	As an example, we consider the case where $n=2$ and $k=i$. The matrix $B$ then reads
	\begin{align*}
		B=
		\begin{pmatrix}
		  	& & & & A_{1,1} & & & & A_{1,2} & & & & & & & & & & & & & & & \\
			& & & & & & & & & & & & A_{2,1} & & & & A_{2,2} & & & & & & & \\
			& & & & & 1 & & & & & & & & 1 & & & & & & & & & & \\
			& & & & & & & & & 1 & & & & & & & & 1 & & & & & & \\
			& & & & 1 & & 1 & & & & & & & & & & & & & & & & & \\
			& & & & & 1 & & 1 & & & & & & & & & & & & & & & & \\
			& & & & i & & 1 & & & & & & & & & & & & & & & & & 1 \\
			& & & & & i & & 1 & & & & & & & & & & & & & & 1 & & \\
			& & & & & & & & 1 & & 1 & & & & & & & & & & & & & \\
			& & & & & & & & & 1 & & 1 & & & & & & & & & & & & \\
			& & & & & & & & i & & 1 & & & & & & & & & & & & & 1 \\
			& & & & & & & & & i & & 1 & & & & & & & & & 1 & & & \\
			& & & & & & & & & & & & 1 & & 1 & & & & & & & & & \\
			& & & & & & & & & & & & & 1 & & 1 & & & & & & & & \\
			& & & & & & & & & & & & i & & 1 & & & & & & & & 1 & \\
			& & & & & & & & & & & & & i & & 1 & & & & & & 1 & & \\
			& & & & & & & & & & & & & & & & 1 & & 1 & & & & & \\
			& & & & & & & & & & & & & & & & & 1 & & 1 & & & & \\
			& & & & & & & & & & & & & & & & i & & 1 & & & & 1 & \\
			& & & & & & & & & & & & & & & & & i & & 1 & 1 & & & \\
			1 & & & & & & & & & & & & & & & & & & & & & & & \\
			& & & 1 & & & & & & & & & & & & & & & & & & & & \\
			& & 1 & & & & & & & & & & & & & & & & & & & & & \\
			& 1 & & & & & & & & & & & & & & & & & & & & & & \\
		\end{pmatrix}.
	\end{align*}

\section{Entanglement-based architecture for implementing universal quantum particle sampling}
\label{sec:entanglement_encoded_implementation}
The entanglement-based architecture implements universal quantum particle sampling by mapping the exchange-phase-weighted permutation sum onto correlated measurements of a high-dimensional entangled state evolving through identical copies of the transformation. It provides an implementation of projection-resolved sampling, where the projection settings are measured separately, and their probabilities can subsequently be combined to obtain the distinguishable-particle distribution.

\subsection{General entanglement-encoded architecture}
\label{sec:general_entanglement_scheme}

For simplicity, consider a collision-free ordered input configuration $\bm{i} =(i_1,\ldots,i_N) $ and $N$ identical copies
$U^{(1)},\ldots,U^{(N)}$ of the programmed transformation $U$. Each copy contains
one photon, while the assignment of the input modes to the labelled copies is
coherently superposed over all permutations $\sigma\in S_N$.

For a fixed internal-state projection setting
$\boldsymbol{\xi}=(\xi_1,\ldots,\xi_N)$, define
\begin{equation}
\alpha_{\sigma}^{(\boldsymbol{\xi})}
=
e^{i\phi\ell(\sigma)}
\prod_{a=1}^{N}
C_{a,\sigma(a)} .
\label{eq:general_encoded_coeff}
\end{equation}
The corresponding unnormalized encoded state is
\begin{equation}
|\widetilde{\Psi}_{\mathrm{enc}}^{(\boldsymbol{\xi})}\rangle
=
\sum_{\sigma\in S_N}
\alpha_{\sigma}^{(\boldsymbol{\xi})}
\prod_{a=1}^{N}
\hat{a}_{i_{\sigma(a)}}^{(a)\dagger}
|0\rangle .
\label{eq:unnormalized_encoded_state}
\end{equation}

For distinct input modes, the permutation components are mutually orthogonal
across the labelled transformation copies, and the state normalization is
\begin{equation}
\mathcal{N}_{\mathrm{enc}}^{(\boldsymbol{\xi})}
=
\sum_{\sigma\in S_N}
\left|
\alpha_{\sigma}^{(\boldsymbol{\xi})}
\right|^2 .
\label{eq:general_encoded_norm}
\end{equation}
The normalized statistics-encoded state is therefore
\begin{equation}
|\Psi_{\mathrm{enc}}^{(\boldsymbol{\xi})}\rangle
=
\frac{1}{
\sqrt{\mathcal{N}_{\mathrm{enc}}^{(\boldsymbol{\xi})}}
}
\sum_{\sigma\in S_N}
e^{i\phi\ell(\sigma)}
\prod_{a=1}^{N}
C_{a,\sigma(a)}
\hat{a}_{i_{\sigma(a)}}^{(a)\dagger}
|0\rangle .
\label{eq:normalized_encoded_state}
\end{equation}

After each copy implements the same transformation $U$, a copy-resolved
$N$-fold correlated detection is performed at $\bm{s} = (s_1,\ldots,s_N)$, with one photon detected at output mode $s_a$ of copy $a$. The measured
correlated detection probability is
\begin{equation}
P_{\mathrm{enc}}^{(\boldsymbol{\xi})}
(\bm{s}|\bm{i},U)
=
\frac{1}{
\mathcal{N}_{\mathrm{enc}}^{(\boldsymbol{\xi})}
}
\left|
\sum_{\sigma\in S_N}
e^{i\phi\ell(\sigma)}
\prod_{a=1}^{N}
C_{a,\sigma(a)}
U_{s_a,i_{\sigma(a)}}
\right|^2 .
\label{eq:encoded_measured_probability}
\end{equation}

Comparison with
Eq.~(\ref{eq:projection_resolved_probability}) gives
\begin{equation}
P_{\phi,\boldsymbol{\xi}}^{\mathrm{proj}}
(\bm{s}|\bm{i},U)
=
\frac{
\mathcal{N}_{\mathrm{enc}}^{(\boldsymbol{\xi})}
}{N!}
P_{\mathrm{enc}}^{(\boldsymbol{\xi})}
(\bm{s}|\bm{i},U) .
\label{eq:encoded_to_projected_probability}
\end{equation}

The rescaling factor is independent of the output configuration and is
determined entirely by the programmed input state. In the fully
indistinguishable matched-projection limit, $C_{a\alpha}=1$ and $\mathcal{N}_{\mathrm{enc}}=N!$, so the ordered target distribution can be obtained by the measured correlated probability.

The relative phases of the entangled-state coefficients encode the exchange phase $\phi$, whereas their magnitudes encode the internal-state projection weights determined by the coefficients $C_{a\alpha}$, and hence the projection-resolved distinguishability structure. The scheme therefore emulates the target exchange statistics and the selected projection-resolved distinguishability structure without modifying the intrinsic bosonic statistics of the photons. 
One programmed entangled state realizes one projection-resolved setting. To
implement a conventional unresolved Gram-matrix-determined distribution, the experiment
is repeated over a complete orthonormal set of internal-state projection
settings, and the reconstructed projection-resolved probabilities are summed according to
Eq.~(\ref{eq:partial_as_sector_sum}).

\subsection{Compressed implementation for three-particle sampling}
\label{sec:compressed_three_particle}

A direct implementation of the general entanglement-based architecture for three particles would
require a three-photon entangled state and three physical copies
of $U$. Our processor instead reproduces the same ordered three-particle amplitudes
using a two-photon six-dimensional entangled state and two physical copies of
$U$.

\emph{\textbf{Three-particle sampling at full indistinguishability.}} 
Let $\bm{i} =(i_1,i_2,i_3)$ denote the three distinct input modes and let $(r,p,q)$ be an ordered output
triple. We adopt the permutation ordering
\begin{equation}
\begin{aligned}
\sigma_1&=(1,2,3),&
\sigma_2&=(1,3,2),&
\sigma_3&=(2,1,3),\\
\sigma_4&=(2,3,1),&
\sigma_5&=(3,1,2),&
\sigma_6&=(3,2,1).
\end{aligned}
\label{eq:S3_permutation_order}
\end{equation}

For a fixed output index $q$, corresponding to the output of the third copy in
the ideal three-copy scheme, define
\begin{equation}
\beta_{\sigma}^{(q,\boldsymbol{\xi})}
=
e^{i\phi\ell(\sigma)}
\left[
\prod_{a=1}^{3}
C_{a,\sigma(a)}
\right]
U_{q,i_{\sigma(3)}} .
\label{eq:q_slice_compiled_coeff}
\end{equation}
The chip generates the normalized two-photon state
\begin{equation}
|\Psi_q^{(\boldsymbol{\xi})}\rangle
=
\frac{1}{
\sqrt{\mathcal{N}_q^{(\boldsymbol{\xi})}}
}
\sum_{\sigma\in S_3}
\beta_{\sigma}^{(q,\boldsymbol{\xi})}
\hat{a}_{i_{\sigma(1)}}^{(1)\dagger}
\hat{a}_{i_{\sigma(2)}}^{(2)\dagger}
|0\rangle ,
\label{eq:q_slice_two_photon_state}
\end{equation}
where
\begin{equation}
\mathcal{N}_q^{(\boldsymbol{\xi})}
=
\sum_{\sigma\in S_3}
\left|
\beta_{\sigma}^{(q,\boldsymbol{\xi})}
\right|^2 .
\label{eq:q_slice_norm}
\end{equation}

The two photons then propagate through two identical physical copies
$U^{(1)}$ and $U^{(2)}$. The correlated coincidence probability at output
mode $r$ of the first copy and output mode $p$ of the second copy is
\begin{equation}
Q_q^{(\boldsymbol{\xi})}(r,p)
=
\frac{1}{
\mathcal{N}_q^{(\boldsymbol{\xi})}
}
\left|
\sum_{\sigma\in S_3}
e^{i\phi\ell(\sigma)}
\prod_{a=1}^{3}
C_{a,\sigma(a)}
U_{r,i_{\sigma(1)}}
U_{p,i_{\sigma(2)}}
U_{q,i_{\sigma(3)}}
\right|^2 .
\label{eq:q_slice_conditional_probability}
\end{equation}

The ordered projection-resolved three-particle probability is reconstructed as
\begin{equation}
P_{\mathrm{ord}}^{(\boldsymbol{\xi})}(r,p,q)
=
\frac{
\mathcal{N}_q^{(\boldsymbol{\xi})}
}{3!}
Q_q^{(\boldsymbol{\xi})}(r,p).
\label{eq:q_slice_ordered_reconstruction}
\end{equation}

Repeating the measurement over all $q$ reconstructs the complete ordered
distribution. The Fock-basis probability for an unordered output
configuration $\{a,b,c\}$, with $a\leq b\leq c$, is obtained by summing all
distinct ordered triples with the same sorted occupation pattern,
\begin{equation}
P_{\mathrm{Fock}}^{(\boldsymbol{\xi})}
(\{a,b,c\})
=
\sum_{\operatorname{Sort}(r,p,q)=\{a,b,c\}}
P_{\mathrm{ord}}^{(\boldsymbol{\xi})}(r,p,q).
\label{eq:Fock_from_ordered_sector}
\end{equation}

In the fully indistinguishable matched-projection limit, $C_{a\alpha}=1$, and
\begin{equation}
\mathcal{N}_q
=
2\sum_{\alpha=1}^{3}
|U_{q,i_\alpha}|^2 .
\label{eq:fully_indistinguishable_q_norm}
\end{equation}

\emph{\textbf{Three-particle sampling with tunable indistinguishability.}} 
For the indistinguishability experiments, the target internal states are taken
from the three-dimensional family defined in
Eq.~(\ref{eq:three_internal_states_Gamma}). We use the orthonormal basis
\begin{equation}
\mathcal{B}
=
\{|1\rangle,|2\rangle,|3\rangle\}
\end{equation}
and label a resolved output projection setting by
\begin{equation}
\boldsymbol{\mu}
=
(\mu_1,\mu_2,\mu_3),
\qquad
\mu_1, \mu_2, \mu_3\in
\{1,2,3\} .
\label{eq:three_level_sector}
\end{equation}

For a target value of $\Gamma$, the projection coefficient associated with
this projection setting is
\begin{equation}
C_{a\alpha}^{(\boldsymbol{\mu},\Gamma)}
=
\langle\mu_a|\eta_\alpha(\Gamma)\rangle .
\label{eq:Gamma_sector_projection_coeff}
\end{equation}
For every fixed $q$-slice and projection setting $\boldsymbol{\mu}$, the compiled
coefficient becomes
\begin{equation}
\beta_{\sigma}^{(q,\boldsymbol{\mu},\Gamma)}
=
e^{i\phi\ell(\sigma)}
\left[
\prod_{a=1}^{3}
\langle\mu_a|\eta_{\sigma(a)}(\Gamma)\rangle
\right]
U_{q,i_{\sigma(3)}} .
\label{eq:Gamma_sector_compiled_coeff}
\end{equation}

These six coefficients are encoded into the programmable two-photon
six-dimensional entangled state. The measured conditional distribution
$Q_{q,\boldsymbol{\mu}}(r,p)$ is converted to the corresponding unnormalized
projection-setting contribution according to
\begin{equation}
P_{\phi,\Gamma}^{(\boldsymbol{\mu})}(r,p,q)
=
\frac{
\mathcal{N}_{q,\boldsymbol{\mu}}^{(\Gamma)}
}{3!}
Q_{q,\boldsymbol{\mu}}(r,p),
\label{eq:Gamma_sector_probability}
\end{equation}
where
\begin{equation}
\mathcal{N}_{q,\boldsymbol{\mu}}^{(\Gamma)}
=
\sum_{\sigma\in S_3}
\left|
\beta_{\sigma}^{(q,\boldsymbol{\mu},\Gamma)}
\right|^2 .
\label{eq:Gamma_sector_norm}
\end{equation}

The unresolved partially distinguishable distribution is reconstructed by the
incoherent sum
\begin{equation}
P_{\phi,\Gamma}(r,p,q)
=
\sum_{\boldsymbol{\mu}\in\{1,2,3\}^3}
P_{\phi,\Gamma}^{(\boldsymbol{\mu})}(r,p,q).
\label{eq:Gamma_unresolved_sector_sum}
\end{equation}

The factor
$\mathcal{N}_{q,\boldsymbol{\mu}}^{(\Gamma)}/3!$ is essential because
different projection settings generally have different total probabilities.
Experimentally, projection-settings with zero programmed weight are omitted, while all
nonzero settings are measured and combined at the probability level.

This procedure sequentially emulates the projection-resolved settings of a
virtual three-dimensional internal space. It therefore realizes the full
range $0\leq\Gamma\leq1$ without requiring each physical photon to carry or be
measured in a genuine three-level internal degree of freedom.

\section{Chip and system}
This section presents further details of our silicon quantum photonic chip and the full-stack system, and the calibration results. 

\subsection{Chip}

Our silicon photonic chip is designed and fabricated on the silicon-on-insulator (SOI) platform. The layer stack of the platform from bottom to top comprises a silicon substrate, a 2-$\mu m$-thick buried silicon dioxide layer, a 220-nm-thick top silicon waveguide layer, and a 2-$\mu m$-thick silicon dioxide cladding layer covering the silicon waveguides.
Thermo-optic phase shifters are patterned on a TiN metal layer on top of the waveguide layer with the size of 160 $\mu m$ long and 2.5 $\mu m$ wide, and they work based on thermal-optical effect. Grating couplers are used for injecting laser beam and collecting the photons generated on the chip. 
The effective footprint of the device is approximately $15 \times 5 ~{\mbox{mm}}^2$, with more than 2100 components integrated. These components include 6 SWFM sources, 6 tunable filters which separate idler and signal photons, 297 thermo-optic phase shifters, 298 multimode interferometer (MMI) beamsplitters, 74 grating couplers, 104 waveguide crossers, 106 length-matching delay lines, 198 low-loss wide-waveguides, 510 electronical pads and 510 electronical transmission lines. 

The propagation loss for channel waveguide on our chip is 1.5 $\mbox{dB}/\mbox{cm}$. The grating couplers have the insertion loss 3.9 dB$/$facet in average. The multi-mode interferometers (MMI) on our chip have around 0.2 dB insertion loss. 
All the thermo-optic phase shifters each have $\sim${900} $\Omega$ resistance and the applied voltage for $2\pi$ phase shift is less than {7} {V}. 
The six on-chip filters are implemented based on asymmetric Mach–Zehnder interferometers, which are designed to separate the signal (1539.7 nm) and idler (1552.5 nm) photons, directing them into two networks respectively. With fine voltages applied, the filters yield an extinction ratio of $\sim$18 dB for both photons.

\subsection{System}

\begin{figure*}[htbp!] 
\centering    
\includegraphics[width=0.65\textwidth]{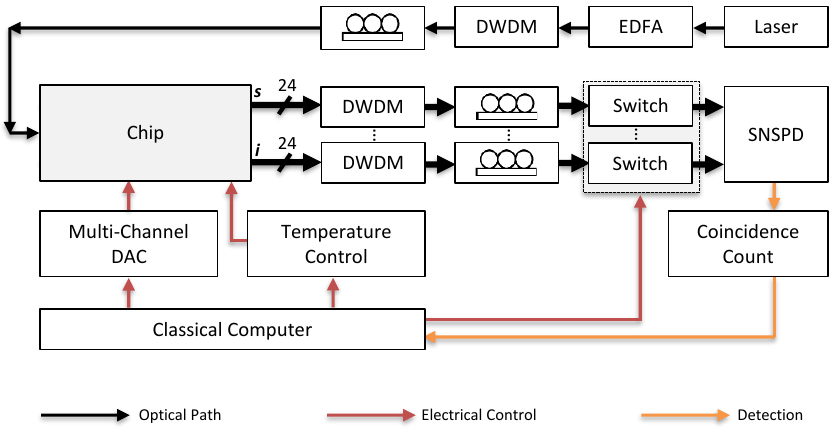}
\caption[SystemHardware]{\textbf{System hardware architecture.} Light from a tunable laser is routed to the photonic quantum chip through an erbium-doped fibre amplifier (EDFA), dense wavelength-division multiplexer (DWDM), polarization controller (PC). The signal and idler outputs are separately routed through DWDMs, PCs, and optical switches, detected by superconducting nanowire single-photon detectors (SNSPDs), and recorded by coincidence counting logic (CC). External multichannel voltage drivers configure the on-chip thermo-optic phase shifters, while a temperature controller stabilizes the chip. A central computer coordinates the photon source, routing, electrical control, detection and data acquisition.
}
\label{figSM_SystemHardware}
\end{figure*}

\subsubsection{System hardware architecture}

The schematic diagram of the whole system is shown in Fig~\ref{figSM_SystemHardware}. Bright light with the wavelength of 1546.15 nm is collected from a tunable laser and further amplified using an EDFA, up to {65 mW}. The amplified spontaneous emission (ASE) noise is suppressed using a DWDM module filter of 200 GHz. The bright light is then injected into the silicon chip via a 74-channel V-Groove fiber array (VGA) with 127 $\mu$m spacing and 8-degree angle polished from top to bottom. An in-line polarization controller is used to ensure that TE-polarized light is launched into the chip through on-chip TE grating couplers. 

Photons emerging from different outputs are collected by the same VGA, and detected by 24 fiber-coupled superconducting nanowire single-photon detectors (SNSPDs) according to the configuration of 24 two-channel optical switches (OS). The switches are divided into two groups, and each group has 12 switches. The two groups of switches are connected to the upper ($\mathrm{LON}_\mathrm{A}$) and lower ($\mathrm{LON}_\mathrm{B}$) networks respectively. 
Before the optical switches, 48 off-chip DWDM module filters (200 GHz), divided into two groups, are used to collect the signal photons from $\mathrm{LON}_\mathrm{A}$ and the idler photons from $\mathrm{LON}_\mathrm{B}$. 48 in-line polarization controllers are further used in each output channel for optimizing the photon detection efficiencies. 
By using 24 switches, 24 SNSPDs and a 24-channel coincidence counting logic (CC), all the 576 (=24$\times$24) combination of two-photon coincidence events from the two networks $\mathrm{LON}_\mathrm{A}$ and $\mathrm{LON}_\mathrm{B}$ can be obtained with 4 configurations of the optical switches. For each optical switch configuration, 144 combination of two-photon coincidence events can be measured. 

The insertion losses for the signal photon channel and the idler photon channels over full passband are {2.7 dB and 2.1 dB} respectively. Each of the DWDM filters have $\ge 45$ dB isolation over full passband for non-adjacent channels. 
After the polarization of photons being optimized by using the in-line polarization controllers, the photon detection efficiency by SNSPDs can be maximized up to 85$\%$. The two-photon coincidences are recorded in a 30 ns integration window, which was chosen according to the jitter-time of SNSPDs. During the experiments, the dark count for single photon detection is around 100 per second, which is negligible relative to the two-photon coincidence events. 

A central computer controls a digital-to-analog converter (DAC) through which all the on-chip thermal phase shifters are configured, and controls the optical switches for configuring the photon detection channel, and also connects to the coincidene counting logic for photon detection data acquisition. A Peltier cell controlled by a proportional integrative derivative (PID) controller is used to actively keep the chip temperature constant.

\subsubsection{System control software}

\begin{figure*}[htbp!] 
\centering    
\includegraphics[width=0.65\textwidth]{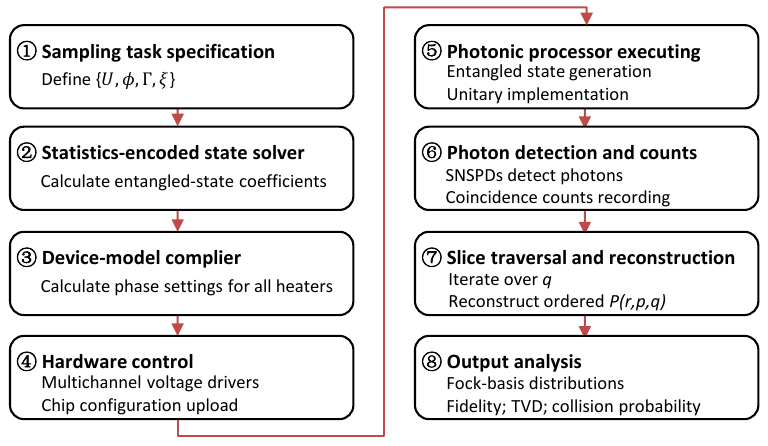}
\caption[SystemSoftware]{\textbf{System control software.} The software stack compiles a multiparticle sampling task into an experimental chip configuration. From the target problem, input modes $I$, exchange phase $\phi$, global indistinguishability $\Gamma$ and projection setting $\xi$, the software derives the target transformation $U$ and calculates the corresponding statistics-encoded entangled-state coefficients. The target network and state parameters are further converted into drive voltages for the on-chip thermo-optic phase shifters and configured to the entangled-state-generation and programmable-interferometer modules. Coincidence events are detected using SNSPDs and time-tagging coincidence counting electronics. By automatically iterating over the slice index $q$, the software reconstructs the ordered three-particle probabilities $P(r,p,q)$, obtains the Fock-basis distribution $P_{\mathrm{Fock}}$, and evaluates the statistical fidelity, total variation distance, collision probability. Calibration procedures, hardware interfaces, and state-projection and fidelity-characterization modules support the main execution workflow.}
\label{figSM_SystemSoftware}
\end{figure*}

We develop a software stack for the system, as shown in Fig.~\ref{figSM_SystemSoftware}, implementating the complete workflow from multiparticle sampling task specification to experimental distribution reconstruction. For each sampling task, the software first derives the target linear-optical transformation $U$ from the prescribed evolution. Using the particle input modes, exchange phase $\phi$, global indistinguishability $\Gamma$, and selected internal-state projection setting $\xi$, it then calculates the required statistics-encoded entangled input state. For the compressed three-particle implementation used here, this procedure automatically generates the complex amplitudes of the six-dimensional two-photon entangled state for each fixed $q$-slice, with the transfer amplitudes of one ideal network copy compiled into the state-preparation coefficients.

The target transformation and entangled-state parameters are subsequently translated into physical chip settings. The software first converts the required amplitudes and phases into phase settings for the on-chip thermo-optic phase shifters and then maps these settings to the corresponding drive voltages. The voltages are uploaded through hardware-control interfaces to external multichannel electronic drivers, which configure both the entangled-state-generation module and the programmable linear-optical networks. During operation, SNSPDs record the output photon events, while time-tagging and coincidence-counting electronics extract correlated detection events across the output channels.

The software automatically iterates over all $q$-slices, coordinating chip reconfiguration, data acquisition and processing. The complete ordered three-particle sampling distribution $P(r,p,q)$ is reconstructed from the measured two-photon coincidence distributions and their slice-dependent normalization factors. Ordered outcomes corresponding to the same occupation pattern are subsequently combined to obtain the full three-particle Fock-basis distribution. The complete sequence of configuration, acquisition and reconstruction, enables efficient and reproducible operation over large sets of sampling tasks with different networks, input configurations and particle-statistics parameters.

In addition to the task-level execution workflow, the software stack integrates the supporting modules required for reliable system operation. These include interfaces to the optical and electrical hardware, voltage-to-phase calibration of the thermo-optic phase shifters, transfer-matrix characterization of the programmable networks, calibration of the photon sources, projection measurements and fidelity characterization of the generated entangled states. 
Together, these modules provide a closed experimental loop spanning device calibration, task compilation, hardware configuration, photon-count acquisition and experimental data analysis.

\subsection{Calibration of the chip}

\subsubsection{Performance of on-chip linear optical networks}

An arbitrary unitary transformation can be implemented in a linear optical network as described by Reck et al.~\cite{reck1994experimental}, which consists of only 50:50 beam splitters and phase shifters. It requires $N(N-1)/2$ phase shifters to implement an $N \times N$ universal linear transformation. 
Therefore, a 24-mode universal linear optical network that implements an arbitrary 24-dimensional unitary operation for the optical input from any port requires 276 phase shifters in total. On our device, photons are only physically injected into ports $(1,2,3)$ of the network and thus the unnecessary phase shifters and beam-splitters were omitted in fabricated chip. We also removed the phase shifters at the end of network, since they do not contribute to the experimentally measured results. 
Although the physical input ports are fixed to $(1,2,3)$, arbitrary input configurations $(i,j,k)$ are realized by compiling a mode-permutation matrix $P_{
(1,2,3)\to (i,j,k)}$ into the programmed network. Then the implemented transformation is $U_{\mathrm{config}} = U_{\mathrm{ sampling}} P_{
(1,2,3)\to (i,j,k)} $, making the fixed-port injection operationally equivalent to direct injection into modes $(i,j,k)$ of the target sampling network.
 
\begin{figure*}[htbp!] 
\centering    
\includegraphics[width=1\textwidth]{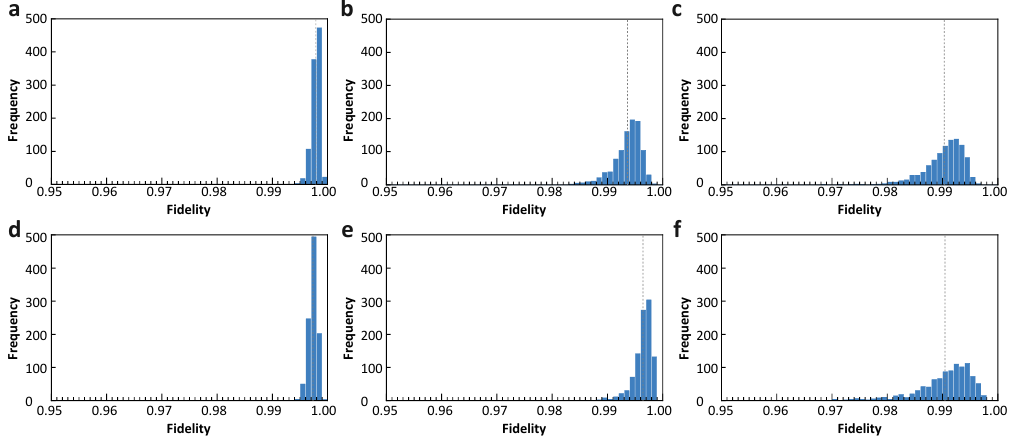}
\caption[Network]{\textbf{Performance of on-chip networks.} Histograms of the measured fidelities for 1000 randomly chosen unitary operations. \textbf{a}-\textbf{c}, random unitary operations implemented in $\mathrm{LON}_\mathrm{A}$ with $\ket{\phi_1}$, $\ket{\phi_2}$, $\ket{\phi_3}$, respectively. \textbf{d}-\textbf{f}, random unitary operations implemented in $\mathrm{LON}_\mathrm{B}$ with $\ket{\phi_1}$, $\ket{\phi_2}$, $\ket{\phi_3}$, respectively. }
\label{figSM_NetworkPerformance}
\end{figure*}

There are two copies of the same linear optical network fabricated on our device as shown in Fig. 2 in the main text, denoted as $\mathrm{LON}_\mathrm{A}$ (top) and $\mathrm{LON}_\mathrm{B}$ (bottom) respectively. 
For showcasing the performance of two on-chip linear optical networks, we programmed each of the networks to implement 1,000 randomly chosen 24-dimensional unitary operations, with different input states as $\ket{\phi_1}=[1,0,0,0,\cdots, 0]^\intercal$, $\ket{\phi_2}=[0,1,0,0,\cdots, 0]^\intercal$, $\ket{\phi_3}=[0,0,1,0,\cdots, 0]^\intercal$ respectively, and measured the corresponding optical outputs. For each instance, we calculated the fidelity $F$ between experimental obtained and ideal theoretical results as 
\begin{align}
F = \sum_{i=1}^{24} {\sqrt{P_{\mathrm{th},i}P_{\mathrm{exp},i}}}
\end{align} 
where $P_{\mathrm{th}}$ and $P_{\mathrm{exp}}$ represent the normalised theoretical and experimental optical output values. Histograms of the measured fidelities for each implemented instance are shown in Fig.~\ref{figSM_NetworkPerformance}. The mean statistical similarities of $\mathrm{LON}_\mathrm{A}$ with $\ket{\phi_1}$, $\ket{\phi_2}$, $\ket{\phi_3}$ inputs are 99.79$\pm$0.07\%, 99.36$\pm$0.25\%, and 99.03$\pm$0.36\%. 
The mean statistical similarities of $\mathrm{LON}_\mathrm{B}$ with $\ket{\phi_1}$, $\ket{\phi_2}$, $\ket{\phi_3}$ inputs are 99.74$\pm$0.08\%, 99.64$\pm$0.17\%, and 99.04$\pm$0.52\%. 

\FloatBarrier

\subsubsection{Performance of on-chip entangled photon state generation}

\begin{figure*}[htbp!] 
\centering    
\includegraphics[width=1\textwidth]{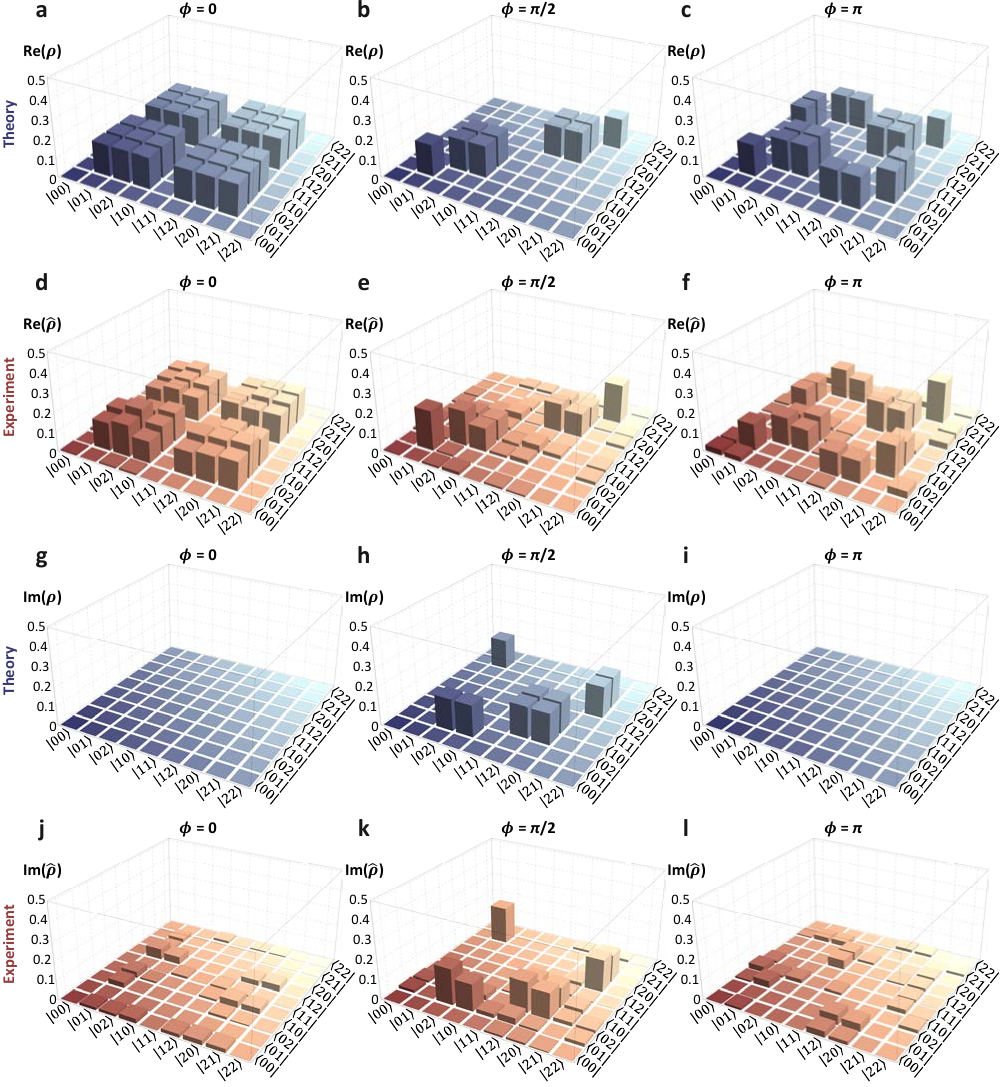}
\caption[StateGenDensity]{\textbf{Quantum state tomography of representative on-chip two-photon entangled states.} Real and imaginary parts of the target and reconstructed density matrices for the maximally entangled states that are corresponding to the six-component statistics-encoded states used in fully indistinguishable bosonic ($\phi=0$, first column) , anyonic ($\phi = \pi/2$, second column) and fermionic ($\phi = \pi$, third column) sampling. The reconstructed states exhibit fidelities of $95.10 \pm 0.42\%$,  $95.81 \pm 0.24\%$ and $91.01 \pm 0.38\%$, respectively. Density matrices were obtained from an informationally complete set of local projection measurements using maximum-likelihood estimation. 
}
\label{figSM_EngStateDensityMat}
\end{figure*}

\begin{figure*}[htbp!] 
\centering    
\includegraphics[width=1\textwidth]{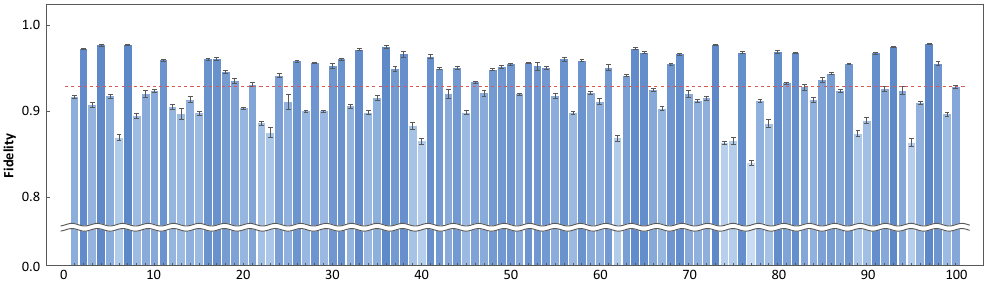}
\caption[StateGenHist]{\textbf{Fidelity benchmark of the programmable entangled-state generator.} Histogram of the reconstructed state fidelities for 100 statistics-encoded two-photon states spanning multiple exchange phases, indistinguishability, and interferometer configurations. The dashed line denotes the mean fidelity $\overline{F}=92.85 \pm 3.30\%$.}
\label{figSM_EngStateGenHistogram}
\end{figure*}

On our chip, a signal and idler photon pair is generated via the elastic scattering of two photons from a bright continuous-wave pump beam inside 1.5 cm long spiral waveguide sources, by using the SFWM nonlinear effect. The injected pump is split across six sources using a tree-structure reconfigurable interferometer. The interferometer consists of 11 thermo-optic phase shifters and 50:50 MMI beam splitters, enabling programmable configuration of the power splitting ratio and relative phase of the light injected into the six sources, with the corresponding coefficients denoted as $\alpha_0$, $\alpha_1$, $\alpha_2$, $\alpha_3$, $\alpha_4$, $\alpha_5$. Here $\sum_{i=0}^5 |\alpha_i|^2 =1$. The photon pairs are generated and coherently bunched in either one of the six sources.

The generated signal photons and idler photons are first separated by on-chip asymmetric-MZI filters, 
then undergo interference and combine via the MZIs, and are subsequently routed into $\mathrm{LON}_\mathrm{A}$ and $\mathrm{LON}_\mathrm{B}$, respectively. At the output modes of the chip, off-chip DWDM modules are employed for post-selection, to realize two-photon path entanglement. 
The two-photon entangled state is obtained as
\begin{align}
\ket{\psi} = (\alpha_0 a_k^\dag b_j^\dag+\alpha_1 a_j^\dag b_k^\dag+\alpha_2 a_l^\dag b_j^\dag+\alpha_3 a_j^\dag b_l^\dag+\alpha_4 a_l^\dag b_k^\dag+\alpha_5 a_k^\dag b_l^\dag )\ket{0}
\end{align}
where $(j,k,l)$ represents $(1,2,3)$ that are the physical input ports of the networks.

To benchmark the on-chip generation of the entangled states required for universal quantum particle sampling, we prepared a representative set of two-photon path-entangled states $\ket{\psi}$ and performed quantum state tomography. 
The two programmable networks were configured as independent local analyzers implementing an informationally complete set of projection measurements. Physical density matrices were then reconstructed from the measured coincidence counts using maximum-likelihood estimation. Figure \ref{figSM_EngStateDensityMat} shows tomography results for the representative maximally entangled states that are corresponding to the six-component statistics-encoded states used in fully indistinguishable bosonic, anyonic and fermionic sampling experiments. The reconstructed real and imaginary parts reproduce the programmed amplitude and phase structures, yielding state fidelities of $95.10 \pm 0.42\%$,  $95.81 \pm 0.24\%$ and $91.01 \pm 0.38\%$, respectively.

We further characterized 100 statistics-encoded states spanning different exchange phases, interferometers, input configurations and slice indices. Fig.~\ref{figSM_EngStateGenHistogram} shows the resulting distribution of state fidelities, with a mean fidelity of $\overline{F}=92.85 \pm 3.30\%$. These measurements verify stable and programmable generation of the high-dimensional entangled states used throughout the particle-sampling experiments.

\FloatBarrier

\section{Additional experimental results for universal quantum particle sampling}

This section provides additional experimental results for universal quantum particle sampling. As same as shown in the main text, these results include several sets of experiments: continuous programming of the exchange phase $\phi$ in the fully indistinguishable regime, independent tuning of the global indistinguishability $\Gamma$, scaling to high-dimensional random networks with dimensions up to 24, and statistical summary of hundreds of sampling instances using our processor.

\subsection{Phase-programmable sampling in the fully indistinguishable regime}

\begin{figure*}[htbp!] 
\centering    
\includegraphics[width=1\textwidth]{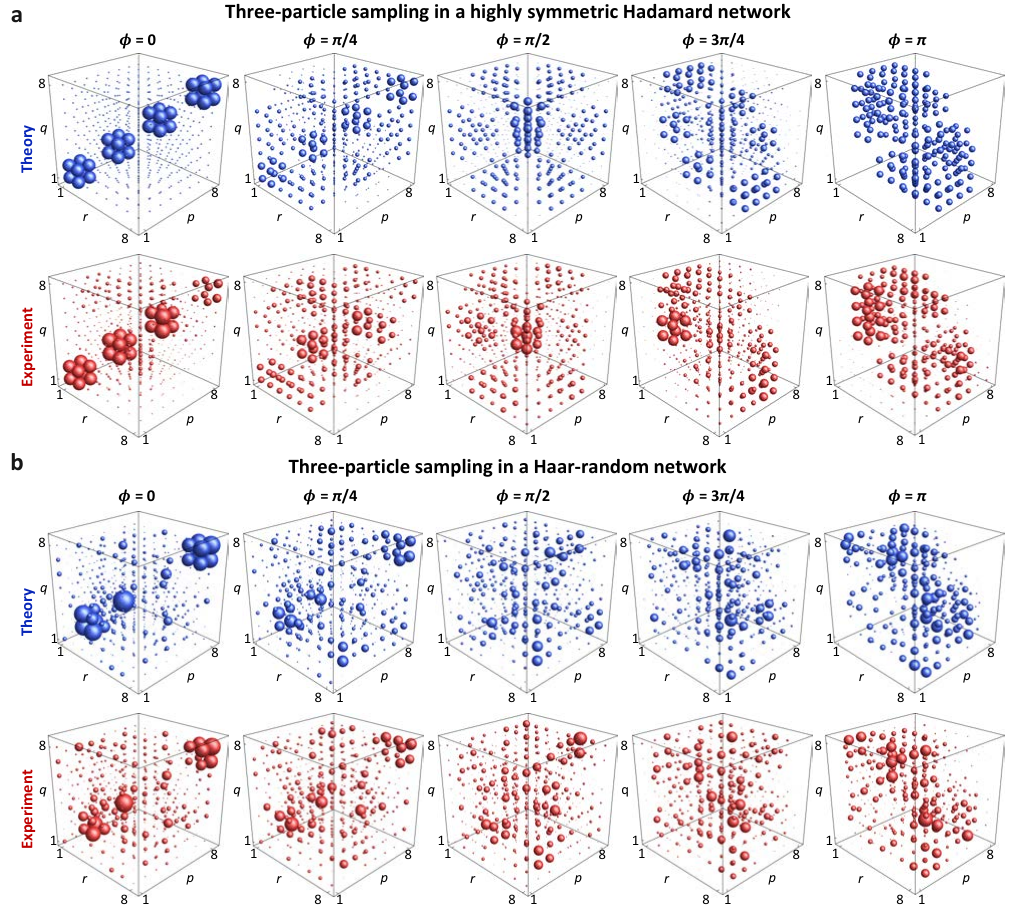}
\caption[StateGenHist]{\textbf{Phase-programmable universal quantum particle sampling in 8-dimensional interferometers.} \textbf{a}, Experimental and theoretical three-particle sampling distributions obtained in the 8-mode Hadamard interferometer $H_{8}$. \textbf{b}, Experimental and theoretical three-particle sampling distributions obtained in the a 8-mode Haar-random interferometer $U_{\mathrm{Haar},8}$. For each transformation, the global indistinguishability is fixed at $\Gamma=1$, while the exchange phase is programmed to $\phi=0$, $\pi/4$, $\pi/2$, $3\pi/4$ and $\pi$. Probabilities are shown by coloured markers. The statistical fidelity $F$ and total variation distance $D$, evaluated over the complete output distribution, are listed in Table ~\ref{tab:additional-sampling-results} for each setting.
}
\label{figSM_S5FigA1}
\end{figure*}

\begin{figure*}[htbp!] 
\centering    
\includegraphics[width=1\textwidth]{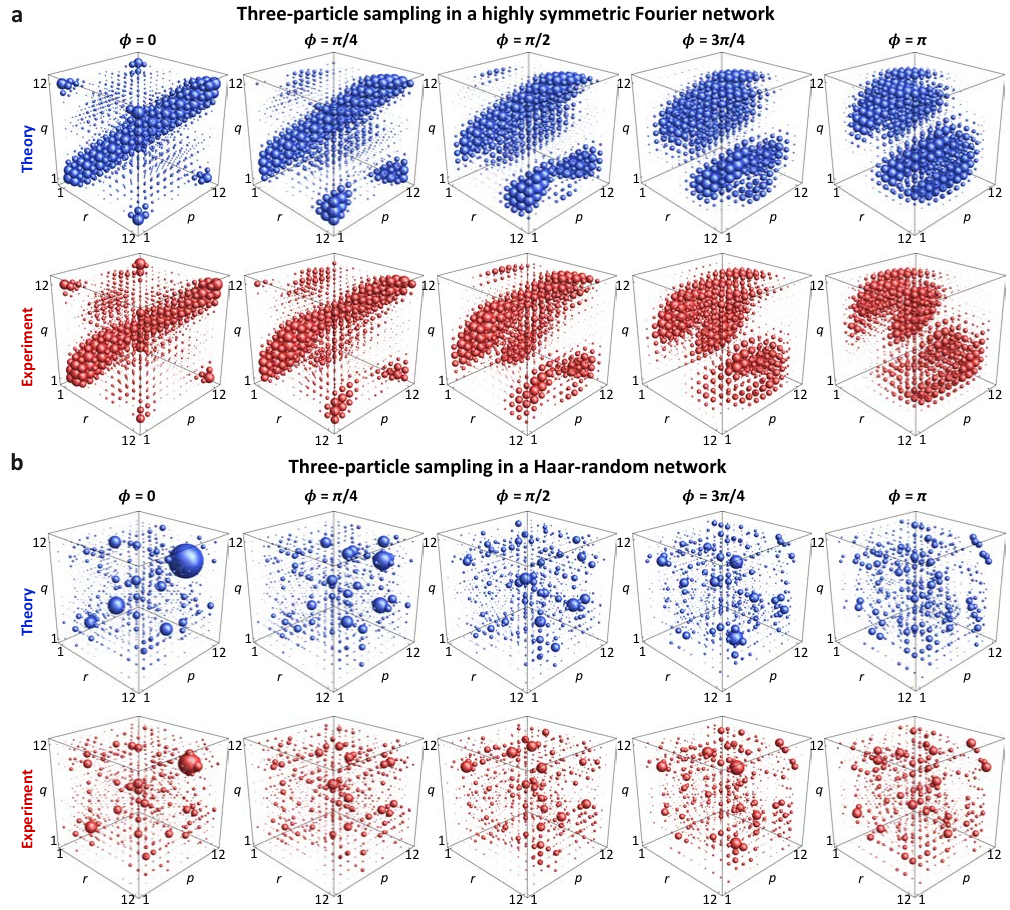}
\caption[StateGenHist]{\textbf{Phase-programmable universal quantum particle sampling in 12-dimensional interferometers.} \textbf{a}, Experimental and theoretical three-particle sampling distributions obtained in the 12-mode Fourier interferometer $F_{12}$. \textbf{b}, Experimental and theoretical three-particle sampling distributions obtained in the a 12-mode Haar-random interferometer $U_{\mathrm{Haar},12}$. For each transformation, the global indistinguishability is fixed at $\Gamma=1$, while the exchange phase is programmed to $\phi=0$, $\pi/4$, $\pi/2$, $3\pi/4$ and $\pi$. Probabilities are shown by coloured markers. The statistical fidelity $F$ and total variation distance $D$, evaluated over the complete output distribution, are listed in Table ~\ref{tab:additional-sampling-results} for each setting.
}
\label{figSM_S5FigA2}
\end{figure*}

\begin{figure*}[htbp!] 
\centering    
\includegraphics[width=1\textwidth]{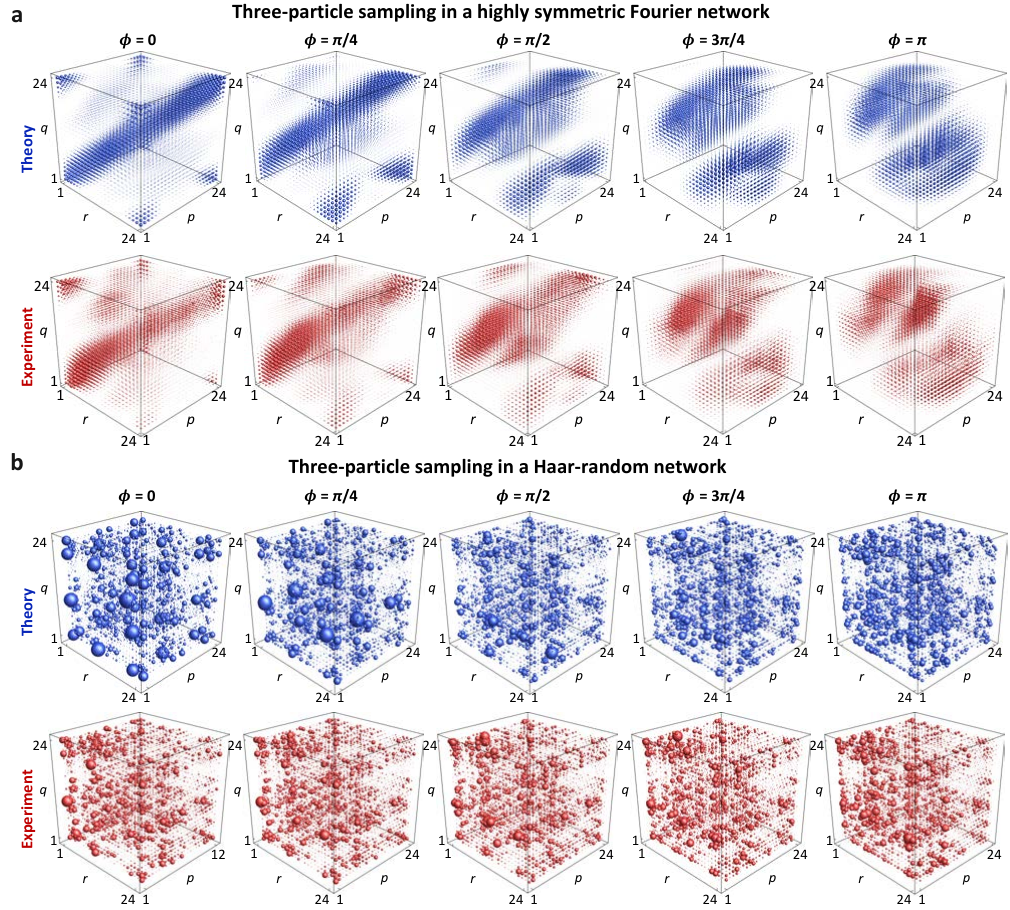}
\caption[StateGenHist]{\textbf{Phase-programmable universal quantum particle sampling in 24-dimensional interferometers.} \textbf{a}, Experimental and theoretical three-particle sampling distributions obtained in the 24-mode Fourier interferometer $F_{24}$. \textbf{b}, Experimental and theoretical three-particle sampling distributions obtained in the a 24-mode Haar-random interferometer $U_{\mathrm{Haar},24}$. For each transformation, the global indistinguishability is fixed at $\Gamma=1$, while the exchange phase is programmed to $\phi=0$, $\pi/4$, $\pi/2$, $3\pi/4$ and $\pi$. Probabilities are shown by coloured markers. The statistical fidelity $F$ and total variation distance $D$, evaluated over the complete output distribution, are listed in Table ~\ref{tab:additional-sampling-results} for each setting.
}
\label{figSM_S5FigA3}
\end{figure*}

To resolve the phase-dependent evolution of the complete output distribution, the three-particle sampling tasks in two representative 16-mode interferometers, a highly symmetric Hadamard network $H_{16}$ and a Haar-random unitary $U_{\rm Haar,16}$ have been shown in the main text. Here we show additional results in structured and Haar-random interferometers of 8, 12 and 24 modes. For each instance, the global indistinguishability was fixed at $\Gamma=1$, and the exchange phases were chosen as $\phi \in \{0, \frac{\pi}{4}, \frac{\pi}{2}, \frac{3\pi}{4}, \pi \}$. Figures \ref{figSM_S5FigA1}-\ref{figSM_S5FigA3} compare the experimental obtained ordered three-particle probability distributions with the corresponding theoretical predictions. 
As $\phi$ increases, the probability landscape evolves continuously from bosonic bunching through structured anyonic interference to fermionic
antibunching. The same evolution is observed in both the symmetric and random networks of differnt dimensions, showing that the programmable exchange statistics are not tied to a particular interferometer symmetry. The statistical fidelities and total variation distances (TVD) for all measured distributions are summarized in Table~\ref{tab:additional-sampling-results}.

\begin{table}[htp!]
\centering
\caption{
\textbf{Agreement between the theoretical and experimentally obtained 
sampling distributions for different unitary transformations and exchange 
phases.}
The agreement is quantified by the statistical fidelity $F$ and the total
variation distance $D$. The uncertainties are estimated by assuming photon Poissonian
counting statistics.
}
\label{tab:additional-sampling-results}

\footnotesize
\renewcommand{\arraystretch}{1.15}

\begin{tabular}{l|c|c|c|c|c}
\hline
\multicolumn{6}{c}{\textbf{(a) Statistical fidelity $F$}}\\
\hline
\diagbox[width=7em,trim=l]{$U$}{$\phi$}
& $0$
& $\frac{\pi}{4}$
& $\frac{\pi}{2}$
& $\frac{3\pi}{4}$
& $\pi$\\
\hline

$H_8$
& $0.97401 \pm 0.00007$
& $0.95786 \pm 0.00008$
& $0.96538 \pm 0.00008$
& $0.97132 \pm 0.00007$
& $0.92674 \pm 0.00010$ \\

$U_{\mathrm{Haar},8}$ 
& $0.96692 \pm 0.00007$
& $0.94678 \pm 0.00009$
& $0.94876 \pm 0.00009$
& $0.95030 \pm 0.00008$
& $0.92323 \pm 0.00011$ \\

$F_{12}$ 
& $0.95642 \pm 0.00007$
& $0.94868 \pm 0.00008$
& $0.94762 \pm 0.00009$
& $0.94734 \pm 0.00007$
& $0.94380 \pm 0.00008$ \\

$U_{\mathrm{Haar},12}$
& $0.95532 \pm 0.00007$
& $0.95373 \pm 0.00007$
& $0.95653 \pm 0.00007$
& $0.96047 \pm 0.00006$
& $0.94227 \pm 0.00008$ \\

$H_{16}$
& $0.96670 \pm 0.00006$
& $0.94398 \pm 0.00007$
& $0.95293 \pm 0.00008$
& $0.95031 \pm 0.00007$
& $0.88904 \pm 0.00009$ \\

$U_{\mathrm{Haar},16}$
& $0.94443 \pm 0.00007$
& $0.92774 \pm 0.00008$
& $0.92794 \pm 0.00008$
& $0.92713 \pm 0.00007$
& $0.90911 \pm 0.00009$ \\

$F_{24}$
& $0.93862 \pm 0.00006$
& $0.91779 \pm 0.00007$
& $0.91884 \pm 0.00007$
& $0.91820 \pm 0.00008$
& $0.91485 \pm 0.00008$ \\

$U_{\mathrm{Haar},24}$
& $0.94754 \pm 0.00006$
& $0.94515 \pm 0.00006$
& $0.94968 \pm 0.00005$
& $0.94704 \pm 0.00005$
& $0.92658 \pm 0.00007$ \\
\hline
\end{tabular}

\par\vspace{0.8em}

\begin{tabular}{l|c|c|c|c|c}
\hline
\multicolumn{6}{c}{\textbf{(b) Total variation distance $D$}}\\
\hline
\diagbox[width=7em,trim=l]{$U$}{$\phi$}
& $0$
& $\frac{\pi}{4}$
& $\frac{\pi}{2}$
& $\frac{3\pi}{4}$
& $\pi$\\
\hline
$H_8$ 
& $0.16367 \pm 0.00029$
& $0.21604 \pm 0.00029$
& $0.18927 \pm 0.00031$
& $0.19117 \pm 0.00027$
& $0.18716 \pm 0.00033$ \\

$U_{\mathrm{Haar},8}$ 
& $0.18854 \pm 0.00030$
& $0.22495 \pm 0.00029$
& $0.24387 \pm 0.00031$
& $0.23801 \pm 0.00026$
& $0.23011 \pm 0.00028$ \\

$F_{12}$
& $0.21102 \pm 0.00026$
& $0.23357 \pm 0.00022$
& $0.23039 \pm 0.00024$
& $0.23043 \pm 0.00023$
& $0.21922 \pm 0.00023$ \\

$U_{\mathrm{Haar},12}$
& $0.22255 \pm 0.00026$
& $0.22827 \pm 0.00024$
& $0.21962 \pm 0.00021$
& $0.21184 \pm 0.00024$
& $0.21263 \pm 0.00023$ \\

$H_{16}$
& $0.21027 \pm 0.00021$
& $0.26082 \pm 0.00020$
& $0.23187 \pm 0.00021$
& $0.24821 \pm 0.00020$
& $0.26112 \pm 0.00022$ \\

$U_{\mathrm{Haar},16}$
& $0.24955 \pm 0.00023$
& $0.28381 \pm 0.00023$
& $0.28779 \pm 0.00020$
& $0.29098 \pm 0.00019$
& $0.28480 \pm 0.00022$ \\

$F_{24}$
& $0.24388 \pm 0.00018$
& $0.28903 \pm 0.00019$
& $0.29114 \pm 0.00017$
& $0.29430 \pm 0.00018$
& $0.28418 \pm 0.00020$ \\

$U_{\mathrm{Haar},24}$
& $0.24181 \pm 0.00016$
& $0.24566 \pm 0.00018$
& $0.23415 \pm 0.00016$
& $0.23730 \pm 0.00017$
& $0.24752 \pm 0.00018$ \\
\hline
\end{tabular}

\end{table}

\FloatBarrier

\subsection{Quantum particle sampling with complete $(\phi, \Gamma)$ settings in 16-mode networks}
\begin{figure*}[htp!]
\centering
\includegraphics[width=1\textwidth]{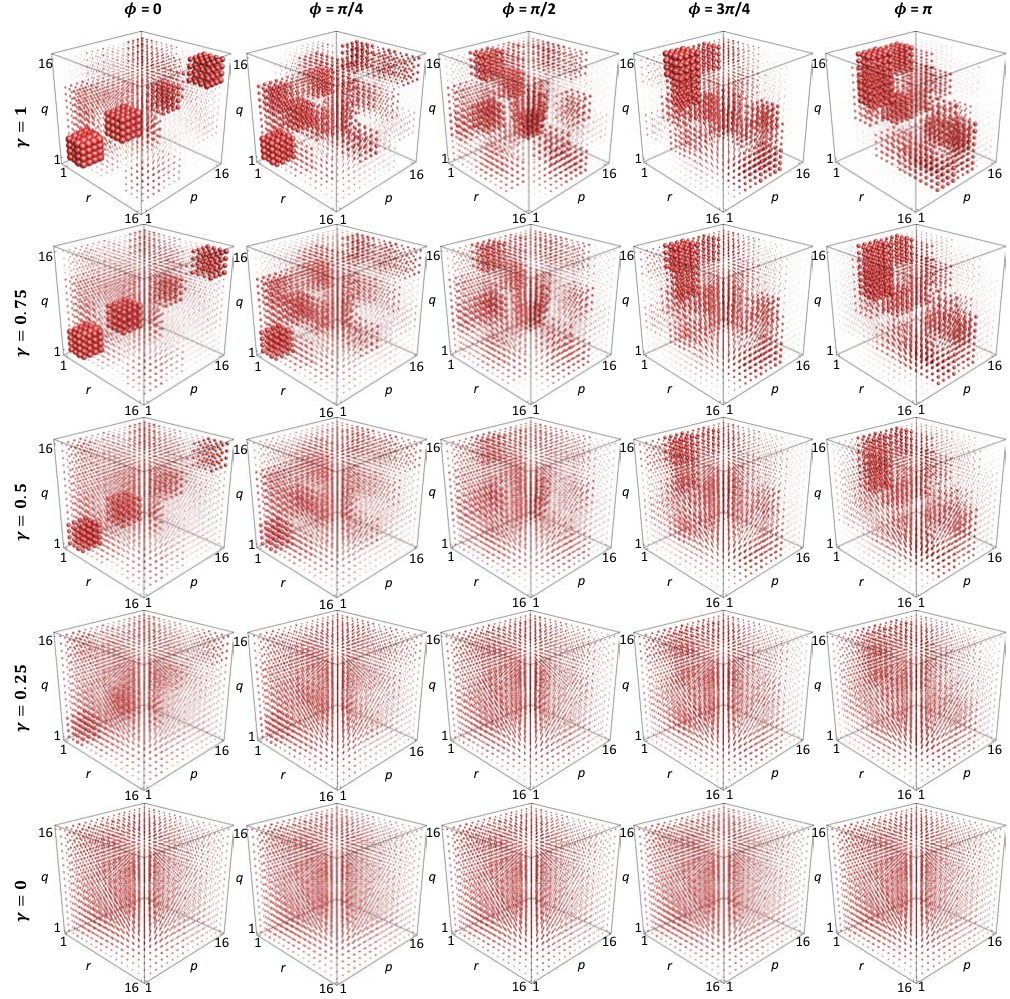}
\caption{
\textbf{Complete exchange-phase--indistinguishability sampling map
in the 16-mode Hadamard interferometer.}
Complete experimental three-particle output distributions measured in $H_{16}$ over
a $5\times5$ grid of exchange phases and
global-indistinguishability settings.
Columns correspond to
$\phi=0,\pi/4,\pi/2,3\pi/4$ and $\pi$, and rows correspond to the
programmed values
$\Gamma=0, 0.25, 0.5, 0.75, 1$, ordered from high to low
indistinguishability. 
At $\Gamma=1$, the probability landscape exhibits a pronounced
dependence on the exchange phase.
As $\Gamma$ is reduced, interference between different permutation
amplitudes is progressively suppressed and the phase-dependent
distributions approach a common distinguishable-particle limit.
}
\label{fig:H8-complete-phi-gamma-map}
\end{figure*}

\begin{figure*}[htp!]
\centering
\includegraphics[width=1\textwidth]{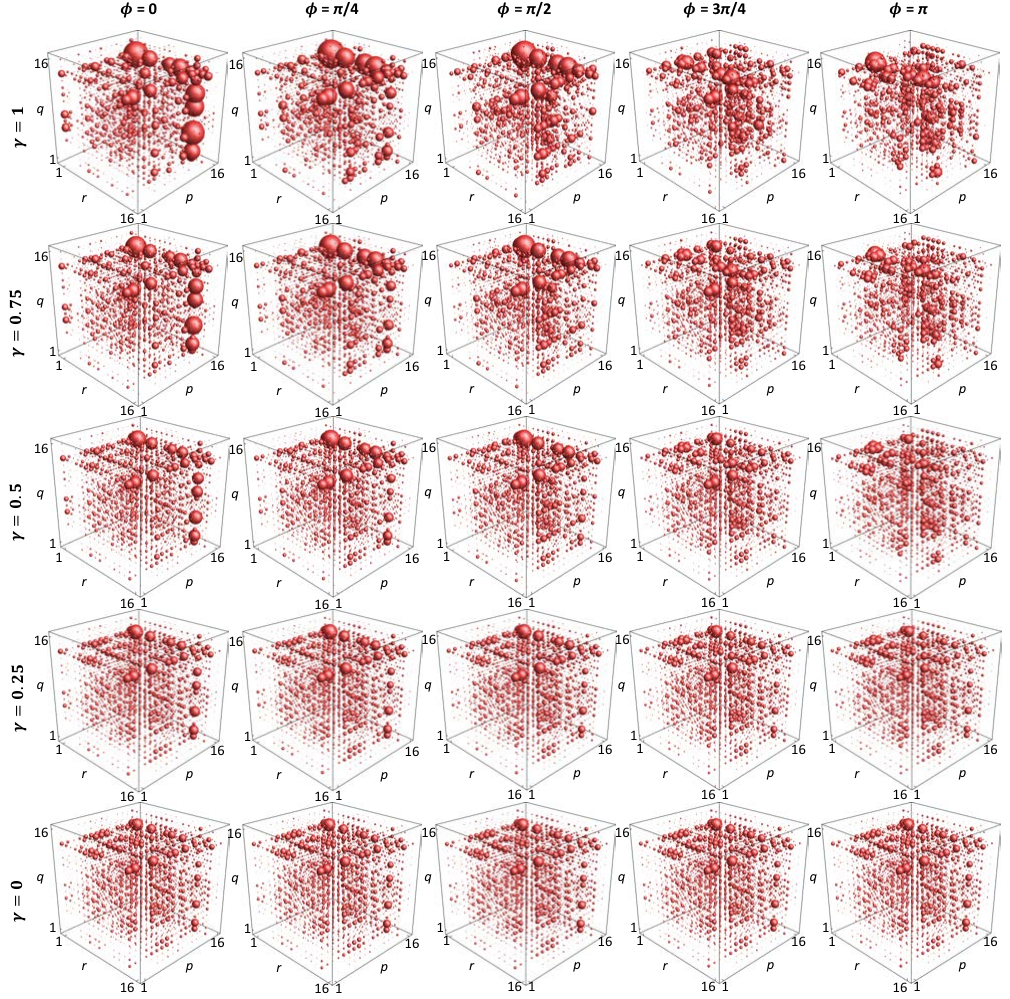}
\caption{
\textbf{Complete exchange-phase--indistinguishability sampling map
in a 16-mode Haar-random interferometer.}
Complete experimental three-particle output distributions measured in
$U_{\mathrm{Haar},16}$ over the same $5\times5$ parameter grid used
in Fig.~\ref{fig:H8-complete-phi-gamma-map}.
Columns correspond to the five programmed exchange phases, and rows
correspond to decreasing global indistinguishability. 
Despite the absence of the symmetries present in $H_{16}$, the
Haar-random transformation exhibits the same systematic transition:
exchange-phase-dependent interference is strongest at $\Gamma=1$,
is progressively reduced as $\Gamma$ decreases, and vanishes in the
distinguishable-particle limit.
}
\label{fig:Haar8-complete-phi-gamma-map}
\end{figure*}

In the main text, we show the collision probablities of three-particle quantum sampling in the 16-mode networks $H_{16}$ and $U_{\rm Haar, 16}$ with the complete settings of $(\phi, \Gamma)$. Here we present the corresponding experimental obtained ordered three-particle probability distributions for each $(\phi, \Gamma)$ setting. Fig.~\ref{fig:H8-complete-phi-gamma-map} and \ref{fig:Haar8-complete-phi-gamma-map} show the results of sampling in $H_{16}$ and $U_{\rm Haar, 16}$, respectively. 
At $\Gamma =1$, the distributions exhibit their strongest dependence on the exchange phase $\phi$. As $\Gamma$ is reduced, the coherent contributions between different permutation amplitudes are progressively suppressed, suppressing the exchange-statistics-dependent features. In the low-indistinguishability regime, distributions associated with different exchange phases converge towards a common distinguishable-particle baseline. The statistical fidelity and TVD for each parameter setting are summarized in Table~\ref{tab:phi-gamma-eight-mode}.

\begin{table}[htp!]
\centering
\caption{
\textbf{Agreement between the theoretical and experimentally measured
three-particle sampling distributions under joint programming of the
exchange phase $\phi$ and global indistinguishability $\Gamma$ in
16-mode interferometers.}
The results are shown for the Hadamard network $H_{16}$ and the
Haar-random network $U_{\mathrm{Haar},16}$. The agreement is quantified
using the statistical fidelity $F$ and the total variation distance
$D$. The uncertainties are estimated by assuming photon Poissonian
counting statistics.
}
\label{tab:phi-gamma-eight-mode}

\scriptsize
\renewcommand{\arraystretch}{1.15}
\setlength{\tabcolsep}{5pt}

\begin{tabular}{l|c|c|c|c|c}
\hline
\multicolumn{6}{c}{
\textbf{(a) Statistical fidelity $F$ for $H_{16}$}
}\\
\hline
\diagbox[width=7em]{$\Gamma$}{$\phi$}
& $0$
& $\frac{\pi}{4}$
& $\frac{\pi}{2}$
& $\frac{3\pi}{4}$
& $\pi$\\
\hline
$0$ 
& $0.99387 \pm 0.00001$
& $0.99382 \pm 0.00001$
& $0.99386 \pm 0.00001$
& $0.99380 \pm 0.00001$
& $0.99375 \pm 0.00001$ \\

$0.25$ 
& $0.99200 \pm 0.00001$
& $0.99202 \pm 0.00001$
& $0.99169 \pm 0.00001$
& $0.99203 \pm 0.00001$
& $0.99224 \pm 0.00001$ \\

$0.5$ 
& $0.98808 \pm 0.00001$
& $0.98682 \pm 0.00001$
& $0.98608 \pm 0.00001$
& $0.98652 \pm 0.00001$
& $0.98656 \pm 0.00001$ \\

$0.75$
& $0.98183 \pm 0.00002$
& $0.97576 \pm 0.00002$
& $0.97547 \pm 0.00003$
& $0.97474 \pm 0.00003$
& $0.97310 \pm 0.00003$ \\

$1$ 
& $0.96670 \pm 0.00005$
& $0.94396 \pm 0.00008$
& $0.95294 \pm 0.00006$
& $0.95031 \pm 0.00007$
& $0.88902 \pm 0.00012$ \\
\hline
\end{tabular}

\par\vspace{0.8em}

\begin{tabular}{l|c|c|c|c|c}
\hline
\multicolumn{6}{c}{
\textbf{(b) Total variation distance $D_{\mathrm{TV}}$ for $H_{16}$}
}\\
\hline
\diagbox[width=7em]{$\Gamma$}{$\phi$}
& $0$
& $\frac{\pi}{4}$
& $\frac{\pi}{2}$
& $\frac{3\pi}{4}$
& $\pi$\\
\hline
$0$ 
& $0.09017 \pm 0.00006$
& $0.09043 \pm 0.00006$
& $0.08997 \pm 0.00006$
& $0.09066 \pm 0.00005$
& $0.09115 \pm 0.00005$ \\

$0.25$ 
& $0.10323 \pm 0.00004$
& $0.10208 \pm 0.00004$
& $0.10427 \pm 0.00004$
& $0.10372 \pm 0.00004$
& $0.10185 \pm 0.00004$ \\

$0.5$ 
& $0.12562 \pm 0.00008$
& $0.13284 \pm 0.00007$
& $0.13178 \pm 0.00006$
& $0.13414 \pm 0.00006$
& $0.13507 \pm 0.00006$ \\

$0.75$ 
& $0.15679 \pm 0.00012$
& $0.18425 \pm 0.00010$
& $0.17352 \pm 0.00011$
& $0.18265 \pm 0.00013$
& $0.18955 \pm 0.00012$ \\

$1$ 
& $0.21026 \pm 0.00021$
& $0.26084 \pm 0.00021$
& $0.23187 \pm 0.00021$
& $0.24821 \pm 0.00022$
& $0.26114 \pm 0.00024$ \\
\hline
\end{tabular}

\par\vspace{1.0em}

\begin{tabular}{l|c|c|c|c|c}
\hline
\multicolumn{6}{c}{
\textbf{(c) Statistical fidelity $F$ for
$U_{\mathrm{Haar},16}$}
}\\
\hline
\diagbox[width=7em]{$\Gamma$}{$\phi$}
& $0$
& $\frac{\pi}{4}$
& $\frac{\pi}{2}$
& $\frac{3\pi}{4}$
& $\pi$\\
\hline
0 
& $0.99481 \pm 0.00001$
& $0.99477 \pm 0.00001$
& $0.99485 \pm 0.00001$
& $0.99481 \pm 0.00001$
& $0.99494 \pm 0.00001$ \\

0.25 
& $0.99330 \pm 0.00001$
& $0.99329 \pm 0.00001$
& $0.99314 \pm 0.00001$
& $0.99276 \pm 0.00001$
& $0.99250 \pm 0.00001$ \\

0.5
& $0.98676 \pm 0.00001$
& $0.98573 \pm 0.00001$
& $0.98542 \pm 0.00001$
& $0.98445 \pm 0.00001$
& $0.98392 \pm 0.00001$ \\

0.75 
& $0.97216 \pm 0.00003$
& $0.96697 \pm 0.00003$
& $0.96674 \pm 0.00003$
& $0.96562 \pm 0.00003$
& $0.96525 \pm 0.00003$ \\

1 
& $0.94442 \pm 0.00006$
& $0.92775 \pm 0.00009$
& $0.92794 \pm 0.00008$
& $0.92714 \pm 0.00009$
& $0.90911 \pm 0.00009$ \\
\hline
\end{tabular}

\par\vspace{0.8em}

\begin{tabular}{l|c|c|c|c|c}
\hline
\multicolumn{6}{c}{
\textbf{(d) Total variation distance $D_{\mathrm{TV}}$ for
$U_{\mathrm{Haar},16}$}
}\\
\hline
\diagbox[width=7em]{$\Gamma$}{$\phi$}
& $0$
& $\frac{\pi}{4}$
& $\frac{\pi}{2}$
& $\frac{3\pi}{4}$
& $\pi$\\
\hline
0
& $0.07927 \pm 0.00007$
& $0.07958 \pm 0.00007$
& $0.07876 \pm 0.00008$
& $0.07952 \pm 0.00007$
& $0.07807 \pm 0.00007$ \\

0.25 
& $0.09094 \pm 0.00005$
& $0.09104 \pm 0.00005$
& $0.09239 \pm 0.00005$
& $0.09461 \pm 0.00005$
& $0.09562 \pm 0.00005$ \\

0.5
& $0.12863 \pm 0.00007$
& $0.13524 \pm 0.00008$
& $0.13810 \pm 0.00007$
& $0.14344 \pm 0.00007$
& $0.14215 \pm 0.00008$ \\

0.75 
& $0.18604 \pm 0.00012$
& $0.20416 \pm 0.00014$
& $0.20792 \pm 0.00013$
& $0.21297 \pm 0.00011$
& $0.20829 \pm 0.00013$ \\

1 
& $0.24956 \pm 0.00020$
& $0.28381 \pm 0.00025$
& $0.28781 \pm 0.00020$
& $0.29094 \pm 0.00022$
& $0.28475 \pm 0.00021$ \\
\hline
\end{tabular}

\end{table}

\FloatBarrier

\subsection{Quantum particle sampling with complete $(\phi, \Gamma)$ settings in other networks}

\begin{figure*}[htp!]
\centering
\includegraphics[width=1\textwidth]{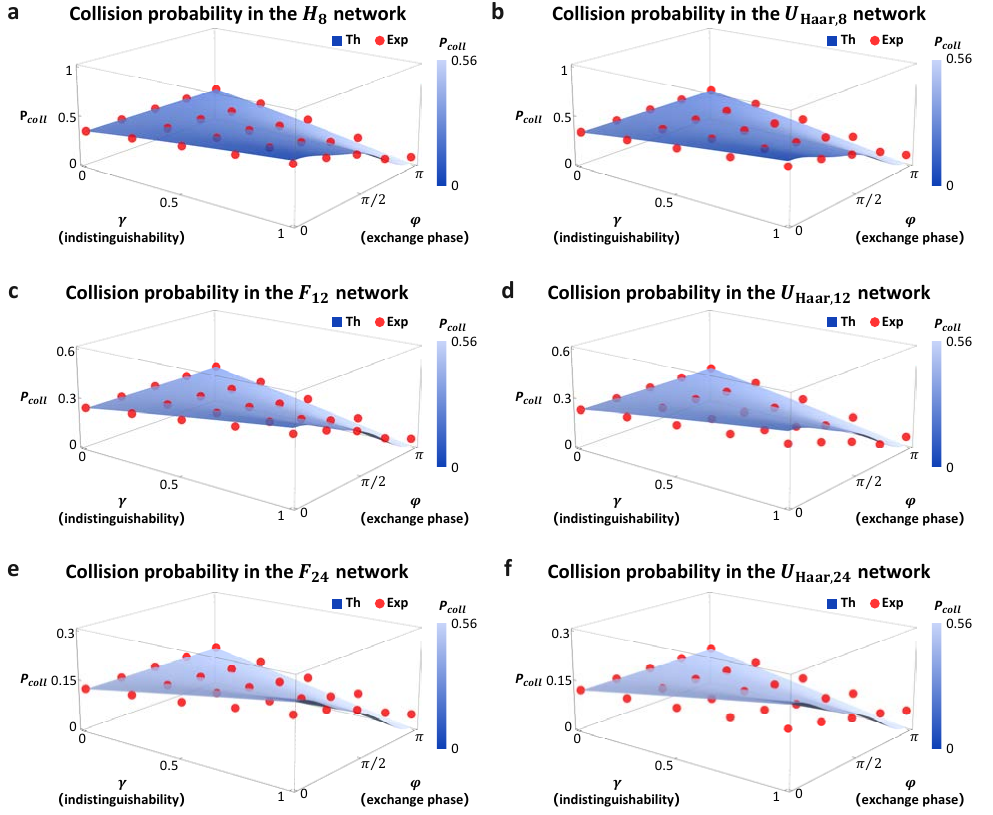}
\caption{
\textbf{Joint control of exchange phase and global
indistinguishability in higher-dimensional networks.}
\textbf{a--f}, Collision-probability landscapes
$P_{\mathrm{coll}}(\phi,\Gamma)$ measured in $H_{8}$
(\textbf{a}), $U_{\mathrm{Haar},8}$ (\textbf{b}), $F_{12}$
(\textbf{c}), $U_{\mathrm{Haar},12}$ (\textbf{d}), $F_{24}$
(\textbf{e}) and $U_{\mathrm{Haar},24}$ (\textbf{f}).
The collision probability is defined as the total probability of
output configurations in which at least two particles occupy the
same spatial mode.
Theoretical predictions are shown as continuous surfaces and
experimental results as discrete markers.
At high indistinguishability, the collision probability varies
strongly with the exchange phase.
As $\Gamma$ decreases, the phase-dependent contrast is reduced, and
the results converge towards the distinguishable-particle baseline. 
Summary agreement metrics for all networks are reported in
Table~\ref{tab:higher-dimensional-joint-control}.
}
\label{fig:higher-dimensional-joint-control}
\end{figure*}

We further investigated the joint programmability of exchange phase and indistinguishability in the quantum particle sampling tasks of structured and Haar-random interferometers of 8, 12, and 24 modes. For each instance, we measured the collision probability $P_{\mathrm{coll}}(\phi, \Gamma)$ over the 25 settings in total, where $\phi$ is chosen from $ \{0, \frac{\pi}{4}, \frac{\pi}{2}, \frac{3\pi}{4}, \pi \}$ and $\Gamma$ is chosen from $\{0, 0.25, 0.5, 0.75,1 \}$. Fig.~\ref{fig:higher-dimensional-joint-control} compares the theoretical and measured collision-probability landscapes. All networks show the same behaviours: the collision probability depends strongly on $\phi$ in the fully indistinguishable regime, the phase-dependent contrast decreases as $\Gamma$ is reduced, and all exchange phases converge to a common classical baseline in the distinguishable-particle limit. These measurements show that the independent programming of $\phi$ and $\Gamma$ is maintained across different network dimensions and transformation classes.
Table~\ref{tab:higher-dimensional-joint-control} summaries all the experimental results.

\begin{table*}[htp!]
\centering
\caption{
\textbf{Summary of the joint control of exchange phase and global
indistinguishability in higher-dimensional networks.}
For each transformation, the number of parameter settings
$N_{\mathrm{set}}=25$ corresponds to five exchange phases and five
global-indistinguishability settings.
The quantities $\overline{F}$ and $\overline{D}_{\mathrm{TV}}$
denote the mean statistical fidelity and mean total variation distance,
respectively, evaluated over all parameter settings, whereas
$F_{\min}$ denotes the minimum fidelity. 
The uncertainties are estimated by assuming photon Poissonian
counting statistics. 
The quantities $\langle\delta F\rangle$ and
$\langle\delta D_{\mathrm{TV}}\rangle$ denote the mean
one-standard-deviation uncertainties of the fidelity and total
variation distance, respectively.
}
\label{tab:higher-dimensional-joint-control}

\footnotesize
\renewcommand{\arraystretch}{1.20}
\setlength{\tabcolsep}{4pt}

\resizebox{\textwidth}{!}{%
\begin{tabular}{c|c|c|c|c|c|c|c|c}
\hline

\textbf{Dimension}
&
\textbf{Transformation}
&
\textbf{Network type}
&
\shortstack{\textbf{Number of}\\
\textbf{settings}\\
$N_{\mathrm{set}}$}
&
\shortstack{\textbf{Mean fidelity}\\
$\overline{F}$}
&
\shortstack{\textbf{Minimum fidelity}\\
$F_{\min}$}
&
\shortstack{\textbf{Mean TVD}\\
$\overline{D}_{\mathrm{TV}}$}
&
\shortstack{\textbf{Mean fidelity}\\
\textbf{uncertainty}\\
$\langle\delta F\rangle$}
&
\shortstack{\textbf{Mean TVD}\\
\textbf{uncertainty}\\
$\langle\delta D_{\mathrm{TV}}\rangle$}
\\

\hline

8  & $H_8$        & Hadamard    & 25 & 0.98613 & 0.92673 & 0.10870 & 0.01653 & 0.05034 \\
8  & $U_{\mathrm{Haar},8}$   & Haar-random & 25 & 0.98294 & 0.92323 & 0.11853 & 0.02040 & 0.06715 \\  
12 & $F_{12}$       & Fourier     & 25 & 0.98020   & 0.94385 & 0.14119 & 0.01711 & 0.05095 \\
12 & $U_{\mathrm{Haar},12}$  & Haar-random & 25 & 0.98293 & 0.94231 & 0.12845 & 0.01626 & 0.05570 \\
16 & $H_{16}$        & Hadamard     & 25 & 0.98957   &0.90454 &0.09262 & 0.01819 & 0.03014 \\
16 & $U_{\mathrm{Haar},16}$  & Haar-random & 25 & 0.99054 & 0.94409 & 0.09146 & 0.01191 &0.04161\\
24 & F(24)       & Fourier     & 25 & 0.97131 & 0.91504 & 0.16412  & 0.02734 & 0.07313 \\
24 & $U_{\mathrm{Haar},24}$ & Haar-random & 25 & 0.97855  & 0.92673 & 0.14331 & 0.01990  & 0.06084 \\

\hline
\end{tabular}%
}

\end{table*}

\FloatBarrier

\subsection{Three-particle qauntum sampling in 24-mode Haar-random interferometer}

\begin{figure*}[htp!]
\centering
\includegraphics[width=1\textwidth]{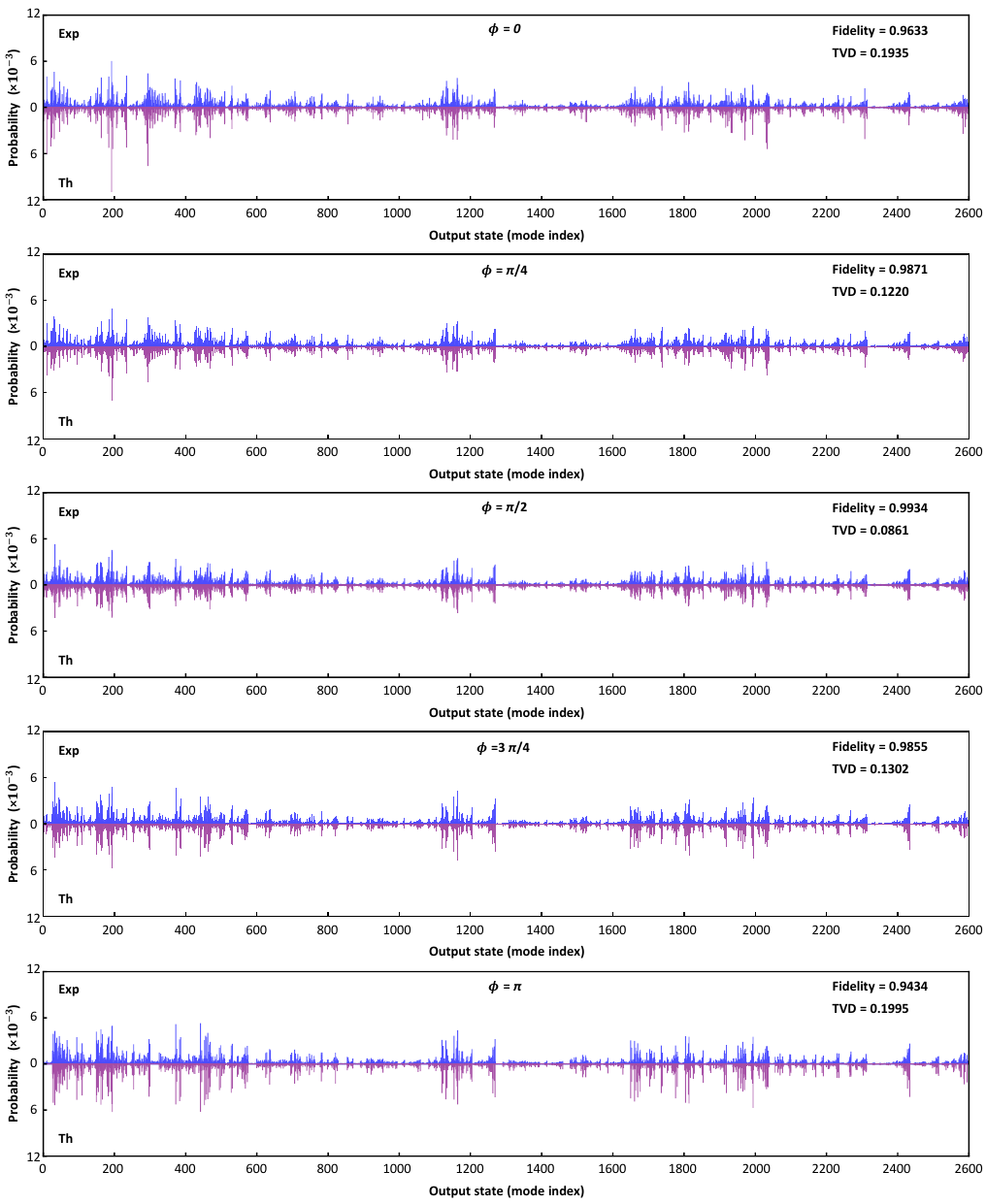}
\caption{
\textbf{Detailed three-particle sampling results in a 24-mode
Haar-random interferometer.} 
Measured and ideal theoretical output probability distributions for three-particle sampling in a 24-dimensional Haar-random network at the ideal quantum limit ($\Gamma = 1$), are shown for exchange phases $\phi =0, \pi/4, \pi/2, 3\pi/4, \pi$, respectively. Statistical fidelity $F$ and total variation distance are listed for each case.
}
\label{fig:detailed-24-mode-samplingS5FigD1}
\end{figure*}

\begin{figure*}[htp!]
\centering
\includegraphics[width=1\textwidth]{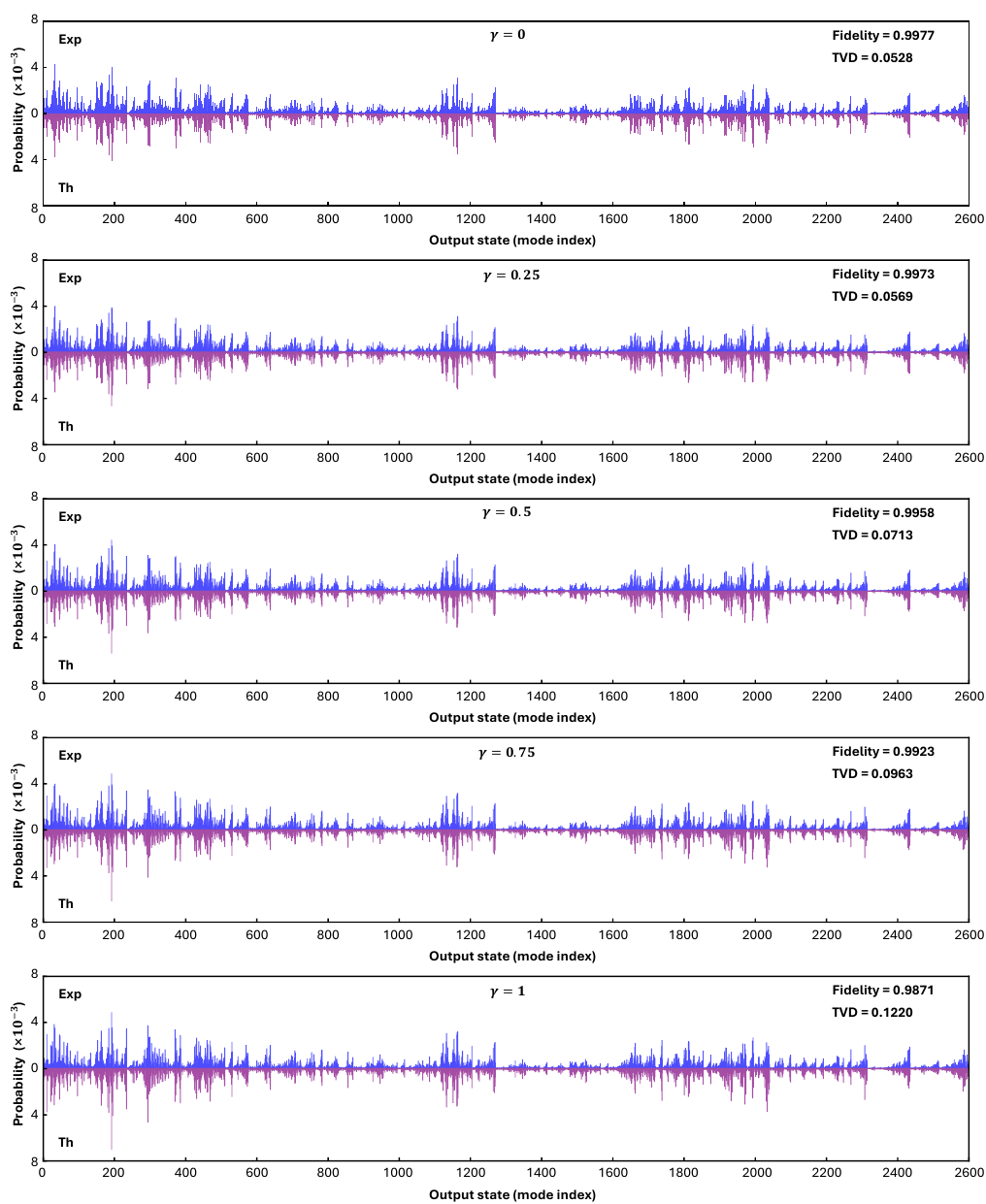}
\caption{
\textbf{Detailed three-particle sampling results in a 24-mode
Haar-random interferometer.}
Measured and ideal theoretical output probability distributions for three-particle sampling in a 24-dimensional Haar-random network with the exchange phase $\phi = \pi/4$, at different global indistinguishability $\Gamma = 0, 0.25, 0.5, 0.75, 1$, respectively. Statistical fidelity $F$ and total variation distance are listed for each case.
}
\label{fig:detailed-24-mode-samplingS5FigD2}
\end{figure*}

In addition to the result of one representative three-particle sampling experiment in a 24-mode Haar-random interferometer, we present here more experimental results of three-particle sampling in the same 24-mode Haar-random interferometer. Each of the instances spans a Fock space of dimension 2600. 
Fig. \ref{fig:detailed-24-mode-samplingS5FigD1} shows the sampling results with different exchange phases as $\phi \in \{0, \pi/4, \pi/2, 3\pi/4, \pi \}$ at $\Gamma=1$. Fig. \ref{fig:detailed-24-mode-samplingS5FigD2} shows the sampling results with different global indistinguishability as $\Gamma \in \{0, 0.25, 0.5, 0.75,1 \}$ at $\phi= \pi/2$. Here all the reported fidelity and TVDs are evaluated are using the complete 2600-dimensional probability distributions. 
The corresponding numerical results and counting uncertainties are listed in Table~\ref{tab:detailed-24-mode-sampling-compact}.

\begin{table*}[htp!]
\centering
\caption{
\textbf{Detailed results for three-particle sampling in a 24-mode
Haar-random interferometer.}
The statistical fidelity and total variation distance are evaluated
over the complete 2,600-dimensional output distribution. The uncertainties are estimated by assuming photon Poissonian
counting statistics.
}
\label{tab:detailed-24-mode-sampling-compact}

\footnotesize
\renewcommand{\arraystretch}{1.20}
\setlength{\tabcolsep}{5pt}

\resizebox{\textwidth}{!}{%
\begin{tabular}{c|c|c|c|c|c|c|c|c}
\hline

\textbf{Scan}
&
$\boldsymbol{\phi}$
&
$\boldsymbol{\Gamma}$
&
\shortstack{\textbf{Statistical fidelity}\\
$F\pm\delta F$}
&
\shortstack{\textbf{Total variation distance}\\
$D_{\mathrm{TV}}\pm\delta D_{\mathrm{TV}}$}
&
\shortstack{\textbf{Experimental collision}\\
\textbf{probability}\\
$P_{\mathrm{coll}}^{\mathrm{exp}}
\pm\delta P_{\mathrm{coll}}$}
&
\shortstack{\textbf{Theoretical collision}\\
\textbf{probability}\\
$P_{\mathrm{coll}}^{\mathrm{th}}$}
&
\shortstack{\textbf{Coincidence events}\\
$N_{\mathrm{coin}}$}
&
\shortstack{\textbf{Acquisition time}\\
$T_{\mathrm{acq}}$}
\\

\hline
\multicolumn{9}{c}{
\textbf{(a) Exchange-phase scan at $\Gamma=1$}
}\\
\hline

Phase
& $0$
& $1$
& $0.96325 \pm 0.00004$
& $0.19363 \pm 0.00015$
& $0.15740 \pm 0.00016$
& $0.22256$
& $9843240$
& $319.302\,\mathrm{s}$ \\

Phase
& $\frac{\pi}{4}$
& $1$
& $0.98703 \pm 0.00002$
& $0.12214 \pm 0.00017$
& $0.13584 \pm 0.00012$
& $0.17484$
& $1001579$
& $319.303\,\mathrm{s}$ \\

Phase
& $\frac{\pi}{2}$
& $1$
& $0.99340 \pm 0.00002$
& $0.08635 \pm 0.00018$
& $0.10761 \pm 0.00011$
& $0.11981$
& $1032954$
& $324.153\,\mathrm{s}$ \\

Phase
& $\frac{3\pi}{4}$
& $1$
& $0.98549 \pm 0.00003$
& $0.13035 \pm 0.00016$
& $0.08265 \pm 0.00010$
& $0.05625$
& $1055984$
& $320.546\,\mathrm{s}$ \\

Phase
& $\pi$
& $1$
& $0.94340 \pm 0.00006$
& $0.19961 \pm 0.00017$
& $0.05227 \pm 0.00009$
& $0$
& $1042154$
& $319.488\,\mathrm{s}$ \\

\hline
\multicolumn{9}{c}{
\textbf{(b) Global-indistinguishability scan at $\phi=\pi/2$}
}\\
\hline

Indistinguishability
& $\frac{\pi}{2}$
& $\Gamma_{1}$
& $0.99765 \pm 0.00001$
& $0.05288 \pm 0.00006$
& $0.11803 \pm 0.00004$
& $0.11768$
& $120000004$
& $2032.52\,\mathrm{s}$ \\

Indistinguishability
& $\frac{\pi}{2}$
& $\Gamma_{2}$
& $0.99766 \pm 0.00001$
& $0.05295 \pm 0.00004$
& $0.11656 \pm 0.00003$
& $0.11794$
& $269995715$
& $5277.88\,\mathrm{s}$ \\

Indistinguishability
& $\frac{\pi}{2}$
& $\Gamma_{3}$
& $0.99741 \pm 0.00001$
& $0.05609 \pm 0.00006$
& $0.11377 \pm 0.00004$
& $0.11843$
& $245550542$
& $5209.92\,\mathrm{s}$ \\

Indistinguishability
& $\frac{\pi}{2}$
& $\Gamma_{4}$
& $0.99596 \pm 0.00001$
& $0.06964 \pm 0.00011$
& $0.11055 \pm 0.00008$
& $0.11906$
& $246424598$
& $5216.3\,\mathrm{s}$ \\

Indistinguishability
& $\frac{\pi}{2}$
& $\Gamma_{5}$
& $0.99340 \pm 0.00002$
& $0.08635 \pm 0.00018$
& $0.10761 \pm 0.00011$
& $0.11981$
& $103295376$
& $324.153\,\mathrm{s}$ \\

\hline
\end{tabular}%
}

\end{table*}

\FloatBarrier

\subsection{Extended results of quantum particle sampling}

Using our device, we implemented extended instances of quantum particle sampling spanning a broad range of network structures, dimensions, input configurations, exchange phases and global indistinguishability settings. The whole dataset is listed in Table~\ref{tab:benchmark-dataset-composition}, including both structured transformations, such as Hadamard and Fourier networks, and Haar-random transformations with dimensions ranging from 8 to 24 modes. For each instance, we calculated the statistical fidelity and TVD between the experimental measured and theoretical distributions. Across the full benchmark set, our device maintains consistently high fidelities and low total variation distances, demonstrating stable programmable operations across particle statistics, coherence settings, network dimensions and transformation classes.

\begin{table*}[htp!]
\centering
\caption{
\textbf{Composition and performance summary of the 700-instance
universal quantum particle-sampling benchmark dataset.}
The benchmark set contains structured and Haar-random transformations
with dimensions ranging from 8 to 24 modes, multiple exchange phases,
and different global-indistinguishability settings.
For each subset, $N_{\mathrm{inst}}$ denotes the number of sampling
instances, $\overline{F}$ and $F_{\min}$ denote the mean and minimum
statistical fidelities, and $\overline{D}$ denotes the
mean total variation distance.
The quantities $\langle\delta F\rangle$ and
$\langle\delta D \rangle$ are the mean
one-standard-deviation uncertainties.
}
\label{tab:benchmark-dataset-composition}

\footnotesize
\renewcommand{\arraystretch}{1.20}
\setlength{\tabcolsep}{5pt}

\resizebox{\textwidth}{!}{%
\begin{tabular}{c|l|c|c|c|c|c|c}
\hline

\textbf{Category}
&
\textbf{Subset}
&
\shortstack{\textbf{Number of}\\
\textbf{instances}\\
$N_{\mathrm{inst}}$}
&
\shortstack{\textbf{Mean fidelity}\\
$\overline{F}$}
&
\shortstack{\textbf{Minimum fidelity}\\
$F_{\min}$}
&
\shortstack{\textbf{Mean TVD}\\
$\overline{D}_{\mathrm{TV}}$}
&
\shortstack{\textbf{Mean fidelity}\\
\textbf{uncertainty}\\
$\langle\delta F\rangle$}
&
\shortstack{\textbf{Mean TVD}\\
\textbf{uncertainty}\\
$\langle\delta D_{\mathrm{TV}}\rangle$}
\\

\hline
\multicolumn{8}{c}{
\textbf{(a) Composition by exchange phase}
}\\
\hline

Exchange phase
& Bosonic limit, $\phi=0$
& 140 & 0.96750 & 0.92645 & 0.18150 & 0.01771 & 0.05916 \\

Exchange phase
& Fractional statistics, $\phi=\pi/4$
& 140 & 0.96573 & 0.91136 & 0.18548 & 0.01859 & 0.06071 \\

Exchange phase
& Fractional statistics, $\phi=\pi/2$
& 140 & 0.96657 & 0.91316 & 0.18359 & 0.01821 & 0.06036 \\

Exchange phase
& Fractional statistics, $\phi=3\pi/4$
& 140 & 0.96763 & 0.91750 & 0.17993 & 0.01763 & 0.05850 \\

Exchange phase
& Fractional statistics, $\phi=\pi$
& 140 & 0.95430 & 0.89705 & 0.18249 & 0.02569 & 0.06009 \\

\hline
\multicolumn{8}{c}{
\textbf{(b) Composition by network dimension}
}\\
\hline

Dimension
& 8 modes
& 175 & 0.96695 & 0.90237 & 0.17096 & 0.02066 & 0.05972 \\

Dimension
& 12 modes
& 175 & 0.96771 & 0.92603 & 0.17566 & 0.01690 & 0.05104 \\

Dimension
& 16 modes
& 175 & 0.96664 & 0.93184 & 0.17812 & 0.01782 & 0.05502 \\

Dimension
& 24 modes
& 175 & 0.95609 & 0.89705 & 0.20566 & 0.02341 & 0.06592 \\

\hline
\multicolumn{8}{c}{
\textbf{(c) Composition by global indistinguishability}
}\\
\hline

Indistinguishability
& Fully indistinguishable, $\Gamma = 1$
& 500 & 0.95329 & 0.89705 & 0.21708 & 0.01218 & 0.02436 \\

Indistinguishability
& Partially distinguishable, $\Gamma = 0.5$
& 100 & 0.98908 & 0.98147 & 0.11516 & 0.00268 & 0.01359 \\

Indistinguishability
& Fully distinguishable, $\Gamma = 0$
& 100 & 0.99489 & 0.99081 & 0.07762 & 0.00182 & 0.01406 \\

\hline
\multicolumn{8}{c}{
\textbf{(d) Overall benchmark performance}
}\\
\hline

Overall
& Complete benchmark dataset
& 700
& 0.96435
& 0.89705
& 0.18260
& 0.02039
& 0.05963 \\
\hline
\end{tabular}%
}

\end{table*}

\FloatBarrier

\section{Validation tests of quantum particle sampling}

We present in this section the additional results supporting the statistical-validation tests shown in Fig. 4C-G of the main text. A suite of complementary statistical tests is used to reject the most relevant alternative models: uniform sampling, fully distinguishable-particle distribution, the bosonic limiting distribution, the fermionic limiting distribution, and an alternative fractional-statistics distribution characterized by a different exchange phase. Besides the representative example in the main text, here we show more experimental instances with additional network dimensions, transformation matrices, and exchange phases. 

The experimental event sequences and the numerically generated reference samples were processed using identical event-list construction, thresholds, priors and update rules. The theoretical probabilities were calculated from the target transformation matrix programmed on the chip. 
Considering that the measurements are acquired by sequentially scanning the output configurations, an event list was constructed by repeating each output configuration according to its measured count and was randomly shuffled before evaluating the cumulative validation statistic. 
This procedure removes artificial structures arising from the deterministic scanning order without altering the measured event frequencies or the empirical output distribution.
Statistical uncertainties were evaluated independently by Monte Carlo resampling of the measured coincidence counts under Poissonian counting statistics. For each resampled dataset, the event list was reconstructed and randomly permuted before repeating the complete validation analysis. 
The experimental datasets used for the additional statistical-validation tests are summarized in Table~\ref{tab:validation-datasets}.

\begin{table*}[htp!]
\centering
\caption{
\textbf{Experimental datasets used for the additional
statistical-validation tests.}
Each validation instance is specified by the network dimension,
programmed transformation, logical input configuration, exchange
phase, and global indistinguishability. }
\label{tab:validation-datasets}

\footnotesize
\renewcommand{\arraystretch}{1.15}
\setlength{\tabcolsep}{5pt}

\begin{tabular}{c|c|c|c|c|c|c}
\hline
\textbf{ID}
&
\textbf{Dimension}
&
\textbf{Transformation}
&
\shortstack{\textbf{Network}\\\textbf{type}}
&
\shortstack{\textbf{Input}\\$(i,j,k)$}
&
$\boldsymbol{\phi}$
&
$\boldsymbol{\Gamma}$
\\
\hline

1 & 8  & $U_{\mathrm{Haar},8}$     & Haar-random & $(1,2,3)$
  & $\pi/4$  & 1  \\

2 & 8  & $U_{\mathrm{Haar},8}$     & Haar-random & $(1,2,3)$
  & $\pi/2$  & 1  \\

3 & 8  & $U_{\mathrm{Haar},8}$     & Haar-random & $(1,2,3)$
  & $3\pi/4$ & 1  \\

4 & 12  & $U_{\mathrm{Haar},12}$     & Haar-random & $(1,2,3)$
  & $\pi/4$  & 1  \\

5 & 12  & $U_{\mathrm{Haar},12}$     & Haar-random & $(1,2,3)$
  & $\pi/2$  & 1  \\

6 & 12  & $U_{\mathrm{Haar},12}$     & Haar-random & $(1,2,3)$
  & $3\pi/4$ & 1  \\

7 & 16  & $U_{\mathrm{Haar},16}$     & Haar-random & $(1,2,3)$
  & $\pi/4$  & 1  \\

8 & 16  & $U_{\mathrm{Haar},16}$     & Haar-random & $(1,2,3)$
  & $\pi/2$  & 1  \\

9 & 16  & $U_{\mathrm{Haar},16}$     & Haar-random & $(1,2,3)$
  & $3\pi/4$ & 1  \\

10 & 24  & $U_{\mathrm{Haar},24}$     & Haar-random & $(1,2,3)$
  & $\pi/4$  & 1 \\

11 & 24  & $U_{\mathrm{Haar},24}$     & Haar-random & $(1,2,3)$
  & $\pi/2$  & 1  \\

12 & 24  & $U_{\mathrm{Haar},24}$     & Haar-random & $(1,2,3)$
  & $3\pi/4$ & 1  \\

\hline
\end{tabular}
\end{table*}

\FloatBarrier

\subsection{RNE test against the uniform distribution}

We used the RNE test, i.e., the Aaronson--Arkhipov row-norm-estimator test~\cite{aaronson2014bosonsampling}, to reject the uniform distribution. The RNE test computes an efficiently accessible row-norm estimator, and the statistical behaviour of such estimator can deviate significantly from that expected for uniform random outputs. 

The RNE test updates a counter $C$ which is initialized as $C_0 = 0$. For the \(N\)-th experimental event \(S_N\), the counter is updated according to
\begin{equation}
C_N=C_{N-1}
+
\begin{cases}
(+1), & R(S_N)>(n/m)^n,\\
(-1), & R(S_N)\leq (n/m)^n,
\end{cases}
\end{equation}
where \(m\) is the number of network dimensions and \(n\) is the number of input particles. $R(S)$ is the row-norm estimator, defined as 
\begin{equation}
R(S)=
\prod_{i=1}^{n}
\left(
\sum_{j=1}^{n}|A_{ij}|^2
\right),
\end{equation}
where \(A\) is a submatrix extraced from the transformation \(U\) according to the input configuration \(I\) and output configuration \(S\). 
If the experimental RNE counter separates rapidly from the uniform reference trace, it indicates that the measured samples are not from uniform random outputs.

\begin{figure*}[!t]
\centering
\includegraphics[width=\textwidth]{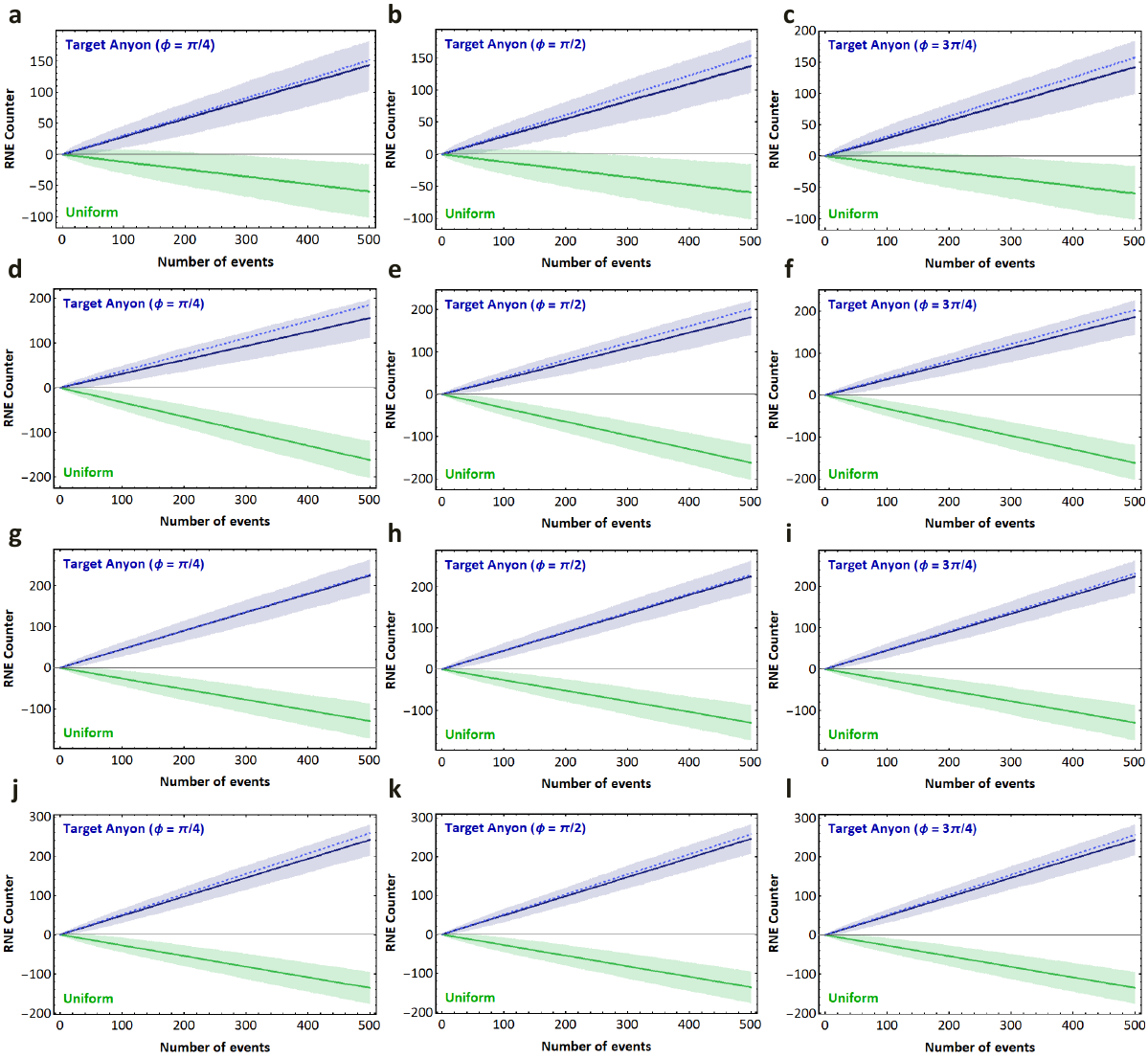}
\caption{
\textbf{Additional RNE tests against uniform sampling.}
\textbf{a--c}, RNE validation of the target anyonic distribution in 8-mode Haar random interferometer against the uniform-distribution hypothesis, at $\phi=\pi/4$, $\phi=\pi/2$, and $\phi=3\pi/4$ respectively. 
\textbf{d--f}, RNE validation of the target anyonic distribution in 12-mode Haar random interferometer against the uniform-distribution hypothesis, at $\phi=\pi/4$, $\phi=\pi/2$, and $\phi=3\pi/4$ respectively. 
\textbf{g--i}, RNE validation of the target anyonic distribution in 16-mode Haar random interferometer against the uniform-distribution hypothesis, at $\phi=\pi/4$, $\phi=\pi/2$, and $\phi=3\pi/4$ respectively. 
\textbf{j--l}, RNE validation of the target anyonic distribution in 24-mode Haar random interferometer against the uniform-distribution hypothesis, at $\phi=\pi/4$, $\phi=\pi/2$, and $\phi=3\pi/4$ respectively. 
For all the tests, the experimental solid curve shows the pointwise median over Poisson-resampled and randomly permuted experimental realizations. The target-anyon dashed curve and the alternative-model solid curve show the corresponding pointwise median trajectories. The shaded experimental and alternative-model regions denote the pointwise $95\%$ quantile intervals. } 
\label{fig:additional-RNE-tests}
\end{figure*}

Beyond the representative RNE trace shown in the main text, we present additional uniformity-rejection tests in Haar-random interferometers of 8, 12, 16 and 24 modes, as shown in Fig.~\ref{fig:additional-RNE-tests}. For each transformation, fractional exchange phases $\phi$=$\pi$/4, $\pi$/2 and 3$\pi$/4 were tested. In all cases, the experimental RNE counter developed a persistent positive drift and separated rapidly from the confidence interval obtained from uniform random samples. The same conclusion was obtained for different random permutations of the measured event sequence, demonstrating that the rejection of uniform sampling is not specific to a particular interferometer, phase setting or output-scanning order.

\FloatBarrier

\subsection{LR test against the fully distinguishable-particle distribution}

We applied the likelihood-ratio test~\cite{spagnolo2014experimental} (denoted as LR test) to reject the fully distinguishable-particle distribution. 
For each output event \(S\), the likelihood ratio between the target quantum particle sampling distribution \(P_{\phi}(S)\) and the fully distinguishable-particle distribution \(P_{\rm dis}(S)\) as
\begin{equation}
L(S)=
\frac{P_{\phi}(S)}
{P_{\rm dis}(S)}.
\end{equation}
A threshold likelihood-ratio counter is initialilzed as $C_0=0$, and it is updated as 
\begin{equation}
C_N=C_{N-1}+w(L(S_N)),
\end{equation}
for the $N$-th event, where
\begin{equation}
w(L)=
\begin{cases}
+2, & L\geq a_2,\\
+1, & 1/a_1\leq L<a_2,\\
0, & a_1<L<1/a_1,\\
-1, & 1/a_2<L\leq a_1,\\
-2, & L\leq 1/a_2.
\end{cases}
\end{equation}
The parameters \(a_1=0.9\) and \(a_2=1.5\) define weak- and strong-evidence thresholds, respectively~\cite{he2017time,spagnolo2014experimental}. 
The same thresholds and update rule are applied to both the experimental data and the simulated data drawn from the fully distinguishable-particle model. 

\begin{figure*}[!t]
\centering
\includegraphics[width=\textwidth]{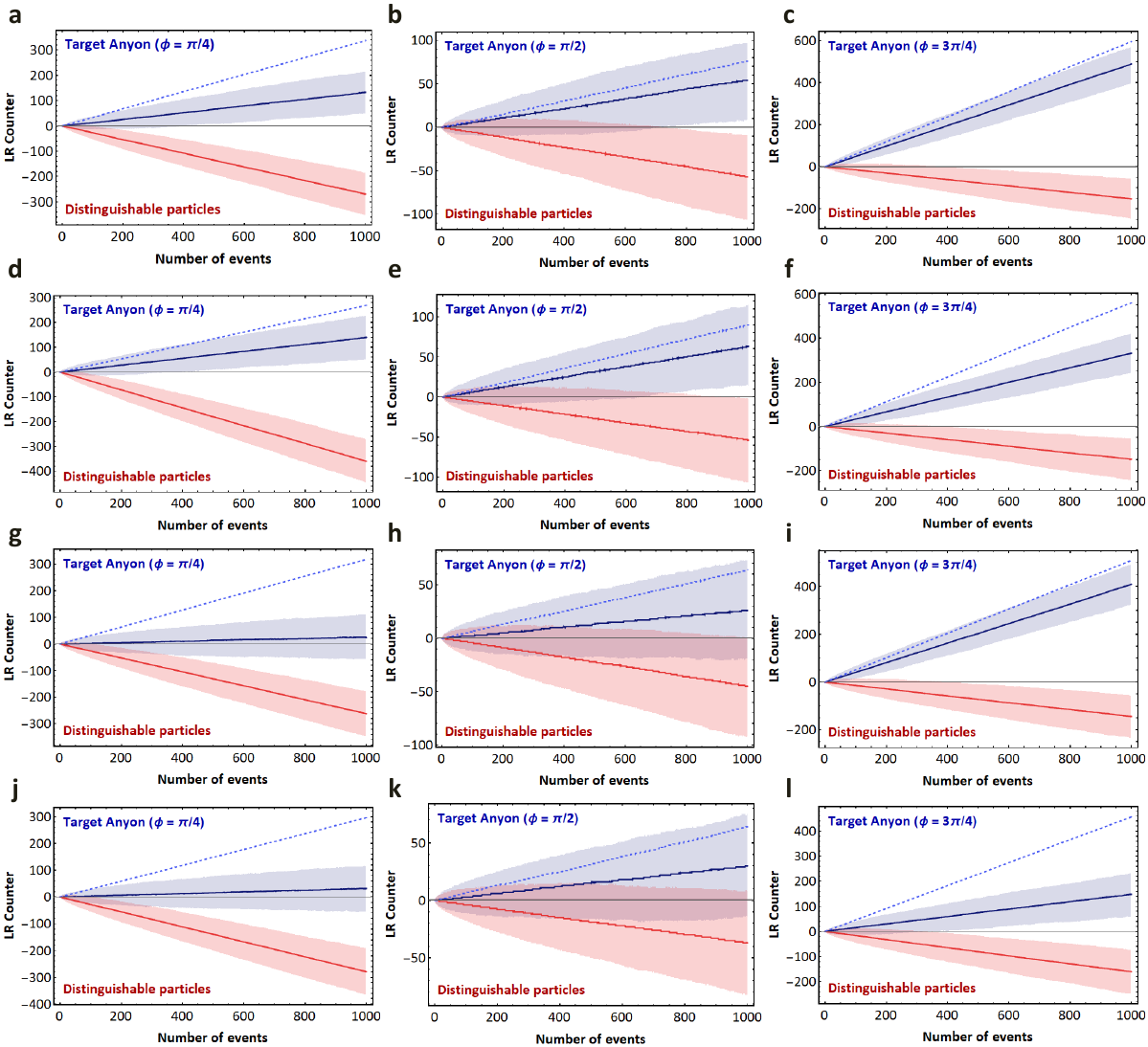}
\caption{
\textbf{Likelihood-ratio tests against the fully
distinguishable-particle distribution.}
\textbf{a--c}, LR validation of the target anyonic distribution in 8-mode Haar random interferometer against the fully distinguishable-particle hypothesis, at $\phi=\pi/4$, $\phi=\pi/2$, and $\phi=3\pi/4$ respectively. 
\textbf{d--f}, LR validation of the target anyonic distribution in 12-mode Haar random interferometer against the fully distinguishable-particle hypothesis, at $\phi=\pi/4$, $\phi=\pi/2$, and $\phi=3\pi/4$ respectively. 
\textbf{g--i}, LR validation of the target anyonic distribution in 16-mode Haar random interferometer against the fully distinguishable-particle hypothesis, at $\phi=\pi/4$, $\phi=\pi/2$, and $\phi=3\pi/4$ respectively. 
\textbf{j--l}, LR validation of the target anyonic distribution in 24-mode Haar random interferometer against the fully distinguishable-particle hypothesis, at $\phi=\pi/4$, $\phi=\pi/2$, and $\phi=3\pi/4$ respectively. 
For all the tests, the experimental solid curve shows the pointwise median over Poisson-resampled and randomly permuted experimental realizations. The target-anyon dashed curve and the alternative-model solid curve show the corresponding pointwise median trajectories. The shaded experimental and alternative-model regions denote the pointwise $95\%$ quantile intervals. 
}
\label{fig:additional-LR-tests}
\end{figure*}

We present additional LR tests in Haar-random interferometers of 8, 12, 16 and 24 modes, as shown in Fig.~\ref{fig:additional-LR-tests}. All tests show the thresholded likelihood-ratio counter consistently favoured the target model as the number of experimental events increased. By contrast, simulated samples drawn from the corresponding $P_{{\rm dis}}(S)$ followed the distinguishable-particle reference traces.

\FloatBarrier

\subsection{Bayesian discrimination between anyonic and bosonic sampling}

We performed Bayesian model discrimination (denoted as Bayesian test) between the target anyonic distribution and the bosonic limiting distribution~\cite{bentivegna2015bayesian,dai2020bayesian}. 
For \(N\) experimental events, $\mathcal{D}_N=\{S_1,S_2,\ldots,S_N\}$, we define the Bayes factor between the target anyonic hypothesis $\{ H_{\phi}: P_{\phi}(S)\}$ and the bosonic hypothesis $\{H_0: P_0(S) \}$: 
\begin{equation}
B_{\phi/0}(N)
=
\prod_{k=1}^{N}
\frac{P_{\phi}(S_k)}
{P_0(S_k)},
\end{equation}
and compute its logarithm, 
\begin{equation}
\log B_{\phi/0}(N)
=
\sum_{k=1}^{N}
\log
\frac{P_{\phi}(S_k)}
{P_0(S_k)}.
\end{equation}
Assuming equal prior probabilities for the two hypotheses, the posterior probability that the data originate from the target anyonic sampler is
\begin{equation}
{\rm Pr}
\left(
H_{\phi}\mid \mathcal{D}_N
\right)
=
\frac{B_{\phi/0}(N)}
{1+B_{\phi/0}(N)}.
\end{equation}
For the the target anyonic sampling experiment, the Bayes factor in favour of the target anyonic model increases rapidly with the number of samples, and the corresponding posterior probability approaches unity.

\begin{figure*}[!t]
\centering
\includegraphics[width=\textwidth]{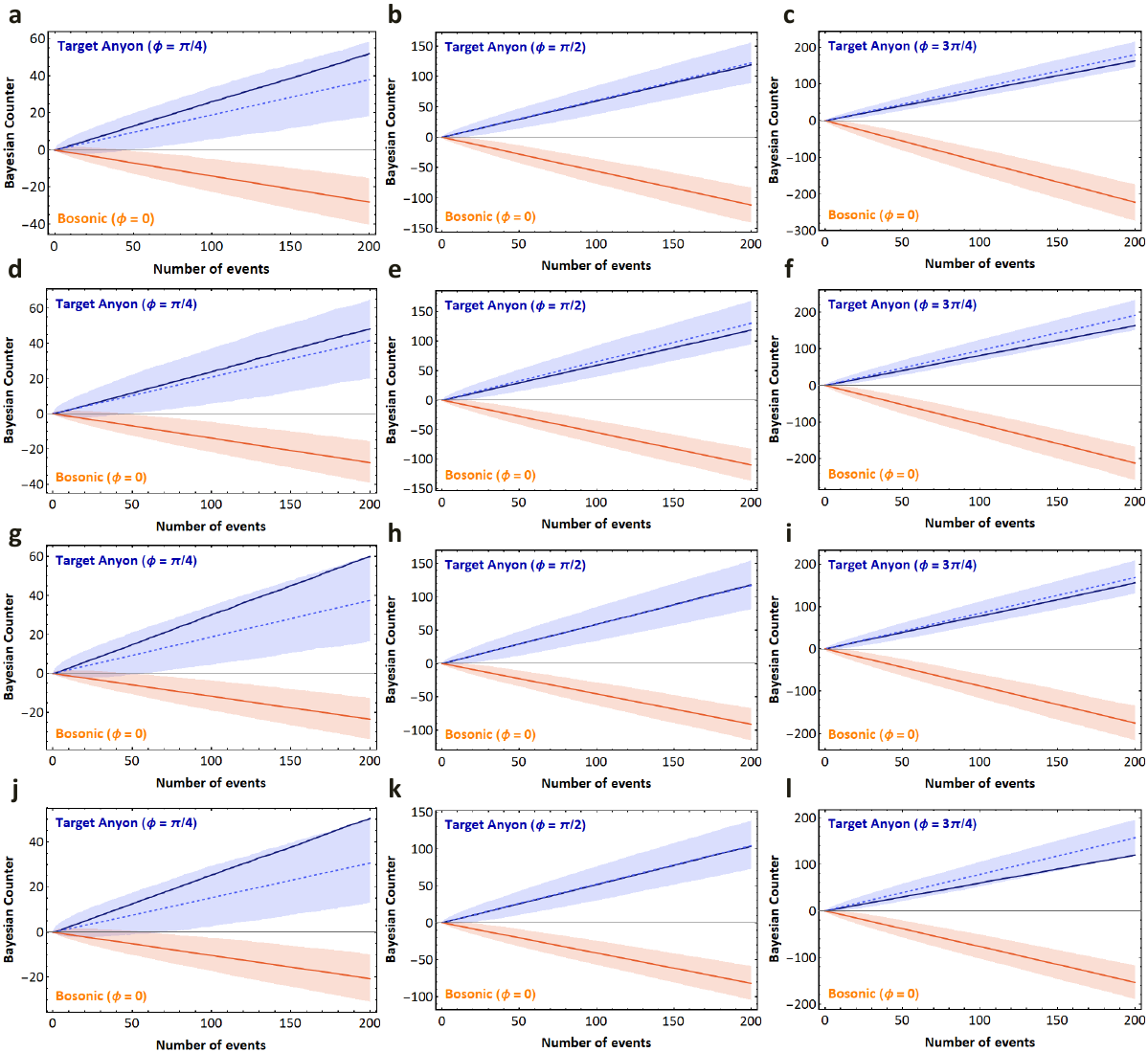}
\caption{
\textbf{Bayesian model comparisons of the target anyonic distribution against the bosonic hypothesis.}
\textbf{a--c}, accumulated Bayes factors comparing the target anyonic distribution in 8-mode Haar random interferometer against the bosonic distribution ($\phi=0$), at $\phi=\pi/4$, $\phi=\pi/2$, and $\phi=3\pi/4$ respectively. 
\textbf{d--f}, accumulated Bayes factors comparing the target anyonic distribution in 12-mode Haar random interferometer against the bosonic distribution ($\phi=0$), at $\phi=\pi/4$, $\phi=\pi/2$, and $\phi=3\pi/4$ respectively. 
\textbf{g--i}, accumulated Bayes factors comparing the target anyonic distribution in 16-mode Haar random interferometer against the bosonic distribution ($\phi=0$), at $\phi=\pi/4$, $\phi=\pi/2$, and $\phi=3\pi/4$ respectively. 
\textbf{j--l}, accumulated Bayes factors comparing the target anyonic distribution in 24-mode Haar random interferometer against the bosonic distribution ($\phi=0$), at $\phi=\pi/4$, $\phi=\pi/2$, and $\phi=3\pi/4$ respectively. 
For all the tests, the experimental solid curve shows the pointwise median over Poisson-resampled and randomly permuted experimental realizations. The target-anyon dashed curve and the alternative-model solid curve show the corresponding pointwise median trajectories. The shaded experimental and alternative-model regions denote the pointwise $95\%$ quantile intervals. }
\label{fig:Bayes-anyon-boson-trace}
\end{figure*}

\begin{figure*}[!t]
\centering
\includegraphics[width=\textwidth]{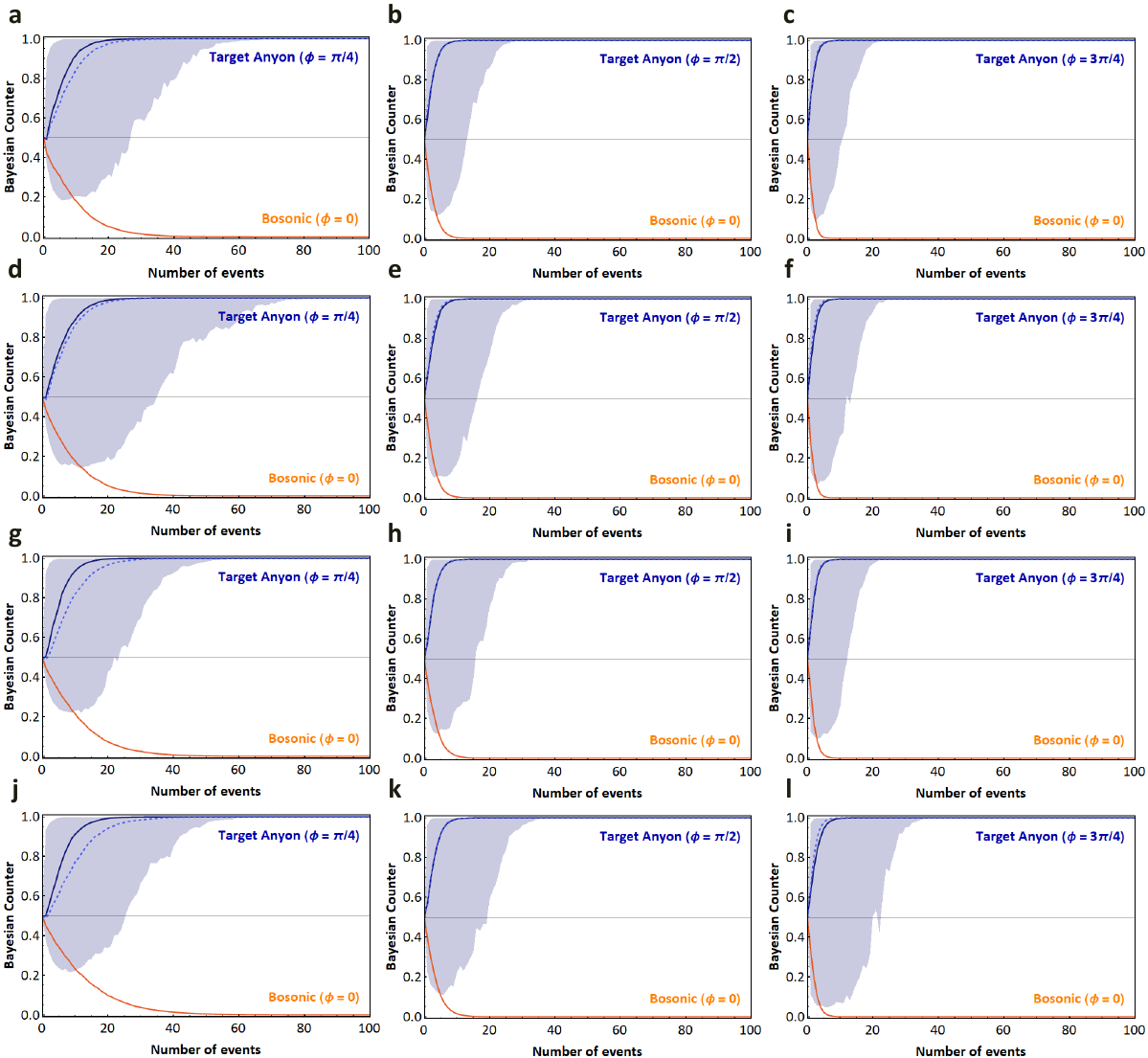}
\caption{
\textbf{Posterior probability assigned to the target anyonic hypothesis.}
\textbf{a--c}, corresponding posterior probability assigned to the target anyonic distribution in 8-mode Haar random interferometer against the bosonic distribution ($\phi=0$), at $\phi=\pi/4$, $\phi=\pi/2$, and $\phi=3\pi/4$ respectively. 
\textbf{d--f}, corresponding posterior probability assigned to the target anyonic distribution in 12-mode Haar random interferometer against the bosonic distribution ($\phi=0$), at $\phi=\pi/4$, $\phi=\pi/2$, and $\phi=3\pi/4$ respectively. 
\textbf{g--i}, corresponding posterior probability assigned to the target anyonic distribution in 16-mode Haar random interferometer against the bosonic distribution ($\phi=0$), at $\phi=\pi/4$, $\phi=\pi/2$, and $\phi=3\pi/4$ respectively. 
\textbf{j--l}, corresponding posterior probability assigned to the target anyonic distribution in 24-mode Haar random interferometer against the bosonic distribution ($\phi=0$), at $\phi=\pi/4$, $\phi=\pi/2$, and $\phi=3\pi/4$ respectively. 
For all the tests, the experimental solid curve shows the pointwise median over Poisson-resampled and randomly permuted experimental realizations. The target-anyon dashed curve and the alternative-model solid curve show the corresponding pointwise median trajectories. The shaded experimental and alternative-model regions denote the pointwise $95\%$ quantile intervals. }
\label{fig:Bayes-anyon-boson-posterior}
\end{figure*}

We present additional Bayesian comparisons between the target anyonic distribution and the bosonic limiting distribution were performed for $\phi$ = $\pi$/4, $\pi$/2 and 3$\pi$/4 in multiple Haar-random interferometers of 8, 12, 16 and 24 modes. We first present the accumulated Bayes factor comparing the target anyonic distribution with the bosonic limiting distribution as a function of the number of detected events $N$ for each test, as shown in Fig.~\ref{fig:Bayes-anyon-boson-trace}. For all tests, the accumulated Bayes factors converged towards the target anyonic hypothesis. 
For a more direct probabilistic interpretation, we convert the accumulated Bayes factor into the posterior probability ${\rm Pr} \left( H_{\phi}\mid \mathcal{D}_N
\right)$, assuming equal prior probabilities for the two comparing models. The corresponding posterior trajectories are shown in Fig.~\ref{fig:Bayes-anyon-boson-posterior}. The results show that the corresponding posterior probability converged towards the target anyonic hypothesis for all tests.

\FloatBarrier

\subsection{Bayesian discrimination between anyonic and fermionic sampling}
We performed Bayesian model discrimination between the target anyonic distribution \(P_{\phi}(S)\) and the fermionic limiting distribution \(P_{\pi}(S)\) in the collision-free output subspace. 
For \(N\) experimental events, $\mathcal{D}_N=\{S_1,S_2,\ldots,S_N\}$, the corresponding Bayes factor is
\begin{equation}
B_{\phi/\pi}(N)
=
\prod_{k=1}^{N}
\frac{P_{\phi}(S_k)}
{P_{\pi}(S_k)},
\end{equation}
and its logarithm is 
\begin{equation}
\log B_{\phi/\pi}(N)
=
\sum_{k=1}^{N}
\log
\frac{P_{\phi}(S_k)}
{P_{\pi}(S_k)}.
\end{equation}
With equal priors, the posterior probability of the target anyonic hypothesis is
\begin{equation}
{\rm Pr}
\left(
H_{\phi}\mid \mathcal{D}_N
\right)
=
\frac{B_{\phi/\pi}(N)}
{1+B_{\phi/\pi}(N)}.
\end{equation}
For the the target anyonic sampling experiment, the experimental data will favour the target anyonic distribution over the fermionic limiting distribution. 

\begin{figure*}[!t]
\centering
\includegraphics[width=\textwidth]{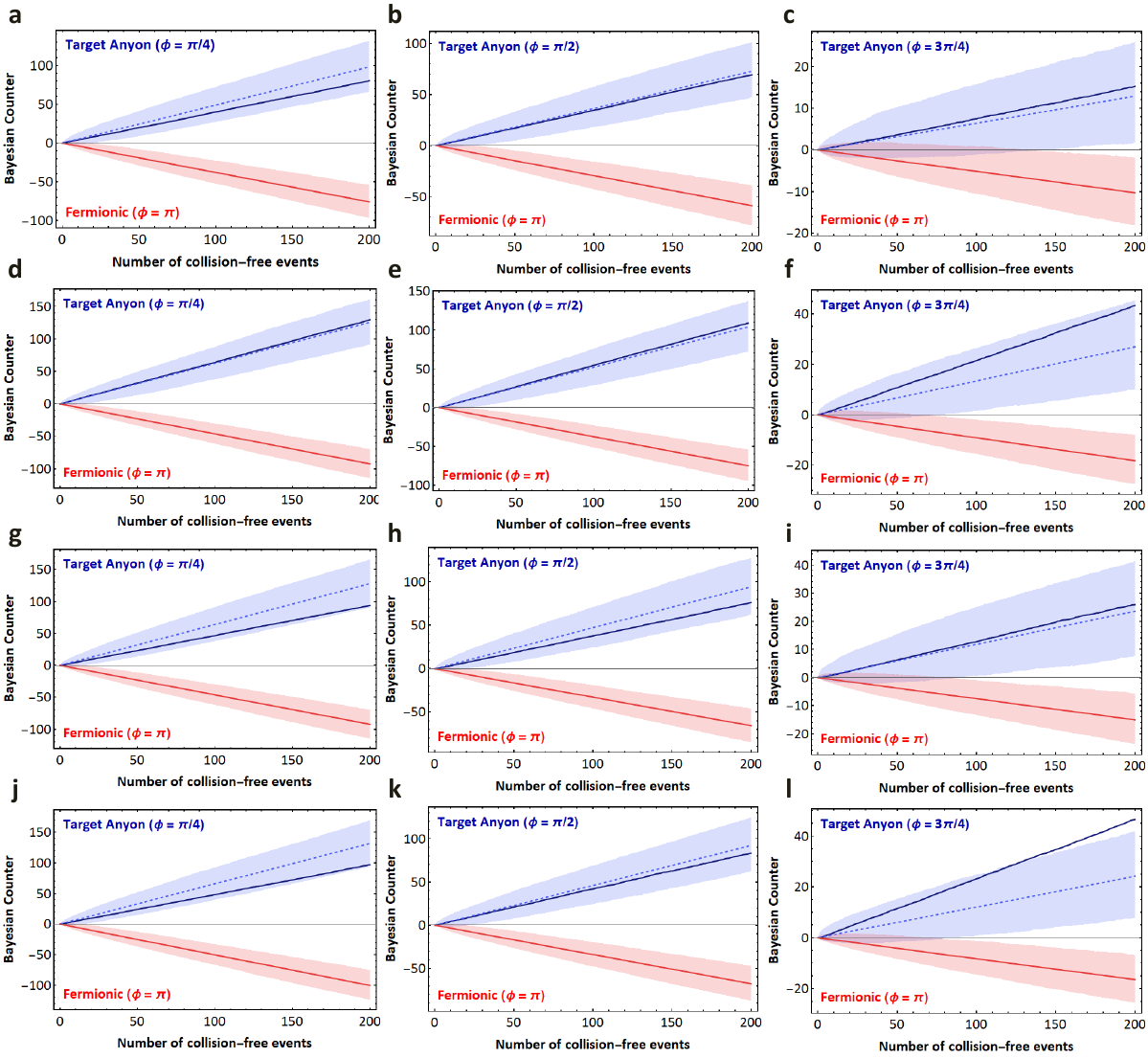}
\caption{
\textbf{Bayesian model comparisons of the target anyonic distribution against the fermionic hypothesis.}
\textbf{a--c}, accumulated Bayes factors comparing the target anyonic distribution in 8-mode Haar random interferometer against the fermionic distribution ($\phi=\pi$), at $\phi=\pi/4$, $\phi=\pi/2$, and $\phi=3\pi/4$ respectively. 
\textbf{d--f}, accumulated Bayes factors comparing the target anyonic distribution in 12-mode Haar random interferometer against the fermionic distribution ($\phi=\pi$), at $\phi=\pi/4$, $\phi=\pi/2$, and $\phi=3\pi/4$ respectively. 
\textbf{g--i}, accumulated Bayes factors comparing the target anyonic distribution in 16-mode Haar random interferometer against the fermionic distribution ($\phi=\pi$), at $\phi=\pi/4$, $\phi=\pi/2$, and $\phi=3\pi/4$ respectively. 
\textbf{j--l}, accumulated Bayes factors comparing the target anyonic distribution in 24-mode Haar random interferometer against the fermionic distribution ($\phi=\pi$), at $\phi=\pi/4$, $\phi=\pi/2$, and $\phi=3\pi/4$ respectively. 
For all the tests, the experimental solid curve shows the pointwise median over Poisson-resampled and randomly permuted experimental realizations. The target-anyon dashed curve and the alternative-model solid curve show the corresponding pointwise median trajectories. The shaded experimental and alternative-model regions denote the pointwise $95\%$ quantile intervals. }
\label{fig:Bayes-anyon-fermion-trace}
\end{figure*}

\begin{figure*}[!t]
\centering
\includegraphics[width=\textwidth]{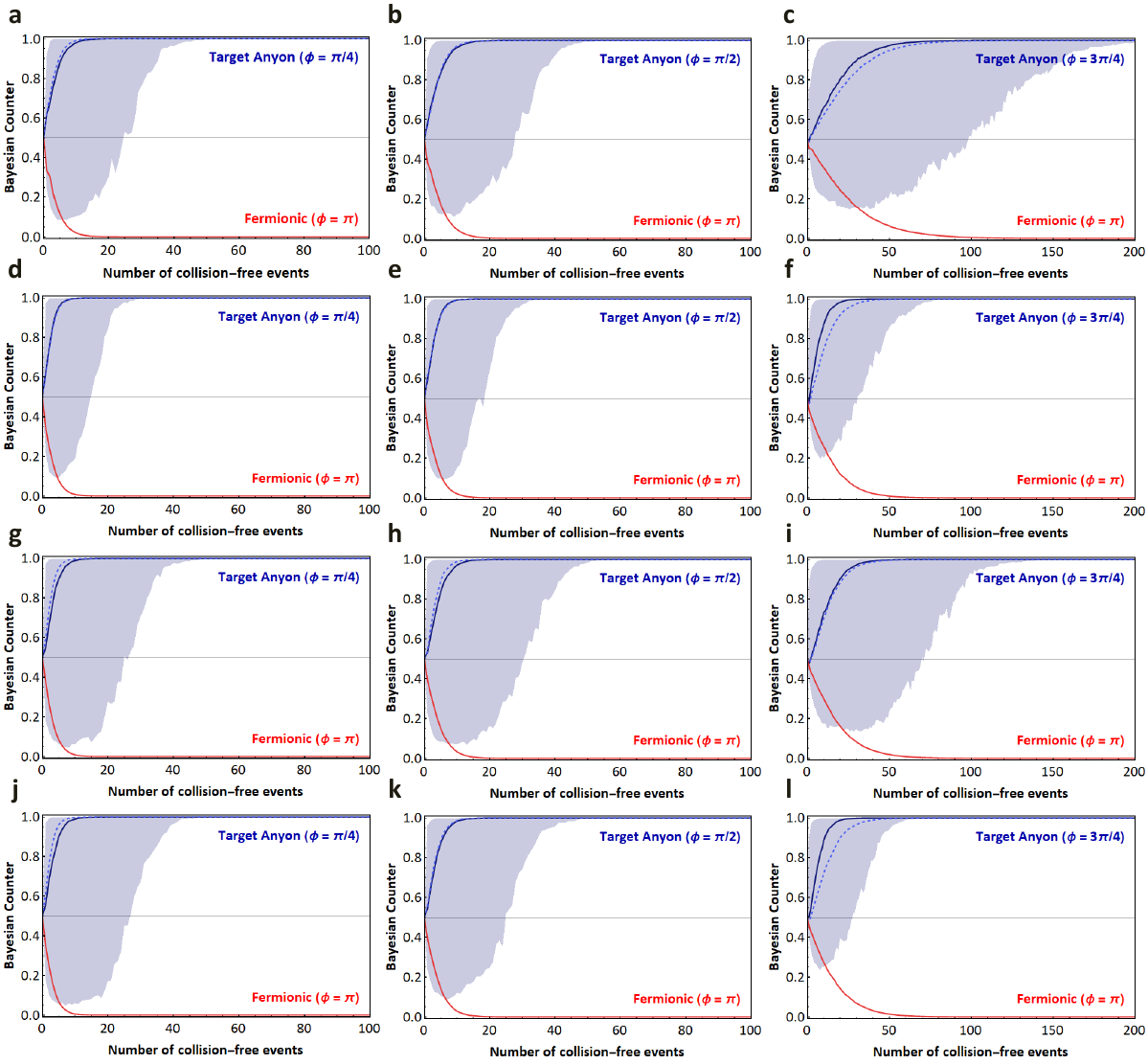}
\caption{
\textbf{Posterior probability assigned to the target anyonic hypothesis.}
\textbf{a--c}, corresponding posterior probability assigned to the target anyonic distribution in 8-mode Haar random interferometer against the fermionic distribution ($\phi=\pi$), at $\phi=\pi/4$, $\phi=\pi/2$, and $\phi=3\pi/4$ respectively. 
\textbf{d--f}, corresponding posterior probability assigned to the target anyonic distribution in 12-mode Haar random interferometer against the fermionic distribution ($\phi=\pi$), at $\phi=\pi/4$, $\phi=\pi/2$, and $\phi=3\pi/4$ respectively. 
\textbf{g--i}, corresponding posterior probability assigned to the target anyonic distribution in 16-mode Haar random interferometer against the fermionic distribution ($\phi=\pi$), at $\phi=\pi/4$, $\phi=\pi/2$, and $\phi=3\pi/4$ respectively. 
\textbf{j--l}, corresponding posterior probability assigned to the target anyonic distribution in 24-mode Haar random interferometer against the fermionic distribution ($\phi=\pi$), at $\phi=\pi/4$, $\phi=\pi/2$, and $\phi=3\pi/4$ respectively. 
For all the tests, the experimental solid curve shows the pointwise median over Poisson-resampled and randomly permuted experimental realizations. The target-anyon dashed curve and the alternative-model solid curve show the corresponding pointwise median trajectories. The shaded experimental and alternative-model regions denote the pointwise $95\%$ quantile intervals. }
\label{fig:Bayes-anyon-fermion-posterior}
\end{figure*}

We present additional Bayesian comparisons between the target anyonic distribution and the fermionic limiting distribution were performed for $\phi$ = $\pi$/4, $\pi$/2 and 3$\pi$/4 in multiple Haar-random interferometers of 8, 12, 16 and 24 modes. 
For each test instance, both candidate distributions were renormalized over the common collision-free output subspace, and only collision-free experimental events were retained in the Bayesian update. 
We first present the accumulated Bayes factor comparing the target anyonic distribution with the fermionic limiting distribution as a function of the number of detected events $N$ for each test, as shwn in Fig.~\ref{fig:Bayes-anyon-fermion-trace}. For all tests, the accumulated Bayes factors converged towards the target anyonic hypothesis. 
The corresponding posterior probability ${\rm Pr} \left( H_{\phi}\mid \mathcal{D}_N \right)$ trajectories are shown in Fig.~\ref{fig:Bayes-anyon-fermion-posterior}. The results show that the corresponding posterior probability converged towards the target anyonic hypothesis rather than fermionic for all tests.

\FloatBarrier

\subsection{Bayesian discrimination between different anyonic phases}

We extended the phase-resolved Bayesian test to the pairs of anyonic distributions characterized by different exchange phases. Consider the two hypotheses $\{ H_{\phi_1}: P_{\phi_1}(S) \}$ and $\{ H_{\phi_2}: P_{\phi_2}(S) \}$, the corresponding Bayes factor is defined as 
\begin{equation}
B_{\phi_1 / \phi_2}(N)
=
\prod_{k=1}^{N}
\frac{P_{\phi_1}(S_k)}
{P_{\phi_2}(S_k)},
\end{equation}
and its logrithm is 
\begin{equation}
\log B_{\phi_1 / \phi_2}(N)
=
\sum_{k=1}^{N}
\log
\frac{P_{\phi_1}(S_k)}
{P_{\phi_2}(S_k)}.
\end{equation}
A positive value of \(\log B_{\phi_1 / \phi_2}\) favours the \(H_{\phi_1}\) hypothesis, whereas a negative value favours the \(H_{\phi_2}\) hypothesis. 
With equal priors, the posterior probability of the target anyonic hypothesis is
\begin{equation}
{\rm Pr}
\left(
H_{\phi}\mid \mathcal{D}_N
\right)
=
\frac{B_{\phi_1/\phi_2}(N)}
{1+B_{\phi_1/\phi_2}(N)}.
\end{equation}
For the the target anyonic sampling experiment, the experimental data will favour the target anyonic distribution over the alternative anyonic distribution.

\begin{figure*}[!t]
\centering
\includegraphics[width=\textwidth]{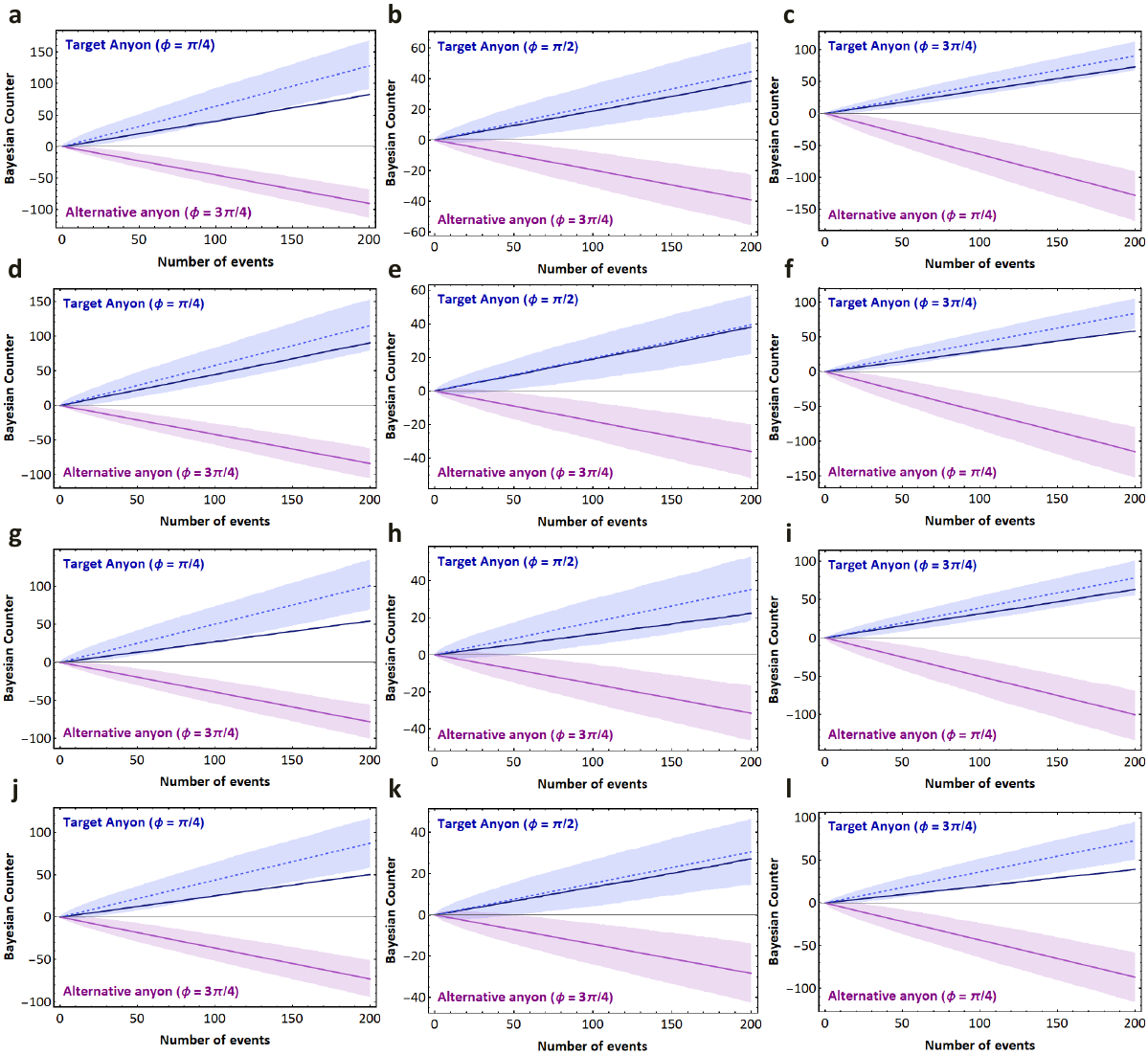}
\caption{
\textbf{Bayesian model comparisons of the target anyonic distribution against the alternative anyonic hypothesis.}
\textbf{a--c}, accumulated Bayes factors comparing the target anyonic distribution in 8-mode Haar random interferometer against an alternative anyonic distribution, at $\phi=\pi/4$, $\phi=\pi/2$, and $\phi=3\pi/4$ respectively. 
\textbf{d--f}, accumulated Bayes factors comparing the target anyonic distribution in 12-mode Haar random interferometer against an alternative anyonic distribution, at $\phi=\pi/4$, $\phi=\pi/2$, and $\phi=3\pi/4$ respectively. 
\textbf{g--i}, accumulated Bayes factors comparing the target anyonic distribution in 16-mode Haar random interferometer against an alternative anyonic distribution, at $\phi=\pi/4$, $\phi=\pi/2$, and $\phi=3\pi/4$ respectively. 
\textbf{j--l}, accumulated Bayes factors comparing the target anyonic distribution in 24-mode Haar random interferometer against an alternative anyonic distribution, at $\phi=\pi/4$, $\phi=\pi/2$, and $\phi=3\pi/4$ respectively. 
For all the tests, the experimental solid curve shows the pointwise median over Poisson-resampled and randomly permuted experimental realizations. The target-anyon dashed curve and the alternative-model solid curve show the corresponding pointwise median trajectories. The shaded experimental and alternative-model regions denote the pointwise $95\%$ quantile intervals. }
\label{fig:Bayes-conjugate-phases-trace}
\end{figure*}

\begin{figure*}[!t]
\centering
\includegraphics[width=\textwidth]{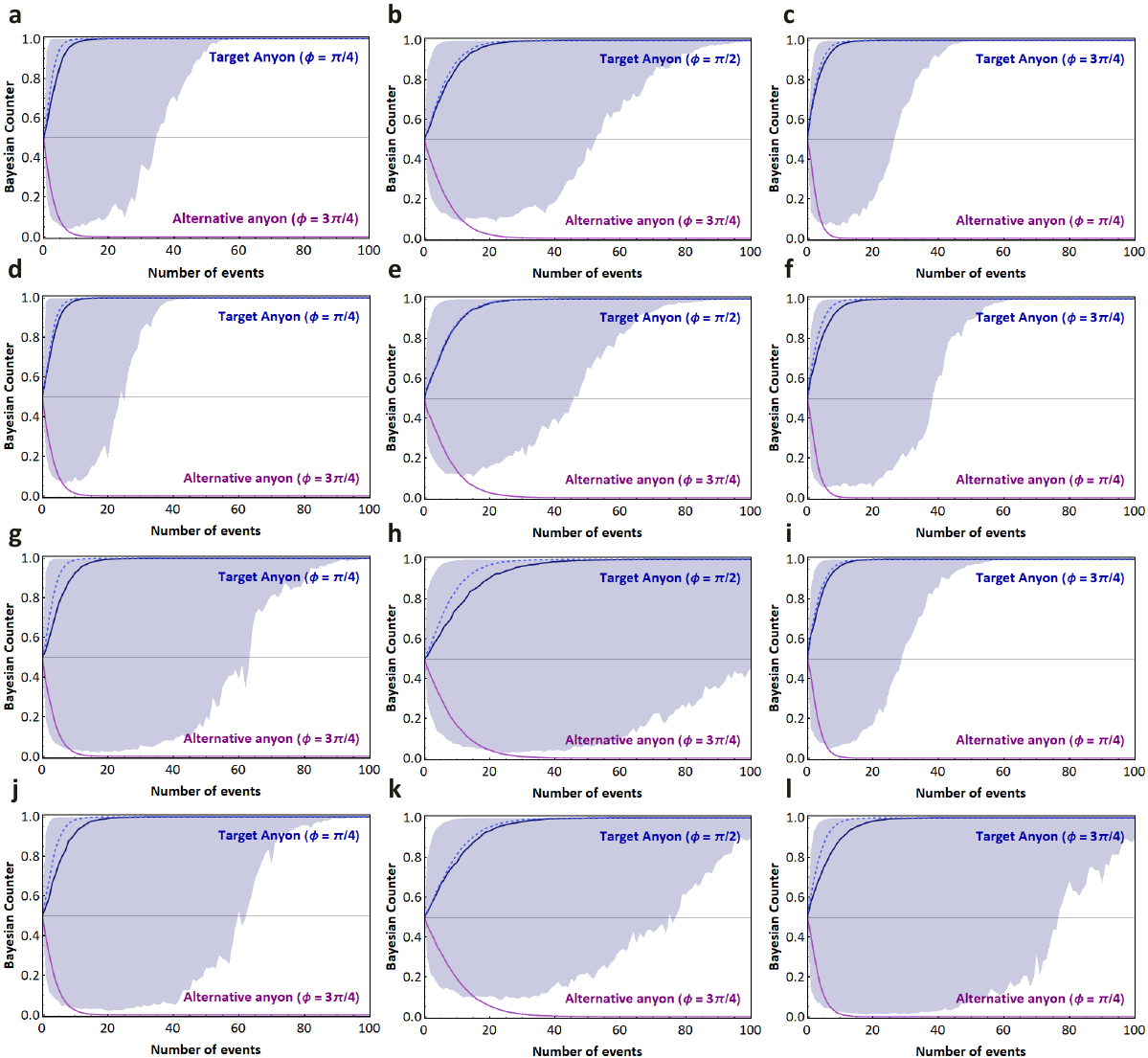}
\caption{
\textbf{Posterior probability assigned to the target anyonic hypothesis.}
\textbf{a--c}, corresponding posterior probability assigned to the target anyonic distribution in 8-mode Haar random interferometer against an alternative anyonic distribution, at $\phi=\pi/4$, $\phi=\pi/2$, and $\phi=3\pi/4$ respectively. 
\textbf{d--f}, corresponding posterior probability assigned to the target anyonic distribution in 12-mode Haar random interferometer against an alternative anyonic distribution, at $\phi=\pi/4$, $\phi=\pi/2$, and $\phi=3\pi/4$ respectively. 
\textbf{g--i}, corresponding posterior probability assigned to the target anyonic distribution in 16-mode Haar random interferometer against an alternative anyonic distribution, at $\phi=\pi/4$, $\phi=\pi/2$, and $\phi=3\pi/4$ respectively. 
\textbf{j--l}, corresponding posterior probability assigned to the target anyonic distribution in 24-mode Haar random interferometer against an alternative anyonic distribution, at $\phi=\pi/4$, $\phi=\pi/2$, and $\phi=3\pi/4$ respectively. 
For all the tests, the experimental solid curve shows the pointwise median over Poisson-resampled and randomly permuted experimental realizations. The target-anyon dashed curve and the alternative-model solid curve show the corresponding pointwise median trajectories. The shaded experimental and alternative-model regions denote the pointwise $95\%$ quantile intervals. }
\label{fig:Bayes-conjugate-phases-posterior}
\end{figure*}

We present additional Bayesian comparisons between the target anyonic distribution and the alternative anyonic distribution in multiple Haar-random interferometers of 8, 12, 16 and 24 modes. For each case, we test three pairs of different exchange phases: \{$\pi$/4, 3$\pi$/4\}, \{$\pi$/2, 3$\pi$/4\}, and \{3$\pi$/4, $\pi$/4\}. For each test, the accumulated Bayes factor comparing the target anyonic distribution with the alternative anyonic distribution as a function of the number of detected events $N$ was shown in Fig.~\ref{fig:Bayes-conjugate-phases-trace}. For all tests, the accumulated Bayes factors converged towards the target anyonic hypothesis. 
The corresponding posterior probability ${\rm Pr} \left( H_{\phi_1}\mid \mathcal{D}_N \right)$ trajectories are shown in Fig.~\ref{fig:Bayes-conjugate-phases-posterior}. The results show that the corresponding posterior probability converged towards the correct anyonic hypothesis rather than the other anyonic for all tests. These measurements demonstrate that the processor resolves phase-specific fractional-statistics interference rather than merely reproducing a coarse interpolation between the bosonic and fermionic limits.

\FloatBarrier

\section{Further details of work distribution estimation via particle sampling}

\subsection{Framework of $N$-particle quantum work distribution}

We consider a quantum system with $N$ particles is driven by an external work parameter $\lambda$ with the time evolution. The system at time $t=0$  is initially prepared  as a thermal equilibrium state at inverse temperature $\beta$ and then is detached from the heat bath. We study the dynamics of the system when the work parameter is changed from the initial value $\lambda_0$ at time $t=0$ to the ending value $\lambda_\tau$ at time $t=\tau$. During this nonequilibrium process, the work distribution of the system can be described by \cite{talkner2007fluctuation,jarzynski2015quantum}
\begin{align}
\begin{split}         \label{wd:eq1}
\rho(W)=&\sum_{i}\sum_{f}P\big(|i^{\lambda_0}: n_{i}^{\lambda_0}\rangle\big)   \\&
\times P\big(|i^{\lambda_0}: n_{i}^{\lambda_0}\rangle\rightarrow|f^{\lambda_\tau}: n_{f}^{\lambda_\tau}\rangle\big)   \\&
\times \delta\big(W-n_{f}^{\lambda_\tau}E_{f}^{\lambda_\tau}+n_{i}^{\lambda_0}E_{i}^{\lambda_0}\big).
\end{split}
\end{align}
Here $i^{\lambda_0}$ ($f^{\lambda_\tau}$) denotes the quantum number at time $t=0$ ($t=\tau$), and $n_{i}^{\lambda_0}$ ($n_{f}^{\lambda_\tau}$) denotes the occupied number of the particles in the $i$th ($f$th) eigenstate at time $t=0$ ($t=\tau$). $E_{i}^{\lambda_0}$ ($E_{f}^{\lambda_\tau}$) denotes the $i$th ($f$th) eigenenergy at initial (final) time. $\sum_{i}n_{i}^{\lambda_0}\equiv \sum_{f}n_{f}^{\lambda_\tau}\equiv N$ due to the conservation of total particle number. The work distribution  described in Eq. (\ref{wd:eq1}) is determined by two factors. The first one is the initial thermal probability distribution $P\big(|i^{\lambda_0}:  n_{i}^{\lambda_0}\rangle\big)=\frac{1}{Z^{\lambda_0}}\text{exp}(-\beta n_{i}^{\lambda_0}E_{i}^{\lambda_0})$, where $Z^{\lambda_0}=\sum_{i}\text{exp}(-\beta n_{i}^{\lambda_0}E_{i}^{\lambda_0})$  is a partition function. The other one is the multi-particle transition probability $P\big(|i^{\lambda_0}: n_{i}^{\lambda_0}\rangle\rightarrow|f^{\lambda_\tau}: n_{f}^{\lambda_\tau}\rangle\big)$ between the initial and the final eigenstates, which can be constructed by the single-particle transition probability.

The transition amplitude between single-particle eigenstates is defined as $\langle f^{\lambda_\tau}|\hat{U}(t)|i^{\lambda_0}\rangle$, where the time evolution operator $\hat{U}(t)$ satisfies the Schr\"{o}dinger equation $i\hbar\partial_t\hat{U}(t)=H(t)\hat{U}(t)$. Due to the interference of multiple identical particles, the identical $N$-particle transition probability can be expressed as \cite{tichy2012many,tichy2014interference,gong2014interference}
\begin{eqnarray}              \label{wd:eq2}
\begin{aligned}
P\big(|i^{\lambda_0}: n_{i}^{\lambda_0}\rangle\rightarrow|f^{\lambda_\tau}: n_{f}^{\lambda_\tau}\rangle\big)=\prod_{f=1}\frac{1}{n_{f}^{\lambda_\tau}!}\prod_{i=1}\frac{1}{n_{i}^{\lambda_0}!}\bigg|\sum_{\sigma\in S_N}e^{i\phi\ell(\sigma)}\prod_{j=1}^N\Lambda^\text{(\textbf{I,F})}_{j,\sigma(j)}\bigg|^2.
\end{aligned}
\end{eqnarray}
Here $\Lambda=(\langle f^{\lambda_\tau}|\hat{U}(t)|i^{\lambda_0}\rangle)$ is a transition amplitude matrix. \textbf{I} is an initial distribution vector of particle number $n_{i}^{\lambda_0}$ in the every eigenstate $i$ ($i=1, 2, \ldots$) and \textbf{F} is an final distribution vector of particle number $n_{f}^{\lambda_\tau}$ in the  every eigenstate $f$ ($f=1, 2, \ldots$). Matrix $\Lambda^\text{(\textbf{I,F})}$ denotes an $N\times N$ sub-matrix of $\Lambda$, whose row entries are from the
$n_{f}^{\lambda_\tau}$ copies of the $f$th row of $\Lambda$ and column entries are from $n_{i}^{\lambda_0}$ copies of the $i$th column of $\Lambda$. $S_N$ is a permutation group with $N$ elements, and $\sigma$ is a permutation of the $S_N$. $\ell(\sigma)$ denotes the minimum number of neighbouring swaps that maps the permutation $\sigma$ to natural order permutation. Phase $\phi$ denotes the exchange statistics of the identical particles. The identical particles in three-dimensional space are either bosons ($\phi=0$) or fermions ($\phi=\pi$), which obey the Bose-Einstein statistics and Fermi-Dirac statistics, respectively. The transition probability of multiple bosons and fermions is related to the \emph{permanent} and \emph{determinant} of a corresponding matrix. While anyons ($0<\phi<\pi$) interposed between bosons and fermions obey the fractional statistics, and whose transition probability is related to the \emph{imminent} of the matrix \cite{matthews2013observing}.

\subsection{Method of grouped probability estimation for simulating work distribution}

In this text, we explain that general particle sampling can efficiently simulate the work distribution of multiple particles. Although the output probability of particle sampling is related to the fermionant of the matrix, this does not imply that the fermionant can be effectively estimated using particle sampling. As the network dimension or the number of particles increases, the probability of each output event decreases exponentially. Estimating such a small probability would require an exponentially large sample size, which is impractical. However, some researches \cite{huh2015boson,wang2020efficient,oh2024quantum} have shown that if these output events are grouped based on certain features and their probabilities are summed, the resulting sum is polynomial rather than exponentially small. This suggests that exponentially large samples may not be necessary. This method of grouped probability estimation opens a new way to simulate some classically difficult problems by using particle sampling.

We next describe the method of grouped probability estimation and demonstrate that the method is feasible for simulating the work distribution
by analyzing the required sample number. According to the delta function in Eq. \eqref{wd:eq1}, we can group the photon output outcomes and sum their probabilities if $W=m_f E_f^{\lambda_\tau} -n_i E_i^{\lambda_0}$. Then the grouped result is given by
\begin{align}
\begin{split}             \label{eq24}
G(W)&=\sum_{\textbf{m}}P(\textbf{m}|\textbf{n})\delta\big(W-m_f E_f^{\lambda_\tau}+n_i E_i^{\lambda_0}\big)\\&
=\sum_{\textbf{m}\in\mathcal{G}(W)}P(\textbf{m}|\textbf{n}).
\end{split}
\end{align}
Here $P(\textbf{m}|\textbf{n})$ denotes the transition probability that input photon distribution is $\textbf{n}=(n_1, \ldots, n_M)$ and output photon distribution is $\textbf{m}=(m_1, \ldots, m_M)$, which is equivalent to $P\big(|i^{\lambda_0}: n_{i}^{\lambda_0}\rangle\rightarrow|f^{\lambda_\tau}: n_{f}^{\lambda_\tau}\rangle\big)$. Each $n_i$ ($m_i$) denotes the input (output) photon number in the $i$th optical mode. $M$ is the dimension of the optical network $U$. The sets $\mathcal{G}(W)=\{\mathbf{m}\in\mathbb{Z}^M_{\geq0} | m_f E_f^{\lambda_\tau} -n_i E_i^{\lambda_0}=W\}$ correspond to each grouped outcome
$W\in\{0, \ldots, W_{\text{max}}\}$. We introduce two integer energy vectors $\mathbf{E_i}\in \mathbb{Z}^M_{\geq0}$ and $\mathbf{E_f}\in \mathbb{Z}^M_{\geq0}$, where each element of the vectors $E_{i}\leq \mathcal{O}\big(\text{poly}(M)\big)$ and $E_{f}\leq \mathcal{O}\big(\text{poly}(M)\big)$  ($i,f=1, \ldots, M$) are at most an integer of a polynomial with respect to $M$. It is not difficult to find that the initial eigenenergy $E_i^{\lambda_0}$ and final eigenenergy $E_f^{\lambda_\tau}$ are floating-point number, which can be written as $E_{i}$ and $E_{f}$ respectively by multiplying by a large number. This means that the total number of different groups is $W_\text{max}+1 \leq \mathcal{O}\big(\text{poly}(M)\big)$. Since the total sum of probabilities is 1, the probabilities for each group is $|G(W)|\geq \mathcal{O}\big(1/\text{poly}(M)\big)$ rather than exponentially small number. Based on grouped outcome, the work distribution in Eq. \eqref{wd:eq1} can be expressed by
\begin{align}
\begin{split}         \label{eq25}
\rho(W)=&\sum_{i}P\big(|i^{\lambda_0}: n_{i}^{\lambda_0}\rangle\big)G(W).
\end{split}
\end{align}
This expression shows that the work distribution can be effectively simulated using at most a polynomial number of samples rather than an exponential number of samples. In summary, compared to the method of estimating individual event probability separately, the grouped probability estimation turns the required number of samples into a polynomial, which confirms the experimental feasibility of the scheme.

\subsection{Identical particles work distribution in a 1D harmonic potential}
\emph{\textbf{Harmonic oscillator.}}
Here we consider an example to explain how to simulate the work distribution by the particle sampling. We consider a one-dimensional (1D) quntum harmonic oscillator with a time-dependent angular frequency $\omega_t$, whose Hamiltonian is described by
\begin{eqnarray}              \label{eq3}
\begin{aligned}
H(x,t)&=\frac{p^2}{2m}+\frac{m}{2}\omega_t^2x^2\\&
=-\frac{\hbar^2}{2m}\frac{d^2}{dx^2}+\frac{m}{2}\omega_t^2x^2.
\end{aligned}
\end{eqnarray}
Here $m$ (setting $m=1$) is the single-particle mass. The angular frequency $\omega_t$ as the work parameter is changed from $\omega_0$ at initial time $t=0$ to $\omega_\tau$ at ending time $t=\tau$. We denote $\psi_n^t$ is instantaneous eigenstate of Hamiltonian $H(x,t)$ at any time $t$ and the corresponding eigenvalue is $E_n^t=\hbar\omega_t(n+\frac{1}{2})$. We assume that the oscillator is initially thermalized at inverse temperature $\beta=1/(k_BT)$. The single-particle time evolution operator is given by \cite{husimi1953miscellanea,deffner2008nonequilibrium},
\begin{eqnarray}              \label{eq4}
\begin{aligned}
U_t(x,x_0)=\sqrt{\frac{m}{2\pi i\hbar X_t}}\text{exp}\bigg(\frac{im}{2\hbar X_t}(\dot{X_t}x^2-2xx_0+Y_tx_0^2)\bigg).
\end{aligned}
\end{eqnarray}
Here $X_t$ and $Y_t$ are the solutions of motion equation
\begin{eqnarray}              \label{wd:eq5}
\begin{aligned}
\ddot{\xi_t}+\omega_t^2\xi_t=0,
\end{aligned}
\end{eqnarray}
which satisfy the boundary conditions $X_{t=0}=0$, $\dot{X}_{t=0}=1$ and $Y_{t=0}=1$, $\dot{Y}_{t=0}=0$.  An explicit analytical solution of Eq. (\ref{wd:eq5}) for the linear quench
\begin{eqnarray}              \label{wd:eq6}
\begin{aligned}
\omega_t^2=\omega_0^2+(\omega_\tau^2-\omega_0^2)t/\tau
\end{aligned}
\end{eqnarray}
is shown as below.

The solutions are given by Airy A and Airy B functions \cite{husimi1953miscellanea,deffner2008nonequilibrium}.
\begin{eqnarray}              \label{A1}
\begin{aligned}
X_t=&\frac{\pi\tau^{1/3}}{[(\omega_0-\omega_\tau)(\omega_0+\omega_\tau)]^{1/3}}\bigg[-\text{Ai}\bigg(\frac{(t-\tau)\omega_0^2-t\omega_\tau^2}{\tau^{1/3}[(\omega_0-\omega_\tau)(\omega_0+\omega_\tau)]^{2/3}}\bigg)
\text{Bi}\bigg(-\frac{\omega_0^2 \tau^{2/3}}{[(\omega_0-\omega_\tau)(\omega_0+\omega_\tau)]^{2/3}}\bigg) \\&
+\text{Ai}\bigg(-\frac{\omega_0^2 \tau^{2/3}}{[(\omega_0-\omega_\tau)(\omega_0+\omega_\tau)]^{2/3}}\bigg)
\text{Bi} \bigg(\frac{(t-\tau)\omega_0^2-t\omega_\tau^2}{\tau^{1/3}[(\omega_0-\omega_\tau)(\omega_0+\omega_\tau)]^{2/3}}\bigg)\bigg],
\end{aligned}
\end{eqnarray}
\begin{eqnarray}              \label{A2}
\begin{aligned}
Y_t=&\pi\bigg[-\text{Ai}'\bigg(-\frac{\omega_0^2 \tau^{2/3}}{[(\omega_0-\omega_\tau)(\omega_0+\omega_\tau)]^{2/3}}\bigg)
\text{Bi}\bigg(\frac{(t-\tau)\omega_0^2-t\omega_\tau^2}{\tau^{1/3}[(\omega_0-\omega_\tau)(\omega_0+\omega_\tau)]^{2/3}}\bigg) \\&
+\text{Ai}\bigg(\frac{(t-\tau)\omega_0^2-t\omega_\tau^2}{\tau^{1/3}[(\omega_0-\omega_\tau)(\omega_0+\omega_\tau)]^{2/3}}\bigg)
\text{Bi}' \bigg(-\frac{\omega_0^2 \tau^{2/3}}{[(\omega_0-\omega_\tau)(\omega_0+\omega_\tau)]^{2/3}}\bigg)\bigg].
\end{aligned}
\end{eqnarray}
Here Ai$'(x)$ and Bi$'(x)$ are the first derivatives with respect to $x$.

The single-particle transition amplitude between initial and final eigenstates is expressed by
\begin{eqnarray}              \label{wd:eq7}
\begin{aligned}
U_{f,i}^\tau=\langle f^{\lambda_\tau}|\hat{U}|i^{\lambda_0}\rangle=\int dx\int dx_0\psi_f^{\tau*}(x)U_t(x,x_0)\psi_i^0(x_0).
\end{aligned}
\end{eqnarray}

\emph{\textbf{Single-particle transition amplitude matrix in a 1D harmonic potential.}}
Here we show a specific expression of the single-particle transition amplitude matrix.
The 1D harmonic potential  has been studied well \cite{husimi1953miscellanea}. Here we summary the single-particle transition amplitude in a 1D harmonic potential  \cite{husimi1953miscellanea,deffner2008nonequilibrium,deffner2010quantum}. The single-particle transition amplitude between initial and final eigenstates can be give  by $U_{f,i}^\tau$ in Eq. \eqref{wd:eq7}. We introduce the complex variables
\begin{eqnarray}              \label{eqB1}
\begin{aligned}
&\zeta=\omega_0\omega_\tau X_\tau-\omega_0i\dot{X}_\tau+\omega_\tau iY_\tau+\dot{Y}_\tau,\\&
|\zeta|^2=2\omega_0\omega_\tau (Q^*-1),\\&
\sigma=\omega_0\omega_\tau X_\tau-\omega_0i\dot{X}_\tau-\omega_\tau iY_\tau-\dot{Y}_\tau,\\&
|\sigma|^2=2\omega_0\omega_\tau(Q^*+1).
\end{aligned}
\end{eqnarray}
Here
\begin{eqnarray}              \label{eqB2}
\begin{aligned}
Q^*=\frac{1}{2\omega_0\omega_\tau}\big[\omega_0^2(\omega_\tau^2X_\tau^2+\dot{X}_\tau^2)+(\omega_\tau^2Y_\tau^2+\dot{Y}_\tau^2)\big],
\end{aligned}
\end{eqnarray}
and $X_t$ and $Y_t$ are the solutions of motion equation
\begin{eqnarray}              \label{eqB3}
\begin{aligned}
\ddot{\xi_t}+\omega_t^2\xi_t=0,
\end{aligned}
\end{eqnarray}
with the boundary conditions $X_{t=0}=0$, $\dot{X}_{t=0}=1$, $Y_{t=0}=1$,  and $\dot{Y}_{t=0}=0$.  The single-particle transition amplitude matrix can be obtained and the matrix elements result in \cite{husimi1953miscellanea,deffner2010quantum}
\begin{eqnarray}              \label{eqB4}
\begin{aligned}
U_{f,i}^\tau=\sqrt[4]{\omega_0\omega_\tau}\sqrt{\frac{i!f!\zeta^i\zeta^*{^f}}{2^{i+f-1}i\sigma^{i+f+1}}}\sum_{l=0}^{\text{min}(f,i)}\frac{[-2i\sqrt{2/(Q^*-1)}]^l}{l![(i-l)/2]![(f-l)/2]!}.
\end{aligned}
\end{eqnarray}
Note that $U_{f,i}^\tau$=0, if $i$ and $f$ are of different parity. Based on the selection rule $f=i\pm2k$ ($k$ is an integer), $l$ runs only if $i$ and $f$ are of same parity. The expression of $U_{f,i}^\tau$ for $i$ and $f$ both are even number is given by
\begin{eqnarray}              \label{eqB5}
\begin{aligned}
U_{2\mu,2\nu}^\tau=\sqrt{\frac{(2\nu)!(2\mu)!}{2^{2\nu+2\mu-1}i}}\sqrt{\frac{\zeta^{2\nu}\zeta^{*2\mu}}{\sigma^{2\nu+2\mu+1}}}\frac{\sqrt[4]{\omega_0\omega_\tau}}{\Gamma(\mu+1)\Gamma(\nu+1)} \times _2F_1\big(-\mu,-\nu;\frac{1}{2};\frac{2}{1-Q^*}\big).
\end{aligned}
\end{eqnarray}
For $i$ and $f$ both are odd number, the expression of $U_{f,i}^\tau$  is given by
\begin{eqnarray}              \label{eqB6}
\begin{aligned}
U_{2\mu+1,2\nu+1}^\tau=-\sqrt{\frac{8i(2\nu+1)!(2\mu+1)!}{(Q^*-1)2^{2\nu+2\mu+1}}}\sqrt{\frac{\zeta^{2\nu+1}\zeta^{*2\mu+1}}{\sigma^{2\nu+2\mu+3}}}\frac{\sqrt[4]{\omega_0\omega_\tau}}{\Gamma(\mu+1)\Gamma(\nu+1)} \times _2F_1\big(-\mu,-\nu;\frac{3}{2};\frac{2}{1-Q^*}\big).
\end{aligned}
\end{eqnarray}
Here  the hypergeometric function $_2F_1$ is introduced to simplify the Eq. \eqref{eqB4}, and $\Gamma(x)$  denotes the Gamma function.

\subsection{Additional experimental results}

\begin{figure*}[t]
   \centering
    \includegraphics[width=0.98\textwidth]{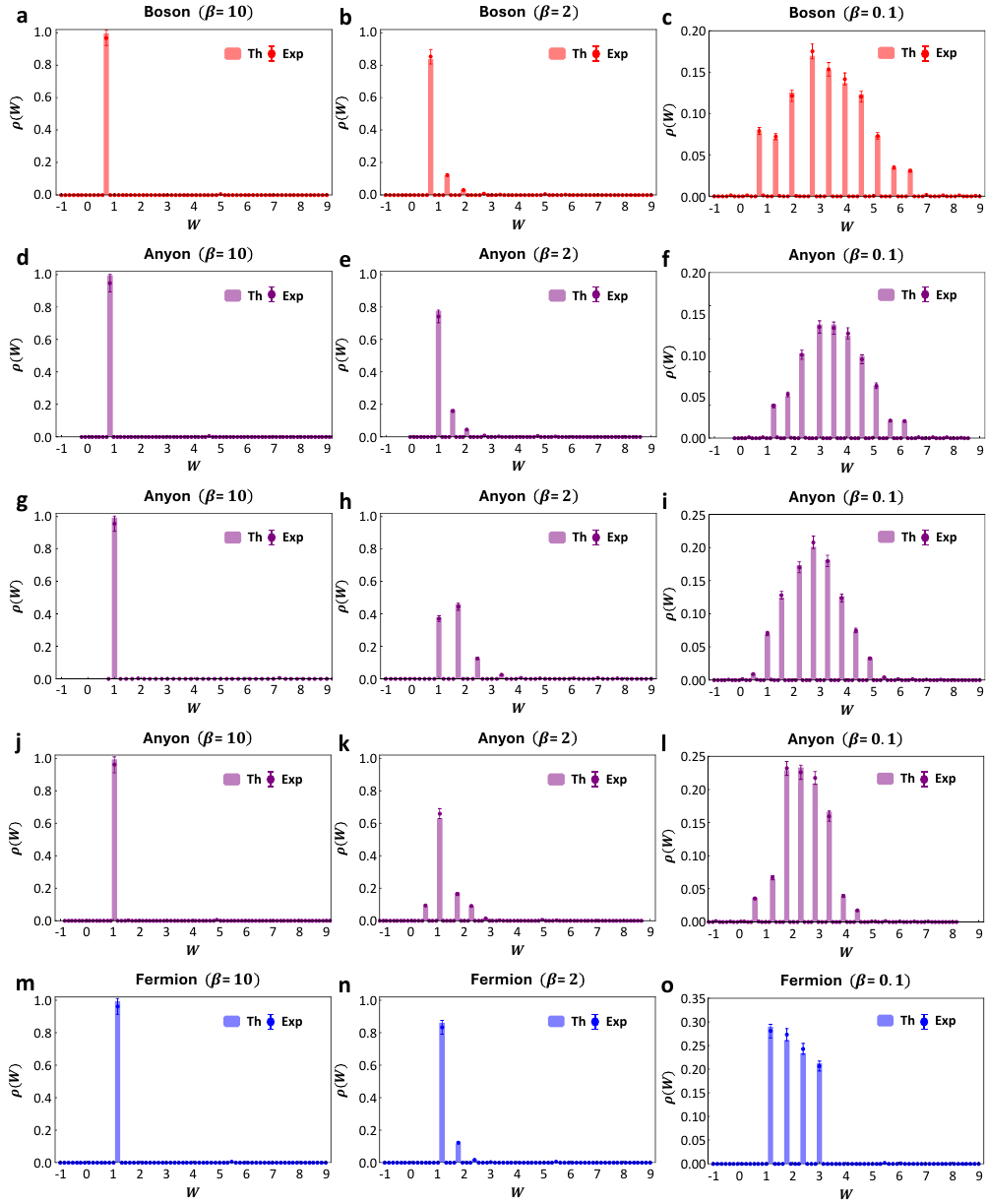}
    \caption{
    \textbf{Quantum work distribution for three particles at different temperatures.}
    \textbf{a--c} Bosons, \textbf{d--f} anyons ($\phi=\pi/4$), \textbf{g--i} anyons ($\phi=\pi/2$), \textbf{j--l} anyons ($\phi=3\pi/4$), and \textbf{g--i} fermions.
    The work parameters $\omega_0=1$, $\omega_1=\sqrt{2}$, and $\tau=2$ are adopted, with $\beta$ ranging from 10 to 0.1.
    Higher temperatures (i.e., smaller $\beta$) lead to a more complex work distribution.
}\label{fig:temperature}
\end{figure*}

The inverse temperature $\beta$ and the driving duration $\tau$ are two key parameters that significantly affect the work distribution of particles. On one hand, within the partition function, $\beta$ governs the initial occupation probabilities of particles across energy eigenstates. As shown in Fig. \ref{fig:temperature}\textbf{a, d, g, j, m}, the work distributions for bosons, anyons, and fermions converge at low temperature in the long-driving-time limit, where the dominant peak corresponds to the ground-state energy throughout the process. This behavior can be interpreted via the quantum adiabatic theorem. As the temperature increases (i.e., as $\beta$ decreases), the work distribution becomes progressively more complex for all particle statistics, ranging from bosons through anyons to fermions. This is because higher temperatures enable particles to populate higher excited states, and transitions between these states become more frequent.

\begin{figure*}[t]
   \centering
    \includegraphics[width=0.98\textwidth]{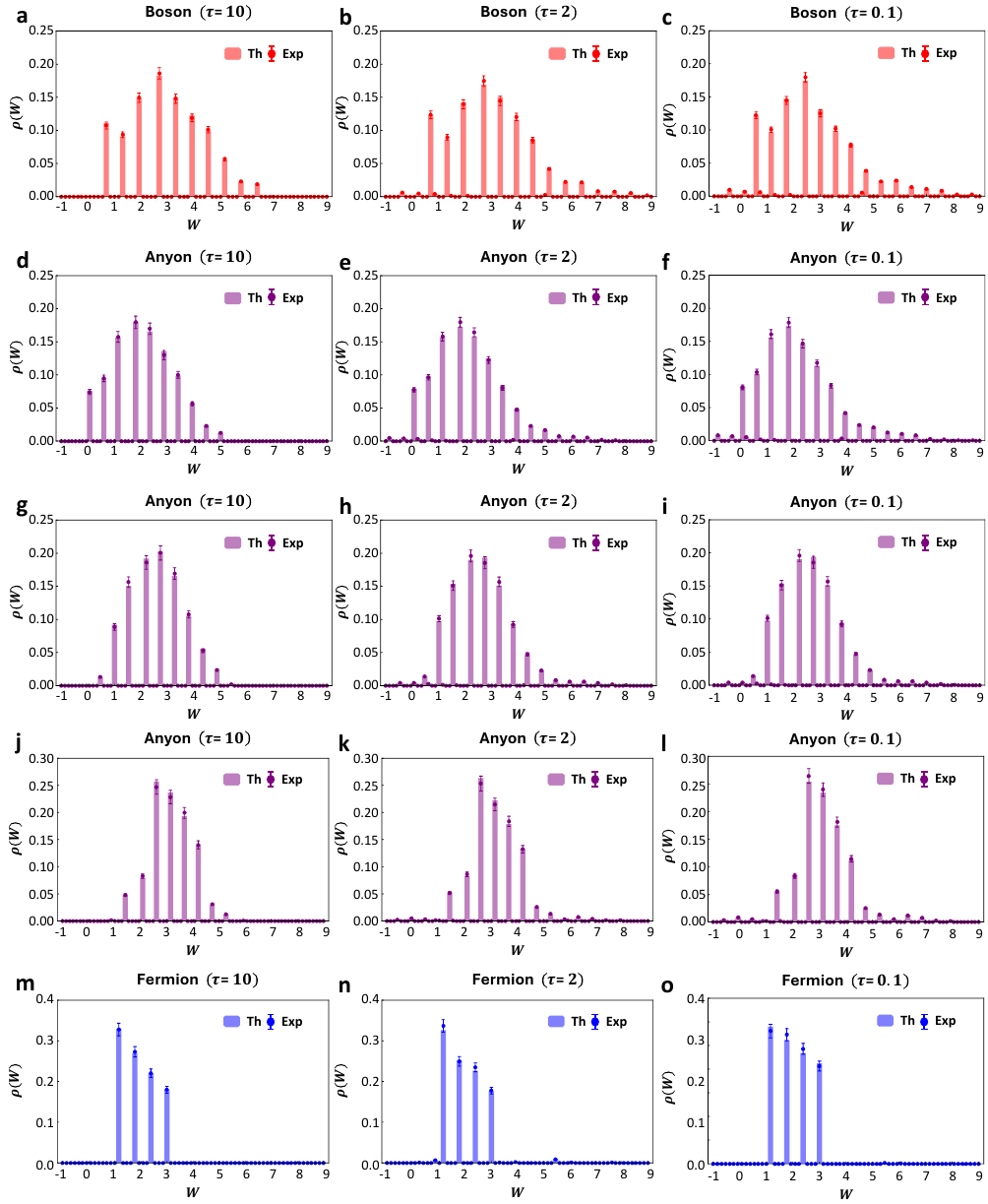}
    \caption{
    \textbf{Quantum work distribution for three particles at different durations.}
    \textbf{a--c} Bosons, \textbf{d--f} anyons ($\phi=\pi/4$), \textbf{g--i} anyons ($\phi=\pi/2$), \textbf{j--l} anyons ($\phi=3\pi/4$), and \textbf{g--i} fermions.
    The work parameters $\omega_0=1$, $\omega_1=\sqrt{2}$, and $\beta=0.2$ are adopted, with driving time $\tau$ ranging from 10 to 0.1.
}\label{fig:time}
\end{figure*}

On the other hand, the driving duration $\tau$ governs both the transition probabilities between many-particle eigenstates and the degree of nonadiabaticity. In the large-$\tau$ adiabatic limit, the distribution concentrates around transitions that follow the instantaneous eigenstates. As $\tau$ decreases, nonadiabatic transitions become more prominent and the work distribution broadens (see Fig. \ref{fig:time}), with this effect being strongest for bosons. This arises because shorter driving times excite the system to higher energy levels and trigger transitions between these excited states.

In Figs. \ref{fig:temperature}--\ref{fig:time}, we observe that fermions extract more work while bosons extract less work, and this distinction becomes more pronounced at high temperatures. This originates from the bunching effect of bosons and the antibunching effect of fermions. At high temperatures, the work distribution of bosons tends to be broader than that of fermions, with the anyon work distribution lying intermediate between the two cases (see $\beta=0.1$ in Fig. \ref{fig:temperature}). We infer that the differences between different particle statistics become more prominent as the particle number increases. Interference among multiple particles plays an important role in determining the work distribution, which renders the calculation of work distributions increasingly challenging with particle number.

\end{document}